\documentclass[aps,preprint]{revtex4}%
\usepackage{amsfonts}
\usepackage{amsmath}
\usepackage{amssymb}
\usepackage{graphicx}%
\providecommand{\U}[1]{\protect\rule{.1in}{.1in}}
\providecommand{\U}[1]{\protect\rule{.1in}{.1in}}
\providecommand{\U}[1]{\protect\rule{.1in}{.1in}}
\providecommand{\U}[1]{\protect\rule{.1in}{.1in}}
\providecommand{\U}[1]{\protect\rule{.1in}{.1in}}
\providecommand{\U}[1]{\protect\rule{.1in}{.1in}}
\providecommand{\U}[1]{\protect\rule{.1in}{.1in}}
\providecommand{\U}[1]{\protect\rule{.1in}{.1in}}
\newtheorem{theorem}{Theorem}

\newtheorem{conjecture}[theorem]{Conjecture}

\newtheorem{definition}[theorem]{Definition}

\newenvironment{proof}[1][Proof]{\noindent\textbf{#1.} }{\ \rule{0.5em}{0.5em}}
\begin{document}
\preprint{ }
\title[ ]{Cosmological Models Without Singularity Based on RW Metric or A New Metric and
their Explanation for Evolution of the Universe}
\author{Shi-Hao Chen}
\affiliation{$^{1}$Department of Physics, Jilin University; $^{2}$Institute of Theoretical
Physics, Northeast Normal University, Changchun 130024, China. shchen@nenu.edu.cn}

\begin{abstract}
A new conjecture is proposed that there are two sorts of matter called
s-matter and v-matter which are symmetric and whose gravitational masses are
opposite to each other, although both masses are positive. Therefore, there
are two sorts of symmetry breaking called V-breaking and S-breaking. In the
S-breaking, s-particles get their masses and form s-galaxies etc., while
v-fermions and v-gauge bosons are still massless and form v-SU(5) singlets.
There is no interaction among the v-SU(5) color-single states except
gravitation so that they distribute loosely in space, cannot be observed and
cause space to expand with an acceleration. When temperature increases to the
critical temperature because space contracts, masses of all particles are zero
so that s-particles and v-particles transform from one to another and the
gravitational mass density becomes negative. Consequently, space stop to
contract and inflation must occur. After reheating, space first expands with a
deceleration and then expands with an acceleration up to now.\ There is no
space-time singularity. There are the critical temperature, the highest
temperature and the least scale in the universe. A formula describing
distance-redshift is obtained. A huge void is not empty, and is equivalent to
a huge concave lens. The densities of hydrogen in the huge voids must be much
less than that predicted by the conventional theory. The gravitation between
two galaxies with distance long enough will be less than that predicted by the
conventional theory. A black hole with its mass and density big enough will
transform into a white hole. Primordial nucleosynthesis and CMB are explained.
It is possible that the universe is composed of infinite cosmic islands. The
problem of energy conservation is discussed.



*Supported by tau-charm physics research KJ(x2-yw-N29); National Natural
Science Foundation of China (No. 11075064).

\end{abstract}
\volumenumber{number}
\issuenumber{number}
\eid{identifier}
\date[Date text]{date}
\received[Received text]{date}

\revised[Revised text]{date}

\accepted[Accepted text]{date}

\published[Published text]{date}

\startpage{1}
\endpage{90}
\maketitle
\tableofcontents

\part{The cosmological model without singularity based on the RW metric}

The cosmological model without singularity can be constructed based on the RW
metric and a new metric. The curvature factor $K$ in the RW metric can be
$-1$, $0$ or $1$. The difference between the new metric and the RW metric is
that $K$ in the new metric is a function of the gravitational mass density in
comoving coordinates. In addition to the major inferences of the first model
(this model based on the RW metric), the second model (this model based on the
new metric) has new inferences. For example, the universe is composed of
infinite s-cosmic islands and v-cosmic islands according to the second model.

The present paper is composed of three parts. The first part describes the
first model, the second part explains primordial nucleosynthesis $CMBR$ and
the large scale structure, the third part describes the second model.

\section{Introduction to the model based on the RW metric}

It is impossible to solve the space-time singularity issue and the
cosmological constant issue in the frame of the conventional theory, hence new
conjectures are necessary. Based on a new essential conjecture, we have
constructed a cosmological model and solve the two issues, well explain the
evolution of the universe, primordial nucleosynthesis, cosmic microwave
background radiation $(CMBR)$ and give new predictions.

As is now well known, there is space-time singularity under certain
conditions$^{[1]}$. These conditions fall into three categories. First, there
is the requirement that gravity shall be attractive. Secondly, there is the
requirement that there is enough matter present in some region to prevent
anything escaping from that region. The third requirement is that there should
be no causality violations. Because of the theorems, there must be space-time
singularity in the conventional theory. On the other hand, there should be no
space-time singularity in physics. Hence this problem must be solved.

Recent astronomical observations show that the universe expanded with a
deceleration earlier while is expanding with an acceleration now. This implies
that there is dark energy. $73$ percent of the total energy density of the
universe is dark energy density$^{[2]}$. What is dark energy? Many possible
answers have been given. One possible interpretation is in terms of the
effective cosmological constant $\lambda_{eff}=$ $\lambda+\rho_{gvac},$ here
$\lambda$ and $\rho_{gvac}$ are the Einstein cosmological constant and the
gravitational mass density of the vacuum state, respectively. According to the
equivalence principle, $\rho_{gvac}=\rho_{vac}$, $\rho_{vac}$ is the energy
density of the vacuum state, hence $\lambda_{eff}$ may be written as
$\lambda+\rho_{vac}.$ $\lambda_{eff}$ cannot be derived from basic
theories$^{[3]}$. and $\rho_{vac}\ggg\lambda_{eff}$. Hence the interpretation
is unsatisfactory. Alternatively, dark energy is associated with the dynamics
of scalar field $\phi\left(  t\right)  $ that is uniform in space$^{[4]}$.
This is a seesaw cosmology$^{[5]}$. Thus, discussion about the universe
expansion with an acceleration is still open to the public.

$\rho_{gvac}=\rho_{vac}\ggg\lambda_{eff}$ originates from the conventional
quantum field theory and the equivalent principle. $\rho_{vac}\ggg
\lambda_{eff}$ and the singularity issue imply that the conventional theory is
not self-consistent. $\rho_{vac}=0$ is a necessary result of our quantum field
theory without divergence$^{[6]}$. In this theory, there is no divergence of
loop corrections as well, and dark matter which can form dark galaxies is
predicted$^{[7]}$. It is different from the supersymmetric quantum field
theory in which $\rho_{vac}=0$ can be obtained in only some models but is not
necessary. Thus, issue of the cosmological constant is open as well, because
$\lambda_{eff}$ is not still determined.

Huge voids in the cosmos have been observed$^{[8]}$. Such a model in which hot
dark matter (e.g. neutrinos) is dominant can explain the phenomenon. However,
it cannot explain the structure with middle and small scales. Why are there
the huge voids? This is an open problem as well.

We consider that all important existing forms of matter (including dark matter
and dark energy) have appeared. Hence the basic problems should be solved. As
mentioned above, we have constructed a quantum field theory without divergence
which predicts that there must be dark matter. We construct a cosmological
model without singularity which can solve the space-time singularity and
cosmological constant issues and explain well the evolution of the universe in
the present paper.

The bases of the present model are the general relativity, the conventional
quantum field theory and grand unified theory ($GUT$).

We consider the following condition to be necessary in order to solve the
space time singularity and the cosmological constant problems on the basis of
the classical cosmology and in the frame of the conventional quantum field theory.

Condition: There are two sorts of matter with positive masses which are
symmetric, whose gravitational masses are opposite to each other, although
whose masses are all positive.

The two sorts of matter are called $s-matter$ and $v-matter$, respectively.
The condition implies that if $\rho_{s}=\rho_{v},$ then $\rho_{gs}=-\rho
_{gv},$ here $\rho_{g}$ denotes a gravitational mass density, $\rho_{s}>0$ and
$\rho_{v}>0$ and there is no negative energy or negative probability at all$.$
The conditions cannot be realized in the conventional theory, but can be
realized in the present model. In order to uniformly solve the above four
problems consistently on one basis, we present a new conjectures equivalent to
the condition and construct two cosmological model, i.e. $\left[  9\right]  $
and this model in the present paper.

The basic idea of the present model is only conjecture 1. The present model
has the following results:

$\mathbf{1.}$ There is no space-time singularity. A theorem related to
singularity is presented.

$\mathbf{2.}$ The evolution of the universe and the relation between the
optical distance and the redshift predicted by the present model are
consistent with the observations up to now.

There are two sorts of spontaneous symmetry breaking in the present model
because of conjecture 1, and they are called $S-breaking$ and $V-breaking$.

According to the present model, the evolving process of space is as follows.

In the $S-breaking$, space contracts so that temperature $T$ rises. When $T$
arrives the critical temperature $T_{cr}$, the universe is in the most
symmetric state with $s-SU(5)\times v-SU(5)$ symmetry. When space continues to
contract so that $T$ arrives the highest temperature $T_{\max}$, space expands
and then inflates. After inflation, the state with the highest symmetry
transits to the state with the $V-breaking$. After reheating, space expands
with a deceleration, then expands with an acceleration up to now.

$\mathbf{3.}$ There are the critical temperature $T_{cr}$, the highest
temperature $T_{\max}\geq T_{cr}$, the least scale $R_{\min}>0$ and the
largest energy density $\rho_{\max}$ in the universe. $R_{\min}$ and $T_{cr}$
are two new important constants, $T_{\max}$ and $\rho_{\max}$ are finite and
determined by $R\left(  T_{cr}\right)  $.

$\mathbf{4.}$ We generalize equations governing nonrelativistic fluid motion
to present model. The equations of structure formation have been derived out.
According to the equations, galaxies can form earlier than that in the
conventional theory.

$\mathbf{5.}$ Three predictions are given.

$\mathbf{6.}$ Primordial nucleosynthesis and cosmic microwave background
radiation are explained.

$\mathbf{7.}$ Dark energy is explained as $s-matter$ in the $V-breaking.$ In
contrast with the dark energy, $\rho_{sg}=-\rho_{s}<0.$

$\mathbf{8.}$ $\rho_{gvac}=\rho_{s,vac}-\rho_{v,vac}=0$ is proven, although
$\rho_{vac}=\rho_{s,vac}+\rho_{v,vac}$ is still very large. Consequently,
$\lambda_{eff}=\lambda=0.$

Problems $4$ and $6$ will be discussed in the following paper.

In section 2, conjectures are presented; In section 3, action, energy-momentum
tensor and field equations are presented; In section 4, spontaneous symmetry
breaking is given, and evolution equations of space are derived; In section 5,
temperature effects are considered; In section 6, contraction of space, the
highest temperature and inflation of space are discussed; In section 7,
evolving process of space after inflation is discussed; In section 8,
expansion of space after inflation is discussed; In section 9, new predictions
and an inference are given; Section 10 is discussion; Section 11 is the conclusions.

\section{Conjectures, action, energy-momentum tensor and field equations}

\subsection{Conjectures}

In order to solve the above problems, we propose the following conjectures:

\begin{conjecture}
There are two sorts of matter with positive masses which are called
$solid-matter$ ($s-matter$) and $void-matter$ ($v-matter$), respectively. Both
are symmetric and their contributions to the Einstein tensor are opposite each
other$.$ There is no other interaction between both except the interaction
$(2.10)$ of $s-Higgs$ $fields$ and $v-Higgs$ $fields$.
\end{conjecture}

\begin{conjecture}
When $SU(5)$ symmetry holds and temperature is low, all particles must exist
in $SU(5)$ color single states.
\end{conjecture}

This conjecture is obviously a direct generalization of $SU(3)$ color single states.

Other premises of the present model are the cosmological principle, the RW
metric ($k=-1$ in the present model) and the conventional $SU(5)$ grand
unified theory $\left(  GUT\right)  .$ But it is easily seen that the present
model does not rely on the special $GUT$. Provided conjecture 1 and such a
coupling as $\left(  2.10\right)  $ are kept$,$ the $GUT$ can be accepted.

All the following inferences hold when $S\rightleftarrows V$ and
$s\leftrightarrows v$ due to the conjecture $1$.

The gravitational properties of matter and the mode of symmetry breaking
determine the features of space-time. We consider that there are only two possibilities.

$\mathbf{A}$. The first possibility can be described by the conventional
theory. There is only one sort of matter so that the equivalence principle
strictly holds. This theory based on the conjecture is simple, but there must
be essential difficulties. For example, there must be the singularity and
cosmological constant issues which cannot be solved in the frame of this
theory because of the Hawking theorems etc.

$\mathbf{B.}$ The basis of the second possibility is conjecture 1.

We explain it in detail as follows:

$\mathbf{1.}$ It must be emphasized that there is no negative mass or negative
probability in the present model at all. Conjecture 1 implies that $\rho
_{sg}=-\rho_{vg}$ when $\rho_{s}=\rho_{v}$. In the $S-breaking$, $\rho
_{sg}=\rho_{s}\geq0$ and $\rho_{vg}=-\rho_{v}\leq0$. Here $\rho_{g}$ denotes a
gravitational mass density. From this we can regard the gravitation charges of
$s-matter$ and $v-matter$ to be opposite, i.e. $\alpha=-\beta=1$, here
$\alpha$\ and $\beta$ are the gravitation charges. The energy-momentum tensor
should be independent of the gravitation charges. Hence it is necessary to
eliminate $\alpha$\ and $\beta$ from the definition of $T_{\mu\nu}$ by the
operator $\left(  \partial/\alpha+\partial/\beta\right)  $. Consequently, both
the $s-energy$ and the $v-energy$ must be positive (see $(2.20)-(2.21)$ in detail).

$\mathbf{2}$. The observation basis of conjecture 1 is that space expands with
an acceleration. One of the two sorts of matter must loosely distribute in
whole space, can cause space to expand with an acceleration and cannot be
observed as so-called dark energy (see 7 below).

$\mathbf{3.}$ Because of conjecture 1, there must be two sorts of symmetry breaking.

Because of conjecture 1, $s-Higgs$ fields and $v-Higgs$ fields must be
symmetric as well. If this symmetry is not broken, $s-matter$ and $v-matter$
will exist in the same form at arbitrary time and place. This implies that
nature is simply duplicate. This is impossible because nature does like
duplicate. Of course, this contradicts experiments and observations.
Consequently the symmetry must be broken, i.e. $\langle\omega_{s}\rangle
_{0}\neq0$ and $\langle\omega_{v}\rangle_{0}=0$ or $\langle\omega_{v}%
\rangle_{0}\neq0$ and $\langle\omega_{s}\rangle_{0}=0$, here $\omega$ denotes
an arbitrary Higgs field. Thus the coupling constant $\Lambda$\ etc. in
$\left(  2.10\right)  $ must be positive so that there must be the two sorts
of breaking.

The existing probability of the $S-breaking$ and the $V-breaking$ must be
equal because the $s-Higgs$ fields and the $v-Higgs$ fields are symmetric.
This equality can be realized by two sorts of modes.

$(1)$ The universe is composed of infinite $s-cosmic$ islands with the
$S-breaking$ and $v-cosmic$ islands with the $V-breaking$; This possibility
will be discussed in the third part (see also Ref. $\left[  9\right]  )$

$(2)$ The whole universe is in the same breaking (e.g. the $S-breaking$). But
one sort of breaking can transform to another as space contracts to the least
scale $R_{\min}$ (see later)$.$ We discuss the case in the present paper. The
RW metric is applicable to the case.

$\mathbf{4.}$ $S-matter$ and $v-matter$ are no longer symmetric after the
symmetry breaking and there is no interaction except the gravitation among
$v-SU(5)$ color-single states.

In the $S-breaking,$ $\langle\omega_{s}\rangle_{0}\neq0$ and $\langle
\omega_{v}\rangle_{0}=0.$ Consequently, $s-SU(5)$ is broken finally to
$s-SU(3)\times U(1)$ and $v-SU(5)$ still strictly holds. Thus, $s-particles$
get their masses and form $s-atoms,$ $s-observers$ and $s-galaxies$ etc.;
while all $v-fermions$ and $v-gauge$ bosons are massless and all $v-particles$
must form $v-SU(5)$ color-single states after reheating because of conjecture 2.

There is no interaction (e.g. electroweak) except the gravitation among
$v-SU(5)$ color-single states, because $SU(5)$ is a simple group. Consequently
the $v-SU(5)$ color-single states cannot form $v-atoms$ etc., and must
distribute loosely in space as the so-called dark energy.

Thus, in the $S-breaking,$ $s-matter$ is identified with the conventional
matter, while $v-matter$ is similar to dark energy, because of the following
reasons $5$ and $6$. In contrast with the dark energy, the gravitational
masses of $v-matter$ is negative$.$

$\mathbf{5.}$ The equivalence principle still strictly holds for the
$s-particles$ ($m_{sg}=m_{s}$), but is violated by the $v-particles$\ $\left(
m_{vg}=-m_{v}\right)  $ in the $S-breaking$. Although it is such, the motion
equations of all $s-particles$ and all $v-particles$ are still independent of
their masses.

In the $S-breaking,$ there are only the $s-observers$ and the $s-galaxies,$
and there is no $v-observer$ and $v-galaxy.$ Hence the gravitational masses of
$s-particles$ must be positive, i.e. $m_{sg}=m_{s}>0,$ while the gravitational
masses of $v-matter$ must be negative relatively to $s-matter,$ i.e.
$m_{vg}=-m_{v}<0,$ because of conjecture 1. Thus, a $s-photon$ falling in a
gravitational field must have purple shift, but a $v-particle$ (there is no
$v-photon$ and there are only $v-SU(5)$ color single states) falling in the
same gravitational field will have `redshift'. Although the equivalence
principle is violated by $v-matter$ in the $S-breaking,$ this result does not
contradict any observation or experiment, because $v-SU(5)$ color-single
states cannot be observed by an $s-observer$ (see $6$ and $7$ below).

$\mathbf{6}$. There is only the repulsion between $s-matter$ and $v-matter$ so
that any bound state is composed of only $s-particles$ or only $v-particles$.

Because of conjecture 1, there is the repulsion between $s-matter$ and
$v-matter$ and the coupling constant of the repulsion is the same as that of
the gravitation.

The interaction $\left(  2.10\right)  $ is repulsive as well. The interaction
$\left(  2.10\right)  $ can be neglected after reheating, because the masses
of the Higgs particles $\Omega_{s}$ and $\Omega_{v}$ are very large after
reheating. Thus, there is no the transformation of $s-particles$ and
$v-particles$ from one into another in low temperatures. The interaction
$\left(  2.10\right)  $ is important when temperature is high enough ($T\sim
T_{cr}$).

$\mathbf{7.}$ The $v-SU(5)$ color-single states cannot be observed by an
$s-observer$ in fact, because of the following reasons.

As mentioned above, the $v-SU(5)$ color-single states cannot form any atom or
any celestial body with a large mass and must distribute loosely in space
because there is no interaction except the gravitation among them. Hence the
repulsion between a $s-body$ and the $v-SU(5)$ color-single states must be too
small for observation.

On the other hand, $\rho_{v}$ must be very small when $\rho_{s}$ is large
because there is the repulsion and there is only the repulsion determined by
conjecture 1 between $s-matter$ and $v-matter$ after reheating.

After reheating, the interaction $\left(  2.10\right)  $ is too small to
observation as well.

In fact, in the $S-breaking$, only the cosmological effects of $v-matter$ are
important and are consistent with the observed data up to now.

It is seen from $5-7$ that although the equivalence principle is violated by
the $SU(5)$ singlets, but there is no contradiction with experiments and
observations up to now.

$\mathbf{8.}$ $\rho_{s}$ and $\rho_{v}$ can transform from one into another by
$\left(  2.10\right)  $ when $T\sim T_{cr},$\ because the expectation values
of all Higgs fields and the masses of all particles are zero in this case.
Consequently, $\rho_{s}$ and $\rho_{v}$ can transform from one into another by
$\left(  10\right)  $ so that $\rho_{s}=\rho_{v}$, $T_{s}=T_{v}\sim T_{cr}$
and the symmetry of $v-SU(5)\times s-SU(5)$\ holds in this case. Thus space
cannot contract to infinite small and inflation must occur.

$\mathbf{9.}$ From $\left(  2.16\right)  -\left(  2.18\right)  $ and $\left(
2.22\right)  $ we see that $T_{g;\nu}^{\mu\nu}=0$ can be derived from the
field equation, but $T_{;\nu}^{\mu\nu}=0$ cannot be derived. This implies that
there is no restriction for $T^{\mu\nu}.$ Thus, it is possible that the
differential law of energy-momentum conservation $T_{,\nu}^{\mu\nu}=0$ holds,
where $T^{\mu\nu}$ may contain the contribution of gravitational field as well.

In summary, in the $S-breaking,$ the$\ v-SU(5)$ color single states cannot be
observed and have only the cosmological effects. Conjecture 1 does not
contradict any experiment and observation up to now.

We will see in the following that the evolution of the universe can be well
explained and the singularity and cosmological constant issues can be solved.

\subsection{Action}

It is impossible that there are simultaneously the $S-breaking$ and the
$V-breaking$ in the same region because of $\left(  2.10\right)  .$ There are
only $s-observers$ and only $I_{S}$\textbf{ }is applicable in the
$S-breaking$, and there are only $v-observers$ and only $I_{V}$\textbf{ }is
applicable in the $V-breaking.$ Hence in any case, the action is unique. But
the universe in the $S-breaking$ can transform to the universe in the the
$V-breaking,$ hence both $I_{S}$ and $I_{V}$ are necessary. Thus, the actions
should be written as two sorts of form, $I_{S}$ in the $S-breaking$ and
$I_{V}$ in the $V-breaking$. Because of conjecture 1, the structures of
$I_{S}$ and $I_{V}$ are the same, i.e. $I_{S}\rightleftarrows I_{V}$ when
$S\rightleftarrows V$ and $s\rightleftarrows v$. Thus, at the zero-temperature
we have%

\begin{align}
I_{V}  &  =I_{g}+I_{VM}=I_{S}\left(  s\leftrightarrows v,S\longrightarrow
V\right)  ,\text{ \ }I_{S}=I_{g}+I_{SM},\tag{2.1}\\
I_{g}  &  =\frac{1}{16\pi G}\left(  \int_{\Sigma}R\sqrt{-g}d^{4}%
x+2\int\nolimits_{\partial\Sigma}K\sqrt{\pm h}d^{3}x\right)  ,\tag{2.2}\\
I_{SM}  &  =\int d^{4}x\sqrt{-g}\mathcal{L}_{SM},\nonumber\\
\mathcal{L}_{SM}  &  =\alpha\left(  \mathcal{L}_{Ss}+V_{0}\right)
+\beta\mathcal{L}_{Sv}+\frac{1}{2}\left(  \alpha+\beta\right)  V_{sv},
\tag{2.3}%
\end{align}

\begin{align}
I_{VM}  &  =\int d^{4}x\sqrt{-g}\mathcal{L}_{VM},\nonumber\\
\mathcal{L}_{VM}  &  =\alpha\left(  \mathcal{L}_{Vv}+V_{0}\right)
+\beta\mathcal{L}_{Vs}+\frac{1}{2}\left(  \alpha+\beta\right)  V_{vs},
\tag{2.4}%
\end{align}
The physics quantities with the subscript `$S$' represent which have meaning
in only the $S-breaking,$ and have unmeaning in the $V-breaking$! It is the
same for `$V$' as that for `$S$', because of conjecture 1. For simplicity, the
subscripts `$S$' and `$V$' are elided in the following when there is not confusion.%

\begin{equation}
\mathcal{L}_{s}=\mathcal{L}_{sM}\left(  \Psi_{s},g_{\mu\nu},g_{\mu\nu
},_{\lambda}\right)  +V_{s}\left(  \omega_{s}\right)  , \tag{2.5}%
\end{equation}%
\begin{equation}
\mathcal{L}_{v}=\mathcal{L}_{vM}\left(  \Psi_{v},g_{\mu\nu},g_{\mu\nu
},_{\lambda}\right)  +V_{v}\left(  \omega_{v}\right)  , \tag{2.6}%
\end{equation}%
\begin{equation}
V_{sv}\left(  \omega_{s},\omega_{v}\right)  =V_{vs}\left(  \omega_{s}%
,\omega_{v}\right)  ; \tag{2.7}%
\end{equation}%
\[
\omega_{s}\equiv\Omega_{s},\;\Phi_{s},\;\chi_{s};\ \ \ \omega_{v}\equiv
\Omega_{v},\;\Phi_{v},\;\chi_{v},
\]
where the meanings of the symbols are as follows: $g=\det(g_{\mu\nu})$, and
$g_{\mu\nu}=diag(-1,1,1,1)$ in flat space. $R$ is the scalar curvature. Here
$\alpha$ and $\beta$ are two parameters and we finally take $\alpha=-\beta=1$.
$V_{0}$ is a parameter which is so taken that $V_{s\min}\left(  \varpi
_{s}\right)  +V_{0}=0$ in the $S-breaking$ at the $zero-temperature,$ here
$\varpi=\langle\omega\rangle$. $\mathcal{L}_{sM}$ ($\mathcal{L}_{vM}$) is the
Lagrangian density of all $s-fields$ ($v-fields$) and their couplings of the
$SU(5)$ $GUT$ except the Higgs potentials $V_{s},$ $V_{v}$ and $V_{sv}$.
$\Psi_{s}$ and $\Psi_{v}$ represents all $s-fields$ and all $v-fields$,
respectively. $\mathcal{L}_{s}$ and $\mathcal{L}_{s}$ do not contain the
contribution of the gravitation energy and the repulsion energy. It may be
seen that the set of equation $(2.1)-(2.7)$ is unchanged when the\quad
subscripts\quad$s\rightleftarrows v$ and $S\rightleftarrows V$. This shows the
symmetry between $s-matter$ and $v-matter.$

Gibbons and Hawking pointed out that in order to get the Einstein field
equations$^{[10]}$, it is necessary that%

\begin{align*}
I_{g}^{\prime}  &  =\frac{1}{16\pi G}\int\nolimits_{\Sigma}R\sqrt{-g}%
d^{4}x\longrightarrow I_{g}\\
&  =\frac{1}{16\pi G}\left(  \int_{\Sigma}R\sqrt{-g}d^{4}x+2\int
\nolimits_{\partial\Sigma}K\sqrt{\pm h}d^{3}x\right)  .
\end{align*}

This is because it is not necessary that $\delta\Gamma_{\mu\nu}^{\alpha}=0$ on
the boundary $\partial\Sigma.$ Hence $I_{g}^{\prime}$ is replaced by $I_{g}$
in $\left(  2.2\right)  .$ $\Sigma$ is a manifold with four dimensions.
$\partial\Sigma$ is the boundary of $\Sigma.$ $K=trK_{j}^{i}.$ $K_{ij}%
=-\nabla_{i}n_{j}$ is the outer curvature on $\partial\Sigma.$ $n_{j}$ is the
vertical vector on $\partial\Sigma.$ $h=\mid h_{ij}\mid,$ and $h_{ij}$ is the
induced outer metric on $\partial\Sigma$. When $\partial\Sigma$ is space-like,
$\sqrt{\pm h}$ takes positive sign. When $\partial\Sigma$ is time-like,
$\sqrt{\pm h}$ takes negative sign.

The Higgs potentials in $\left(  2.5\right)  -\left(  2.7\right)  $ is the following:%

\begin{align}
V_{s}  &  =-\frac{1}{2}\mu^{2}\Omega_{s}^{2}+\frac{1}{4}\lambda\Omega_{s}%
^{4}\nonumber\\
&  -\frac{1}{2}w\Omega_{s}^{2}Tr\Phi_{s}^{2}+\frac{1}{4}a\left(  Tr\Phi
_{s}^{2}\right)  ^{2}+\frac{1}{2}bTr\left(  \Phi_{s}^{4}\right) \nonumber\\
&  -\frac{1}{2}\varsigma\Omega_{s}^{2}\chi_{s}^{+}\chi_{s}+\frac{1}{4}%
\xi\left(  \chi_{s}^{+}\chi_{s}\right)  ^{2}, \tag{2.8}%
\end{align}

\begin{align}
V_{v}  &  =-\frac{1}{2}\mu^{2}\Omega_{v}^{2}+\frac{1}{4}\lambda\Omega_{v}%
^{4}\nonumber\\
&  -\frac{1}{2}w\Omega_{v}^{2}Tr\Phi_{v}^{2}+\frac{1}{4}a\left(  Tr\Phi
_{v}^{2}\right)  ^{2}+\frac{1}{2}bTr\left(  \Phi_{v}^{4}\right) \nonumber\\
&  -\frac{1}{2}\varsigma\Omega_{v}^{2}\chi_{v}^{+}\chi_{v}+\frac{1}{4}%
\xi\left(  \chi_{v}^{+}\chi_{v}\right)  ^{2}, \tag{2.9}%
\end{align}%
\begin{align}
V_{sv}  &  =\frac{1}{2}\Lambda\Omega_{s}^{2}\Omega_{v}^{2}+\frac{1}{2}%
\alpha\Omega_{s}^{2}Tr\Phi_{v}^{2}+\frac{1}{2}\beta\Omega_{s}^{2}\chi_{v}%
^{+}\chi_{v}\nonumber\\
&  +\frac{1}{2}\alpha\Omega_{v}^{2}Tr\Phi_{s}^{2}+\frac{1}{2}\beta\Omega
_{v}^{2}\chi_{s}^{+}\chi_{s}, \tag{2.10}%
\end{align}
where $\Omega_{a}$, $\Phi_{a}=\overset{24}{\underset{i=1}{\sum}}\left(
T_{i}/\sqrt{2}\right)  \varphi_{ai}$ and $\chi_{a}$ are respectively
$\underline{1}$ , $\underline{24}$\ and $\underline{5}$ dimensional
representations of the $SU(5)$ group, $a=s,v$, here the couplings of $\Phi
_{a}$ and $\chi_{a}$ are ignored for short$^{[11]}$. The meaning of $\alpha$
and $\beta$ in \ $\left(  2.10\right)  $ is different from that in $\left(
2.3\right)  -\left(  2.4\right)  .$ Here $\alpha$ and $\beta$ are coupling
constants. The coupling constants in $\left(  2.8\right)  -\left(
2.10\right)  $ are all positive, especially, as mentioned before, $\Lambda
,$\textbf{\ }$\alpha$ and $\beta$ in $\left(  2.10\right)  $ must be positive.

We do not consider the terms coupling to curvature scalar, e.g. $\xi
R\Omega^{2}$, for a time. In fact, $\xi R\left(  \Omega_{s}^{2}-\Omega_{v}%
^{2}\right)  \sim0$ when temperature $T$ is high enough due to the symmetry
between $s-matter$ and $v-matter$.

\subsection{Energy-momentum tensor and field equations}

By the conventional method, from $\left(  2.2\right)  $ we can get
\begin{equation}
\delta I_{g}=\frac{1}{16\pi G}\int\nolimits_{\Sigma}\left(  R_{\mu\nu}%
-\frac{1}{2}g_{\mu\nu}R\right)  \delta g^{\mu\nu}\sqrt{-g}d^{4}x. \tag{2.11}%
\end{equation}
Considering $\alpha=-\beta=1$, from $\left(  2.3\right)  -\left[  2.4\right]
$ we have
\begin{align}
&  \delta I_{SM}\nonumber\\
&  =\int\frac{1}{\sqrt{-g}}[\frac{\partial\mathcal{L}_{SM}\sqrt{-g}}{\partial
g^{\mu\nu}}-(\frac{\partial\mathcal{L}_{SM}\sqrt{-g}}{\partial g_{,\sigma
}^{\mu\nu}})_{,\sigma}]\delta g^{\mu\nu}\sqrt{-g}d^{4}x\nonumber\\
&  =\int\frac{1}{2}\left[  T_{s\mu\nu}-g_{\mu\nu}V_{0}-T_{v\mu\nu}\right]
\delta g^{\mu\nu}\sqrt{-g}d^{4}x, \tag{2.12}%
\end{align}%
\begin{equation}
\delta I_{VM}=\int\frac{1}{2}\left[  T_{v\mu\nu}-g_{\mu\nu}V_{0}-T_{s\mu\nu
}\right]  \delta g^{\mu\nu}\sqrt{-g}d^{4}x, \tag{2.13}%
\end{equation}%
\begin{align}
T_{a\mu\nu}  &  =T_{aM\mu\nu}-g_{\mu\nu}V_{a},\text{ \ }a=s\text{ or
}v,\tag{2.14}\\
T_{aM\mu\nu}  &  =\frac{2}{\sqrt{-g}}\left[  \frac{\partial\left(  \sqrt
{-g}\mathcal{L}_{aM}\right)  }{\partial g^{\mu\nu}}-\left(  \frac
{\partial\left(  \sqrt{-g}\mathcal{L}_{aM}\right)  }{\partial g^{\mu\nu
},_{\sigma}}\right)  ,_{\sigma}\right]  . \tag{2.15}%
\end{align}
From $(2.11)-(2.13)$ we obtain
\begin{equation}
R_{\mu\nu}-\frac{1}{2}g_{\mu\nu}R=-8\pi GT_{Ag\mu\nu},\text{ \ }A=S\text{
\ }or\text{ \ }V. \tag{2.16}%
\end{equation}
In the $S-breaking$,
\begin{align}
T_{Sg\mu\nu}  &  \equiv T_{s\mu\nu}-g_{\mu\nu}V_{0}-T_{v\mu\nu}=T_{SMg\mu\nu
}-g_{\mu\nu}V_{Sg}\nonumber\\
T_{SMg\mu\nu}  &  \equiv T_{sM\mu\nu}-T_{vM\mu\nu},\text{ \ }V_{Sg}%
=V_{s}+V_{0}-V_{v}. \tag{2.17}%
\end{align}
In the $V-breaking$,%
\begin{align}
T_{Vg\mu\nu}  &  \equiv T_{v\mu\nu}-g_{\mu\nu}V_{0}-T_{s\mu\nu}=T_{VMg\mu\nu
}-g_{\mu\nu}V_{Vg}\nonumber\\
T_{VMg\mu\nu}  &  \equiv T_{vM\mu\nu}-T_{sM\mu\nu},\text{ \ }V_{Vg}%
=V_{v}+V_{0}-V_{s}. \tag{2.18}%
\end{align}
It is seen from $\left(  2.17\right)  $-$\left(  2.18\right)  $ that $V_{Ag}$
is independent of $V_{sv}.$ This implies that the potential energy $V_{sv}$ is
different from other energies in essence. There is no contribution of $V_{sv}$
to $R_{\mu\nu}$, i.e., there is no gravitation and repulsion of the potential
energy $V_{sv}$. This does not satisfy the equivalence principle. But this
does not cause any contradiction with all given experiments and astronomical
observations because $V_{sv}=0$ in either of breaking modes.

We will see that, in fact, $V_{v\min}\left(  \varpi_{v}\right)  =0$ because
$\langle\omega_{v}\rangle=0$ in the $S-breaking$, and $V_{s\min}\left(
\varpi_{s}\right)  =0$ because $\langle\omega_{s}\rangle=0$ in the
$V-breaking$. Hence
\begin{equation}
V_{Ag\min}\left(  \varpi_{s},\varpi_{v}\right)  =V_{a\min}+V_{0}. \tag{2.19}%
\end{equation}
$T_{Ag\mu\nu},$ $T_{AMg\mu\nu}$ and $V_{Ag}$ are the gravitational
energy-momentum tensor density, the gravitational energy-momentum tensor
density without the Higgs potential and the gravitational potential density of
the Higgs fields in the $A-breaking$, respectively.

From $(2.1)$ the energy-momentum tensor density which does not contain the
energy-momentum tensor of gravitational and repulsive interactions can be
defined as%
\begin{align}
T_{A\mu\nu}  &  =\frac{2}{\sqrt{-g}}\left(  \frac{\partial}{\partial\alpha
}+\frac{\partial}{\partial\beta}\right) \nonumber\\
&  \cdot\left[  \frac{\partial\left(  \sqrt{-g}\mathcal{L}_{A}\right)
}{\partial g^{\mu\nu}}-\left(  \frac{\partial\left(  \sqrt{-g}\mathcal{L}%
_{A}\right)  }{\partial g^{\mu\nu},_{\sigma}}\right)  ,_{\sigma}\right]
\nonumber\\
&  \equiv T_{As\mu\nu}+T_{Av\mu\nu}-g_{\mu\nu}\left(  V_{sv}+V_{0}\right)
\nonumber\\
&  =T_{AM\mu\nu}-g_{\mu\nu}V_{A}\equiv T_{\mu\nu}, \tag{2.20}%
\end{align}%
\begin{align}
T_{AM\mu\nu}  &  =T_{sM\mu\nu}+T_{vM\mu\nu}\equiv T_{M\mu\nu},\nonumber\\
V_{A}  &  =V_{s}+V_{v}+V_{sv}+V_{0}\equiv V. \tag{2.21}%
\end{align}
Both $\alpha$\ and $\beta$ in $\left(  2.3\right)  -\left(  2.4\right)  $\ may
be regarded as the gravitation charges (the gravitation charges of $T_{s\mu
\nu},$ $T_{v\mu\nu}$ and $g_{\mu\nu}V_{sv}$ are regarded as $1$, $-1$ and $0$
in the $S-breaking$, respectively.). The energy-momentum tensor should be
independent of the gravitation charges. Hence it is necessary to eliminate
$\alpha$\ and $\beta$ from the definition of $T_{\mu\nu}$ by the operator
$\left(  \partial/\alpha+\partial/\beta\right)  $ which is the only difference
between the definition of $T_{\mu\nu}$ in the present model and that in the
conventional theory. This definition does not contradict any basic principle
and it is completely consistent with the conventional theory (the conventional
theory corresponds to one matter type so that $\beta=0$). Both the $s-energy$
and the $v-energy$ must be positive because of the definition $(2.20)-(2.21).$

It should be pointed out that\ only $\left(  2.16\right)  $ and $\left(
2.17\right)  $ is applicable in the $S-breaking,$ and only $\left(
2.16\right)  $ and $\left(  2.18\right)  $ applicable in the $V-breaking.$

It is proved that the necessary and sufficient condition of $T_{;\nu}^{\mu\nu
}=0$ is $I_{M}$ to be a scalar quantity$^{[12]}$. $I_{S}$ and $I_{V}$ are all
scalar quantities, and $T_{;\nu}^{\mu\nu}$ in the conventional theory
corresponds to $T_{g;\nu}^{\mu\nu}$. Hence we have
\begin{equation}
T_{Sg;\nu}^{\mu\nu}=T_{Vg;\nu}^{\mu\nu}=0. \tag{2.22}%
\end{equation}
Of course, $\left(  2.22\right)  $ can been derived from $\left(  2.16\right)
$.

\section{Spontaneous symmetry breaking}

Ignoring the couplings of $\Phi_{s}$ and $\chi_{s\text{ }}$and suitably
choosing the parameters of the Higgs potential, analogously to Ref. $[11]$, we
can prove from $(2.8)-(2.10)$ that there are the following vacuum expectation
values at the zero-temperature and under the tree-level approximation,%

\begin{equation}
\left\langle 0\left\vert \omega_{v}\right\vert 0\right\rangle \equiv
\overline{\omega}_{v0}=0,\text{ \ }\left\langle 0\left\vert \omega
_{s}\right\vert 0\right\rangle \equiv\overline{\omega}_{s0}\neq0, \tag{3.1}%
\end{equation}

\begin{equation}
\left\langle 0\left\vert \Omega_{s}\right\vert 0\right\rangle =\upsilon
_{\Omega0}, \tag{3.2}%
\end{equation}

\begin{equation}
\left\langle 0\left\vert \Phi_{s}\right\vert 0\right\rangle =Diagonal\left(
1,1,1,-\frac{3}{2},-\frac{3}{2}\right)  \upsilon_{\varphi0}, \tag{3.3}%
\end{equation}

\begin{equation}
\left\langle 0\left\vert \chi_{s}\right\vert 0\right\rangle ^{+}%
=\frac{\upsilon_{\chi0}}{\sqrt{2}}\left(  0,0,0,0,1\right)  , \tag{3.4}%
\end{equation}
The breaking satisfied $\left(  3.1\right)  $ is called the S-breaking.
Ignoring the contributions of $\Phi_{s}$ and $\chi_{s}$ to $\left\langle
0\left\vert \Omega_{s}\right\vert 0\right\rangle ,$ at the zero-temperature we
get%
\begin{equation}
\upsilon_{\Omega0}^{2}=\frac{\mu^{2}}{f},\ \ \ f\equiv\lambda-\frac{15w^{2}%
}{\left(  15a+7b\right)  }-\frac{\varsigma^{2}}{\xi}. \tag{3.5}%
\end{equation}

\begin{equation}
\upsilon_{\varphi0}^{2}=\frac{2w}{15a+7b}\upsilon_{\Omega0}^{2}, \tag{3.6}%
\end{equation}

\begin{equation}
\upsilon_{\chi0}^{2}=\frac{2\zeta}{\xi}\upsilon_{\Omega0}^{2}. \tag{3.7}%
\end{equation}

We take $\Lambda>\lambda>15w^{2}/\left(  15a+7b\right)  +\zeta^{2}/\xi.$ From
$\left(  2.9\right)  $-$\left(  2.10\right)  $ and $\left(  3.1\right)
$-$\left(  3.7\right)  $ it can be proved that all $v-Higgs$ bosons can get
their big enough masses. The masses of the Higgs particles exclusive of the
$\Phi_{s}-particles$ and the $\chi_{s}-particles\ $ in the $S-breaking$ are respectively%

\begin{equation}
m^{2}\left(  \Omega_{s}\right)  =2\mu^{2}, \tag{3.8}%
\end{equation}

\begin{equation}
m^{2}\left(  \Omega_{v}\right)  =\Lambda\upsilon_{\Omega0}^{2}-\mu^{2},
\tag{3.9}%
\end{equation}

\begin{equation}
m^{2}\left(  \Phi_{v}\right)  =\frac{1}{2}\alpha\upsilon_{\Omega0}^{2},
\tag{3.10}%
\end{equation}

\begin{equation}
m^{2}\left(  \chi_{v}\right)  =\beta\upsilon_{\Omega0}^{2}. \tag{3.11}%
\end{equation}
We can choose such parameters that
\begin{align}
m\left(  \Omega_{s}\right)   &  \simeq m\left(  \Omega_{v}\right) \nonumber\\
&  \gg m\left(  \varphi_{v}\right)  \sim m\left(  \varphi_{s}\right)  \gg
m\left(  \chi_{v}\right)  \sim m\left(  \chi_{s}\right)  , \tag{3.12}%
\end{align}
e.g., $m\left(  \Omega_{s}\right)  \sim10^{16}Gev,$ $m\left(  \varphi
_{s}\right)  \sim10^{14}Gev$ and $m\left(  \chi_{s}\right)  \sim10^{2}Gev$. It
is easily seen from $(3.8)-(3.11)$ that all real components of $\Phi_{v}$ have
the same mass $m\left(  \Phi_{v}\right)  $, and all real components of
$\chi_{v}$ have the same mass $m\left(  \chi_{v}\right)  $ in the
$S-breaking.$

The $S-breaking$ and the $V-breaking$ are symmetric because $s-matter$ and
$v-matter$ are symmetric. Hence when $s\rightleftarrows v$ and
$S\rightleftarrows V$ in $(3.1)-(3.12),$ the formulas are still kept.

\section{Evolution equations of space in RW metric}

Based on the RW metric metric,
\begin{align}
(ds)^{2}  &  =-\left(  dt\right)  ^{2}\nonumber\\
&  +R^{2}\left(  t\right)  \left\{  \frac{\left(  dr\right)  ^{2}}{1-kr^{2}%
}+\left(  rd\theta\right)  ^{2}+\left(  r\sin\theta d\varphi\right)
^{2}\right\}  . \tag{4.1}%
\end{align}
In the present model, we take $k=-1.$

Matter in the universe may approximately be regarded as ideal gas distributed
evenly in space. Considering the potential energy densities in $(2.14)$, we
can write $T_{a\mu\nu}$ as
\begin{equation}
T_{a\mu\nu}=\left[  \widetilde{\rho}_{a}+\widetilde{p}_{a}\right]  U_{a\mu
}U_{a\nu}+\widetilde{p}_{a}g_{\mu\nu}, \tag{4.2}%
\end{equation}%
\begin{equation}
\widetilde{\rho}_{a}=\rho_{a}+V_{a}\left(  \varpi_{a}\right)  ,\text{
\ }\widetilde{p}_{a}=p_{a}-V_{a}(\varpi_{a}), \tag{4.3}%
\end{equation}
where $U_{a\mu}$ is a 4-velocity, $U_{a\mu}=\delta_{\mu}^{0}=U_{\mu}$, and
$a=s$ or $v$. $\left(  -g_{\mu\nu}V_{0}\right)  $ can be written as
\begin{align}
-g_{\mu\nu}V_{0}  &  =\left(  \widetilde{\rho}\left(  V_{0}\right)
+\widetilde{p}\left(  V_{0}\right)  \right)  U_{\mu}U_{\nu}+g_{\mu\nu
}\widetilde{p}\left(  V_{0}\right)  ,\tag{4.4}\\
\widetilde{\rho}\left(  V_{0}\right)   &  =V_{0},\text{ \ }\widetilde
{p}\left(  V_{0}\right)  =-V_{0.}\nonumber
\end{align}

Considering $U_{\mu}=\delta_{\mu}^{0},$ substituting $(4.2)-\left(
4.4\right)  $ and the RW metric in $\left(  4.1\right)  $ into $(2.16)$, we
get the evolution equations%

\begin{align}
\overset{\cdot}{R}^{2}+k  &  =\eta\left[  \rho_{g}+V_{g}\right]  R^{2},\text{
\ \ }\eta\equiv8\pi G/3,\tag{4.5}\\
\overset{\cdot\cdot}{R}  &  =-\frac{1}{2}\eta\left[  \left(  \rho_{g}%
+3p_{g}\right)  -2V_{g}\right]  R. \tag{4.6}%
\end{align}
In the $S-breaking$,
\begin{equation}
\rho_{g}=\rho_{s}-\rho_{v},\text{ \ }p_{g}=p_{s}-p_{v},\text{ \ }V_{g}%
=V_{s}+V_{0}-V_{v}. \tag{4.7}%
\end{equation}
In the $V-breaking$,
\begin{equation}
\rho_{g}=\rho_{v}-\rho_{s},\text{ \ }p_{g}=p_{v}-p_{s},\text{\ }V_{g}%
=V_{v}+V_{0}-V_{s}. \tag{4.8}%
\end{equation}
Comparing $(4.5)-(4.6)$ with the Friedmann equations, we see that provided
$\rho$, $p$ and $V$ in the Friedmann equations are replaced by $\rho_{g}$,
$p_{g}$ and $V_{g}$, $(4.5)-(4.6)$ are obtained.

In contrast with the conventional theory, $\rho_{g}>0,$ $=0$ and $\rho_{g}<0$
are all possible in the present model.

From $(4.5)-(4.6)$ we have%

\begin{equation}
\overset{\cdot}{\rho}_{g}+3\left(  \rho_{g}+p_{g}\right)  \frac{\overset
{\cdot}{R}}{R}=-\overset{\cdot}{V}_{g}. \tag{4.9}%
\end{equation}

Pressure density is a function of masses of particles and temperature. In
order to determine the pressure at a given temperature, we divide the
particles into three sorts according to their masses. The first sort is
composed of such particles whose masses are larger than $m_{M}\gtrsim1MeV$.
The second sort is composed of such particles with their masses $m_{l},$ and
$1MeV\gtrsim m_{l}\gtrsim1eV$. We suppose $p_{l}=\kappa\rho_{l}$. It is
obvious $1/3\gtrsim\kappa\gtrsim0$. The third sort is composed of photon-like
particles whose masses are less than $m_{l}$ and may be regarded as zero.
Thus, we have%
\begin{equation}
\rho_{g}=\rho_{Mg}+\rho_{\lg}+\rho_{\gamma g}, \tag{4.10}%
\end{equation}
and $p_{\gamma g}=\rho_{\gamma g}/3$. In the $V-breaking$, $\rho_{Mg}%
=\rho_{vMg}-\rho_{sMg},$ $\rho_{\lg}=\rho_{v\lg}-\rho_{s\lg}$. Considering all
$s-particles$ must be in $s-SU(5)$ color single states whose masses are not
zero so that $p_{s\gamma}=0,$ we have $\rho_{\gamma g}=\rho_{v\gamma g}%
-\rho_{s\gamma g}=\rho_{v\gamma g}.$ Thus, when $T$ is so large that all
masses may be neglected, from $\left(  4.9\right)  -\left(  4.10\right)  $ we
have%
\begin{equation}
p_{g}=\rho_{g}/3\text{, \ }\frac{d\left(  \rho_{g}R^{4}\right)  }%
{dt}=-\overset{\cdot}{V}_{g}R^{4}. \tag{4.11}%
\end{equation}
When $m_{M}\gtrsim T\gtrsim m_{l},$ $\rho_{M}\gg p_{M}\sim0$. Letting\ $p_{l}%
=\kappa\rho_{l}$, we have%
\begin{equation}
\frac{d\left(  \rho_{Mg}R^{3}\right)  }{dt}+\frac{d\left(  \rho_{\lg
}R^{3\left(  1+\kappa\right)  }\right)  }{R^{3\kappa}dt}+\frac{d\left(
\rho_{\gamma g}R^{4}\right)  }{Rdt}=-\overset{\cdot}{V}_{g}R^{3}. \tag{4.12}%
\end{equation}
When $T<m_{l}\sim1eV,$ $\rho_{M}\gg p_{M}\sim0$ and $\rho_{l}\gg p_{l}\sim0$
so that
\begin{equation}
\frac{d\left(  \rho_{mg}R^{3}\right)  }{dt}+\frac{d\left(  \rho_{\gamma
g}R^{4}\right)  }{Rdt}=-\overset{\cdot}{V}_{g}R^{3}. \tag{4.13}%
\end{equation}
where $\rho_{mg}=\rho_{Mg}+\rho_{\lg}$.

\section{Temperature effect}

The thermal equilibrium between the $v-particles$ and the $s-particles$ can be
realized by only $\left(  2.10\right)  .$ The Higgs bosons $\Omega_{s}$ and
$\Omega_{v}$ are hardly produced because their masses are all very big after
reheating. Consequently, the interaction between the $v-particles$ and the
$s-particles$ may be ignored so that there is no thermal equilibrium between
the $v-particles$ and the $s-particles.$ Thus, when temperature is low, we
should use two sorts of temperature $T_{v}$ and $T_{s}$ to describe the
thermal equilibrium of $v-matter$ and the thermal equilibrium of $s-matter$,
respectively. Generally speaking, $T_{v}\neq T_{s}$. We will see when
$T_{s}\gtrsim T_{cr}$, $\langle\Omega_{s}\rangle=\langle\Omega_{v}\rangle=0$
so that $\langle\omega_{s}\rangle=\langle\omega_{v}\rangle=0$ because of
$\left(  2.8\right)  -\left(  2.10\right)  $. Thus, the masses of all
particles are zero and $\rho_{s}$ and $\rho_{v}$ can transform from one to an
other by $\left(  2.10\right)  $. In this case, if $\overset{\cdot}%
{R}=\overset{\cdot\cdot}{R}=0,$ there must be $T_{v}=T_{s}$ and $\rho_{v}%
=\rho_{s}$; If $\overset{\cdot}{R}<0$ and $\overset{\cdot\cdot}{R}>0,$ there
must be the least scale $R_{\min}>0.$ When space contracts to $R_{\min},$
$T_{s}=T_{\max}$ and then inflation occurs. There is no such a case
$\overset{\cdot}{R}<0$ and $\overset{\cdot\cdot}{R}\leq0$ in the present model.

\subsection{Effective potentials}

The influence of finite temperature on the Higgs potential in the present
model are consistent with the conventional theory. For short, we consider only
$\Omega_{a}$ and $\varphi_{a},$ $a=s$ or $v$. When $\chi_{a}$ is considered as
well, the following inferences are still qualitatively valid. For
\begin{equation}
V_{s\Omega}=-\frac{\mu^{2}}{2}\Omega_{s}^{2}+\frac{\lambda}{4}\Omega_{s}^{4},
\tag{5.1}%
\end{equation}
to ignore the terms proportional to $\lambda^{n}(n>1),$ the finite-temperature
effective potential approximate to 1-loop in flat space is$^{[13]\left[
14\right]  }$
\begin{equation}
V_{s\Omega,eff}^{(1)T}=-\frac{1}{2}\left(  \mu^{2}-\frac{\lambda}{4}T_{s}%
^{2}\right)  \overline{\Omega}_{s}^{2}+\frac{\lambda}{4}\overline{\Omega}%
_{s}^{4}-\frac{\pi^{2}}{90}\left(  kT_{s}\right)  ^{4}+\frac{\mu^{2}}{24}%
T_{s}^{2}. \tag{5.2}%
\end{equation}
Considering the influence of the expectation values $v_{\Omega v}\left(
T_{s},T_{v}\right)  ,$ $v_{\varphi s}\left(  T_{s},T_{v}\right)  $ and
$v_{\varphi v}\left(  T_{s},T_{v}\right)  $, and ignoring the terms irrelevant
to $\Omega_{s},$ we have%

\begin{equation}
V_{s\Omega,eff}^{(1)T}=-\frac{1}{2}\mu_{s}^{2}\left(  T_{s},T_{v}\right)
\overline{\Omega}_{s}^{2}+\frac{\lambda}{4}\overline{\Omega}_{s}^{4},
\tag{5.3}%
\end{equation}%
\begin{equation}
\mu_{s}^{2}\equiv\mu^{2}-\frac{\lambda}{4}T_{s}^{2}-\Lambda v_{\Omega v}%
^{2}-\frac{15}{2}\left(  \alpha v_{\varphi v}^{2}-wv_{\varphi s}^{2}\right)  .
\tag{5.4}%
\end{equation}
Similarly $\left(  5.1\right)  -\left(  5.4\right)  $, we have%
\begin{equation}
V_{v\Omega,eff}^{(1)T}=-\frac{1}{2}\mu_{v}^{2}\left(  T_{s},T_{v}\right)
\overline{\Omega}_{v}^{2}+\frac{\lambda}{4}\overline{\Omega}_{v}^{4},
\tag{5.5}%
\end{equation}%
\begin{equation}
\mu_{v}^{2}\equiv\mu^{2}-\frac{\lambda}{4}T_{v}^{2}-\Lambda v_{\Omega s}%
^{2}-\frac{15}{2}\left[  \alpha v_{\varphi s}^{2}-wv_{\varphi v}^{2}\right]  .
\tag{5.6}%
\end{equation}
In the $S-breaking$, we can prove $v_{\Omega v}\left(  T_{s},T_{v}\right)
=v_{\varphi v}\left(  T_{s},T_{v}\right)  =0$. This is because $v_{\Omega
v}=v_{\varphi v}=0$ in low temperatures, and $T_{v}\sim T_{s}$ holds when
$T_{s}$ is high enough so that $v_{\Omega v}=v_{\varphi v}=0$ still holds.
Thus
\[
\mu_{s}^{2}\left(  T_{s},T_{v}\right)  =\mu^{2}-\frac{\lambda}{4}T_{s}%
^{2}+\frac{15}{2}wv_{\varphi s}^{2}\left(  T_{s},T_{v}\right)  .
\]%
\begin{align*}
v_{\Omega s}^{2}\left(  T_{s},T_{v}\right)   &  =\mu_{s}^{2}\left(
T_{s},T_{v}\right)  /\lambda,\text{ \ when }\mu_{s}^{2}\left(  T_{s}%
,T_{v}\right)  >0,\\
v_{\Omega s}\left(  T_{s},T_{v}\right)   &  =0,\text{ \ when }\mu_{s}%
^{2}\left(  T_{s},T_{v}\right)  \leq0.
\end{align*}

For
\begin{equation}
V_{s\varphi}=\frac{1}{2}\left(  \alpha\Omega_{v}^{2}-w\Omega_{s}^{2}\right)
Tr\Phi_{s}^{2}+\frac{1}{4}a(Tr\Phi_{s}^{2})^{2}+\frac{1}{2}bTr\Phi_{s}^{4},
\tag{5.7}%
\end{equation}
ignoring the contributions of the Higgs fields and the fermion fields to one
loop correction, and only considering the contribution of the gauge fields,
when $\overline{\varphi}_{s}\ll kT$, here $k$ is the Boltzmann constant, we
get the effective potential at finite-temperature approximate to 1-loop in
flat space$^{[13]\left[  14\right]  }$
\begin{equation}
V_{s\varphi,eff}^{(1)T}=V(\overline{\varphi}_{s})+B\overline{\varphi}_{s}%
^{4}\left(  \ln\frac{\overline{\varphi}_{s}^{2}}{\sigma^{2}}-\frac{25}%
{6}\right)  +CT_{s}^{2}\overline{\varphi}_{s}^{2}-\frac{\pi^{2}}{15}\left(
kT_{s}\right)  ^{4}, \tag{5.8}%
\end{equation}
where $B=\left(  5625/1024\pi^{2}\right)  g^{4},$ and $\overline{\Phi}%
_{s}=Diagonal\left(  1,1,1,-3/2,-3/2\right)  \overline{\varphi}_{s}.$ In
general, $w=\alpha<\lambda\sim g^{4}<C=\left(  75/16\right)  \left(
kg\right)  ^{2}.$ We take $w=\alpha$ for simplicity. Here $\sigma$ is a
parameter at which the renormalization coupling-constant $\lambda$ is defined
as below,
\begin{equation}
\frac{d^{4}V_{s\varphi,eff}^{(1)T}}{d\overline{\varphi}_{s}}\mid
_{\overline{\varphi}_{s}=\sigma}=\lambda. \tag{5.9}%
\end{equation}
Only considering the contribution of the expectation values of $\Omega_{s}$
and $\Omega_{v}$ to $V_{s\varphi,eff}^{(1)T},$ taking $\left(  15/16\right)
\left(  15a+7b\right)  =\left(  11/3\right)  B$, and ignoring the term
$\left(  kT_{s}\right)  ^{4}$ unconnected with $\overline{\varphi}_{v}$ (it
may be added to in $\rho_{rg}$)$,$ from $\left(  2.8\right)  $ and $\left(
5.7\right)  -\left(  5.8\right)  $ we have
\begin{equation}
V_{s\varphi,eff}^{(1)T}=A_{s}^{2}\overline{\varphi}_{s}^{2}+B\overline
{\varphi}_{s}^{4}\left(  \ln\frac{\overline{\varphi}_{s}^{2}}{\sigma^{2}%
}-\frac{1}{2}\right)  , \tag{5.10}%
\end{equation}%
\begin{equation}
A_{s}^{2}\equiv\frac{15}{4}w\left(  v_{\Omega v}^{2}\left(  T_{v}\right)
-v_{\Omega s}^{2}\left(  T_{s}\right)  \right)  +CT_{s}^{2}. \tag{5.11}%
\end{equation}
Similarly, from $\left(  2.9\right)  $ we have
\begin{align}
V_{v\varphi,eff}^{(1)T}  &  =A_{v}^{2}\overline{\varphi}_{v}^{2}%
+B\overline{\varphi}_{v}^{4}\left(  \ln\frac{\overline{\varphi}_{v}^{2}%
}{\sigma^{2}}-\frac{1}{2}\right)  ,\tag{5.12}\\
A_{v}^{2}  &  \equiv\frac{15}{4}w\left(  v_{\Omega s}^{2}\left(  T_{s}\right)
-v_{\Omega v}^{2}\left(  T_{v}\right)  \right)  +CT_{v}^{2}. \tag{5.13}%
\end{align}

\subsection{Critical temperatures $T_{cr}$}

In the following $V_{eff}=V_{eff}^{(1)T}$. $T_{s}$ and $T_{v}$ must rise as
space contracts. From $\left(  5.4\right)  $, $\left(  5.6\right)  $, $\left(
5.11\right)  $ and $\left(  5.13\right)  $ we see that there must be critical
temperatures $T_{s,\varphi cr}$ and $T_{cr}$ so that $\langle\varphi
_{s}\rangle=\langle\varphi_{v}\rangle=0$ when $T_{s}>T_{s,\varphi cr}$ and
$\langle\Omega_{s}\rangle=\langle\Omega_{v}\rangle=0$ when $T_{s}>T_{cr}$. By
suitably choosing the parameters $\mu^{2},$ $\Lambda,$ $\lambda,$ $C$ and $w$,
we can get $T_{cr}>T_{s,\varphi cr}$. Consequently, $\langle\omega_{s}%
\rangle=\langle\omega_{v}\rangle=0$ when $T_{s}>T_{cr}$.

We discuss $T_{s,\varphi cr}$ and $T_{cr}$ in detains as follows.
$V_{eff}=V_{eff}^{(1)T}$ in the following.

1. The critical temperature $T_{s,\varphi cr}$ in the $S-breaking$.

Let $T_{v}=T_{v,\varphi cr}$ when $T_{s}=T_{s,\varphi cr}$ in the
$S-breaking$. $T_{s,\varphi cr}$ is the critical temperature at which the
minima are degenerate, i.e. $V_{eff\min}\left(  \overline{\varphi}%
_{s},T_{s,\varphi cr},T_{v,\varphi cr}\right)  =V_{eff}\left(  v_{s,\varphi
cr},T_{s,\varphi cr},T_{v,\varphi cr}\right)  =V_{eff}\left(  0,T_{s,\varphi
cr},T_{v,\varphi cr}\right)  $. In other words, $\langle\varphi_{s}%
\rangle=v_{\varphi s}\neq0$ when $T_{s}<T_{s,\varphi cr},\ $and $v_{\varphi
s}=0$ when $T_{s}>T_{s,\varphi cr}.$ $T_{s,\varphi cr}$ and $A_{v}^{2}\left(
\overline{\varphi}_{s},T_{s,\varphi cr},T_{v,\varphi cr}\right)  $ can be
determined from $\left(  5.12\right)  -\left(  5.13\right)  $ by that
when\ $\overline{\varphi}_{s}=v_{\varphi scr}$,
\[
V_{eff}\left(  \overline{\varphi}_{s},T_{s,\varphi cr},T_{v,\varphi
cr}\right)  =V_{eff}\left(  0,T_{s,\varphi cr},T_{v,\varphi cr}\right)
,\text{ \ }\frac{\partial}{\partial\overline{\varphi}_{s}}V_{eff}\left(
\overline{\varphi}_{s},T_{s,\varphi cr},T_{v,\varphi cr}\right)  =0.
\]
When $A_{\varphi}^{2}=A_{\varphi}^{2}\left(  \overline{\varphi}_{s}%
,T_{s,\varphi cr},T_{v,\varphi cr}\right)  \equiv A_{cr}^{2}=B\sigma
^{2}e^{-1/2},$ $V_{s,eff\min}$ is degenerate, here $\overline{\varphi}%
_{scr}^{2}\equiv\sigma^{2}e^{-1/2},$ hence we have%
\begin{equation}
T_{s,\varphi cr}^{2}=\frac{A_{cr}^{2}+\left(  15/4\right)  \left(  w\mu
^{2}/\lambda\right)  }{C+15w/16}, \tag{5.14}%
\end{equation}
Here $v_{\Omega v}=v_{\varphi s}=v_{\varphi v}=0$ when $T_{s}>T_{s\varphi cr}$
is considered. When $T_{s}<T_{s\varphi cr},$ i.e., $A_{s}^{2}<A_{cr}^{2},$
$V_{s\varphi,eff\min}<V_{s\varphi,eff}\left(  0\right)  $ and $v_{\varphi
s}\neq0$. When $T_{s}>T_{s\varphi cr},$ i.e., $A_{s}^{2}>A_{cr}^{2},$
$v_{\varphi s}=0$ and $V_{s\varphi,eff\min}=V_{s\varphi,eff}\left(  0\right)
$. Let $\mu_{a}^{2}=\mu_{a\varphi cr}^{2}$ and $A_{v}^{2}=A_{v\varphi cr}^{2}$
when $T_{s}=T_{s\varphi cr},$ then when $T_{s}\geq T_{s\varphi cr},$ from
$\left(  5.6\right)  $ and $\left(  5.13\right)  $ we have%
\begin{align}
\mu_{s}^{2}  &  \equiv\mu^{2}-\frac{\lambda}{4}T_{s}^{2},\tag{5.15}\\
\mu_{s\varphi cr}^{2}  &  =\frac{\mu^{2}-\left(  \lambda/4C\right)  A_{cr}%
^{2}}{1+15w/16C}, \tag{5.16}%
\end{align}%
\begin{align}
\mu_{v}^{2}  &  =\left(  1-\frac{\Lambda}{\lambda}\right)  \mu^{2}%
-\frac{\lambda}{4}\left(  T_{v}^{2}-\frac{\Lambda}{\lambda}T_{s}^{2}\right)
,\tag{5.17}\\
\mu_{v\varphi cr}^{2}  &  =\left(  1-\frac{\Lambda}{\lambda}\right)  \mu
^{2}-\frac{\lambda}{4}\left(  T_{v\varphi cr}^{2}-\frac{\Lambda}{\lambda
}T_{s\varphi cr}^{2}\right)  , \tag{5.18}%
\end{align}%
\begin{equation}
A_{v}^{2}=\frac{15}{4}\frac{w}{\lambda}\mu^{2}+CT_{v}^{2}-\frac{15}{16}%
wT_{s}^{2}, \tag{5.19}%
\end{equation}%
\begin{equation}
A_{v\varphi cr}^{2}=A_{v\varphi}^{2}\left(  T_{s\varphi cr}\right)
=\frac{\left(  15w/4\lambda\right)  \mu^{2}-\left(  15w/16C\right)  A_{cr}%
^{2}}{1+15w/16C}+CT_{v\varphi cr}^{2} \tag{5.20}%
\end{equation}
It is obvious that $v_{a\Omega}\left(  T_{s},T_{v}\right)  =0$ when $\mu
_{a}^{2}\leq0$. Considering $\Lambda\gg\lambda,$ $C>\lambda>w,$ taking
$A_{cr}^{2}<\mu^{2},$\ from $\left(  5.17\right)  -\left(  5.20\right)  $ and
$\left(  5.14\right)  $\ we see that provided $T_{v}\sim T_{s},$ there are
\begin{align}
\mu_{v}^{2}  &  <0,\text{ \ }A_{v\varphi cr}^{2}>A_{cr}^{2}\text{,
\ }\tag{5.21}\\
v_{v\varphi}  &  =v_{v\Omega}=0,\text{ \ \ when }T_{s}\geq T_{s\varphi
cr},\tag{5.22}\\
T_{s\varphi cr}  &  <\frac{2\mu}{\sqrt{\lambda}}. \tag{5.23}%
\end{align}
In fact, $T_{v}\gtrsim T_{s}$ when $T_{s}>T_{s\varphi cr}$. Hence $\left(
5.21\right)  -\left(  5.23\right)  $ must hold when $T_{s}>T_{s\varphi cr}$.
The masses of particles originating from the couplings of particles with
$\langle\omega_{s}\rangle$ and $\langle\omega_{v}\rangle$ will be small,
because $v_{s\varphi}=v_{v\varphi}=v_{v\Omega}=0$ and $v_{s\Omega}$ becomes
small when $T_{s}>T_{s\varphi cr}$. Thus, photon-like particles are dominative
and the transformation of $\rho_{s}$ to $\rho_{v}$ is striking due to
$\overset{\cdot}{V}_{g,eff}>0$ (see section 6.1 in detail). Consequently,%
\begin{equation}
T_{v}\gtrsim T_{s},\text{ }p_{g}\sim\rho_{g}/3,\text{ \ when }T_{s}%
>T_{s\varphi cr}. \tag{5.24}%
\end{equation}
On the other hand, the transformation of $\rho_{s}$ and $\rho_{v}$ from one to
another may be neglected in low temperatures. Thus, it is easily proven that
$v_{v\varphi}=v_{v\Omega}=0$ when $T_{s}<T_{s\varphi cr}$ in the $S-breaking$.
Hence, $v_{\Omega v}=v_{\varphi v}=0$ holds always in the $S-breaking$.

2. Critical temperatures $T_{cr}$

$\left(  5.15\right)  $ holds when $T_{s}>T_{s\varphi cr}$ due to $v_{\varphi
s}=0$ and $\left(  5.22\right)  $. Thus, from $\left(  5.3\right)  ,$ $\left(
5.17\right)  ,$ $\left(  5.19\right)  ,$ $\left(  5.5\right)  $, $\left(
5.10\right)  $ and $\left(  5.12\right)  $\ we have
\begin{align}
v_{s\Omega}  &  =0=v_{s\varphi}=v_{v\varphi}=v_{v\Omega},\tag{5.25}\\
m_{s}  &  =m_{v}=0,\text{ \ when\ }T_{s}\geq T_{cr}\equiv\frac{2\mu}%
{\sqrt{\lambda}}, \tag{5.26}%
\end{align}
where $m_{s}$ and $m_{v}$ denote the masses of all s-particles and the masses
of all v-particles originating from the couplings of particles with
$v_{s\Omega}$, $v_{v\Omega}$, $v_{\varphi s}$ and $v_{\varphi v}.$, respectively.

The symmetry $s-SU(5)\times v-SU(5)$ holds because of $\left(  5.25\right)  $.
The state with this symmetry is called the most symmetric state. $\rho_{s}$
and $\rho_{v}$ can transform from one to another because $\left(  5.26\right)
$ and $\left(  2.10\right)  $. Consequently, if space is static and $T_{s}\geq
T_{cr}$, we have
\begin{align}
T_{s}  &  =T_{v}=T,\text{ }\rho_{g}\left(  T\right)  =0,\text{ }%
p_{g}=0\tag{5.27}\\
\rho_{s}  &  =\rho_{v},\text{ \ }p_{s}=\rho_{s}/3,\text{ }p_{v}=\rho_{v}/3,
\tag{5.28}%
\end{align}%
\begin{equation}
V_{v,eff}=V_{s,eff}=0,\text{ }V_{g,eff}=V_{0}. \tag{5.29}%
\end{equation}
Here $V_{a,eff}=V_{a,eff}\left(  \overline{\varphi}_{a},\overline{\Omega}%
_{a}\right)  $ which does not contain the terms containing $\left(
kT_{a}\right)  ^{4}$ and $\mu^{2}T_{a}^{2}$. The terms are added to $\rho_{a}$
(see $\left(  6.6\right)  $). In fact, space cannot be static, and the
contracting process is not a thermal equilibrium process because
$\overset{\cdot\cdot}{R}>0$ and $\overset{\cdot}{R}<0$. Hence $\left(
5.27\right)  -\left(  5.28\right)  $ does not hold in the contracting process,
but $\left(  5.29\right)  $ possibly holds because $\langle\omega_{s}%
\rangle=\langle\omega_{v}\rangle=0$.

\section{Contraction of space, the highest temperature, inflation of space and
there is no singularity}

On the basis of the cosmological principle, if there is the space-time
singularity, it is a result of space contraction. Thus, we discuss the
contracting process. From the contracting process we will see that there is no
space-time singularity in present model.

We do not consider the couplings of the Higgs fields with the Ricci scalar $R
$ for this time. We will see in the following paper that the following
conclusions still hold when such couplings as $\xi R\Omega_{s}^{2}$ are
considered. In fact, $\xi R\left(  \overline{\Omega}_{s}^{2}-\overline{\Omega
}_{v}^{2}\right)  =0$ because there is the strict symmetry between $s-matter$
and $v-matter$ when $T\gtrsim T_{cr}$.

We chiefly discuss change of $\langle\Omega_{a}\left(  T_{a}\right)  \rangle$
and $\langle\varphi_{a}\left(  T_{a}\right)  \rangle$ as temperature in the
contracting process of space for short. When $\langle\chi_{a}\left(
T_{a}\right)  \rangle$ is considered as well, the inferences are still valid qualitatively.

\subsection{The contracting process of space and the highest temperature}

1. Space contracts in the $S-breakng.$

In the $S-breakng,$ we consider the space-contraction process in which
$\overset{\cdot}{R}<0$. In low temperatures $T_{s}\sim T_{v}\gtrsim0$,
$\langle\omega_{s}\rangle=\langle\omega_{s}\rangle_{0}$ and $\langle\omega
_{v}\rangle=0.$ Consequently $V_{s,eff}=-V_{0},$ $V_{v,eff}=0$
\begin{equation}
V_{g,eff}=V_{s,eff}-V_{v,eff}+V_{0}=0, \tag{6.1}%
\end{equation}
and $\rho_{v}$ and $\rho_{s}$ cannot transform from one into other and
$\overset{\cdot}{V}_{g,eff}\sim0$, because the masses of the Higgs particles
$\Omega_{s}$ and $\Omega_{v}$ are very large. There must be $\rho_{g}=\rho
_{s}-\rho_{v}>0$ because of $\left(  6.1\right)  ,$ $\left(  4.5\right)
-\left(  4.7\right)  $ and $\overset{\cdot}{R}<0$. In the stage, space will
contract faster and faster, i.e. $\overset{\cdot\cdot}{R}<0$ due to $\left(
4.6\right)  .$

2. $\overset{\cdot}{V}_{g,eff}>0$ and $\overset{\cdot}{\rho}_{g}<0.$

The temperature must rise monotonously because space contracts, because the
non-zero momentum of a free particle $p\propto1/R(t),$ $\bigtriangleup
p\bigtriangleup x\gtrsim1$ and $\rho_{m}\propto1/R^{3}(t)$. In the contracting
process, as mentioned section $5.2$, $\langle\omega_{v}\rangle=0$ holds
always. Consequently, $V_{v,eff}=\overset{\cdot}{V}_{v,eff}=0$ holds
always.\ In contrast with $V_{v,eff},$\ $V_{s,eff}$ will rise from
$V_{s,eff}=-V_{0}$ (when $T_{s}\sim0$) to $V_{s,eff}\left(  T_{cr}\right)
=0.$ $\langle\varphi_{s}\rangle\sim0$ when $T_{s}\gtrsim T_{s\varphi cr}$ and
$\langle\Omega_{s}\rangle\sim0$ when $T_{s}\sim T_{cr}.$ It is seen that
$\langle\omega_{s}\rangle$ can strikingly change in the period $T_{s\varphi
cr}\sim T_{s}\lesssim T_{cr}$. Hence $V_{s,eff}$ must strikingly change and
$\overset{\cdot}{V}_{s,eff}>0.$ On the other hand, the transformation of
$\rho_{s}$ into $\rho_{v}$ is striking in the period $T_{s\varphi cr}\sim
T_{s}\lesssim T_{cr}$. This is because in the period, in addition $\omega
_{v}\rangle=0,$ $\langle\varphi_{s}\rangle\sim0$ or $\langle\Omega_{s}%
\rangle\sim0$ so that $m_{s}\sim m_{v}\sim0$. Thus, from $\left(  4.9\right)
$ we have%
\begin{align}
\overset{\cdot}{V}_{g,eff}  &  =\overset{\cdot}{V}_{s,eff}-\overset{\cdot}%
{V}_{v,eff}=\overset{\cdot}{V}_{s,eff}>0,\tag{6.2}\\
\overset{\cdot}{\rho}_{g}  &  =\overset{\cdot}{\rho}_{s}-\overset{\cdot}{\rho
}_{v}<0. \tag{6.3}%
\end{align}

When $T_{s}=T_{cr}$, $V_{s,eff}=V_{v,eff}=0$, because $\langle\omega
_{v}\rangle=\langle\omega_{v}\rangle=0$. Thus, $V_{g,eff}=V_{0}$ and $\left(
4.5\right)  -\left(  4.7\right)  $ is reduced to
\begin{align}
\left(  \overset{\cdot}{R}/R\right)  ^{2}  &  =-\frac{k}{R^{2}}+\eta\left[
\rho_{g}+V_{0}\right]  ,\text{ \ \ }\tag{6.4}\\
\frac{\overset{\cdot\cdot}{R}}{R}  &  =\eta\left[  V_{0}-\rho_{g}\right]  .
\tag{6.5}%
\end{align}%
\begin{align}
\rho_{a}  &  =\frac{\pi^{2}}{30}\left(  g^{\ast}-\frac{7}{3}k^{4}\right)
T_{a}^{4}+\frac{\mu^{2}}{24}T_{a}^{2},\text{ \ }a=s,v,\tag{6.6}\\
\rho_{g}  &  =\rho_{s}-\rho_{v},\text{ \ \ }p_{g}=\rho_{g}/3. \tag{6.7}%
\end{align}
Here $\left(  5.2\right)  $, $\left(  5.8\right)  $, $\left(  5.28\right)  $
and $m_{s}=m_{v}=0$ are considered. $g_{a}^{\ast}=g_{aB}+7g_{aF}/8,$ $a=s,v$,
is the total number of the spin states, and $g_{aB}$ $\left(  g_{aF}\right)  $
is the total number of the spin states of $a-bosons$ $\left(
a-fermions\right)  .$ Considering $s-matter$ and $v-matter$ are symmetric, we
have%
\[
g_{s}^{\ast}=g_{sB}+7g_{sF}/8=g_{v}^{\ast}=g_{vB}+7g_{vF}/8\equiv g^{\ast}.
\]

As a consequence of $\overset{\cdot}{\rho}_{g}<0,$ $\rho_{g}\left(
T_{cr}\right)  <0$ can be obtained. $\rho_{g}\left(  T_{cr}\right)  <0$ can be
realized, in addition to $\left(  6.2\right)  -\left(  6.3\right)  ,$ because
$\overset{\cdot}{R}<0$ and $\overset{\cdot\cdot}{R}>0$ in the contracting
process so that it is not a thermal equilibrium process.

3. To realize $\rho_{g}\left(  T_{cr}\right)  <0$

When $T_{s\varphi cr}\sim T_{s}\lesssim T_{cr},$ $v_{s\varphi}=v_{v\varphi
}=v_{v\Omega}=0$ and $\left(  5.4\right)  $ and $\left(  5.6\right)  $ are
reduced to%
\begin{equation}
\mu_{s}^{2}\equiv\mu^{2}-\frac{\lambda}{4}T_{s}^{2},\text{ \ \ }m_{s\Omega
}^{2}=2\mu_{s}^{2}=2\lambda v_{\Omega s}^{2}, \tag{6.8}%
\end{equation}%
\begin{equation}
\mu_{v}^{2}\equiv\mu^{2}-\frac{\lambda}{4}T_{v}^{2}-\Lambda v_{\Omega s}%
^{2},\text{ \ }m_{v\Omega}^{2}=-2\mu_{v}^{2} \tag{6.9}%
\end{equation}
It is seem from $\left(  6.8\right)  -\left(  6.9\right)  $ when $\mu_{v}%
\sim0$, $m_{s\Omega}\gtrsim0$. Hence
\begin{equation}
m_{s\Omega}\gtrsim m_{v\Omega},\text{ when }m_{v\Omega}\sim0, \tag{6.10}%
\end{equation}
Thus, $\rho_{s}$ can transform to $\rho_{v},$ i.e. $\overset{\cdot}{\rho}%
_{g}<0,$ when $T_{s\varphi cr}\sim T_{s}\lesssim T_{cr}$ due to $\left(
6.10\right)  $ and $\left(  2.10\right)  ,$ i.e.%
\begin{align}
\Omega_{s}+\Omega_{s}  &  \rightleftarrows\Omega_{v}+\Omega_{v},\text{
\ }\Omega_{s}+\Omega_{s}\rightleftarrows\varphi_{v}+\varphi_{v},\nonumber\\
\varphi_{s}+\varphi_{s}  &  \rightleftarrows\Omega_{v}+\Omega_{v}. \tag{6.11}%
\end{align}
The transformation of $\rho_{s}$ to $\rho_{v}$ is very fast because the
coupling constant $\Lambda$ is very large and $m_{s\Omega}$ and $m_{v\Omega}$
are very small in the case. Consequently, $\rho_{s}=\rho_{v}$ can be realized
when $T_{s}<T_{cr}$ or $t\equiv t_{eq}<t_{cr}$. There must be $\overset{\cdot
}{\rho}_{g}\left(  t_{eq}\right)  <0$, because $V_{g,eff}\left(
t_{eq}\right)  <0$ when $T_{s}\left(  t_{eq}\right)  <T_{cr}$. Thus
$\overset{\cdot}{V}_{g,eff}\left(  t_{eq}\right)  >0$ and $\overset{\cdot
}{\rho}_{g}\left(  t_{eq}\right)  =-\overset{\cdot}{V}_{g,eff}\left(
t_{eq}\right)  <0$ because of $\left(  4.9\right)  $ and $\rho_{g}\left(
t_{eq}\right)  =0.$ Consequently,
\begin{equation}
\rho_{g}\left(  t_{cr}\right)  =\rho_{g}\left(  t_{eq}\right)  +\overset
{\cdot}{\rho}_{g}\left(  t_{eq}\right)  \left(  t_{cr}-t_{eq}\right)  <0.
\tag{6.12}%
\end{equation}

4. The boundary condition

Only when the boundary condition is correct, the solution of a partial
differential equation can be correct. we consider the correct physics boundary
condition of the equations $\left(  6.4\right)  -\left(  6.5\right)  $ is
\begin{equation}
\overset{\cdot}{R}=0\text{ when }R>0. \tag{6.13}%
\end{equation}
The condition
\[
\overset{\cdot}{R}<0\text{ \ when \ }R=0
\]
is not correct, because it is impossible that $\overset{\cdot}{R}<0$ when
$R=0$.

The condition $\left(  6.12\right)  $ requires $\rho_{g}<0$ in the contracting
process. The condition $\left(  6.12\right)  $ can be realized provided
$\rho_{g}\left(  T_{cr}\right)  <0$ (see the following theorem)$,$ because
$\rho_{g}\left(  t\right)  \sim R^{-4}$ so that there must be $-k/R^{2}%
+\eta\left(  -C_{g}/R^{4}+V_{0}\right)  =0$ at $R=R_{\min}>0$. Here $C_{g}>0$
is a constant.

If $\overset{\cdot}{V}_{g,eff}>0$ $\left(  \text{i.e.}.\overset{\cdot}{\rho
}_{g}<0\right)  $ when $t\geq t_{cr},$ there are such solutions satisfying
$\rho_{g}\left(  t_{cr}\right)  =0$ (i.e. $T_{s}=T_{v}$), $\left(
6.14\right)  $ and $\left(  6.4\right)  -\left(  6.5\right)  $. We do not
discuss the case for a time.

\subsection{A theorem related to singularity}

\begin{theorem}
Let $\overset{\cdot}{V}_{g}=0$ and $p_{g}=\kappa\rho_{g}$, $1/3\geq\kappa
\geq0$, for a contracting process in which $\overset{\cdot}{R}<0$ at the
initial time $t_{1}$, then in the case $\rho_{g}<0$, all solutions of $\left(
4.5\right)  -\left(  4.6\right)  $ satisfy the boundary condition
$\overset{\cdot}{R}\left(  t_{2}\right)  =0$ and $R\left(  t_{2}\right)  >0,$
here $t_{2}>t_{1};$ In the case $k>0$ and $\rho_{g}>0,$ when $\rho_{g}\left(
t_{2}\right)  +3p_{g}\left(  t_{2}\right)  -2V_{g}\leq0,$ the solutions of
$\left(  4.5\right)  -\left(  4.6\right)  $ satisfy the boundary condition;
When $k>0,$ $\rho_{g}=0$ and $V_{g}>0,$ the solutions of $\left(  4.5\right)
-\left(  4.6\right)  $ satisfy the boundary condition.
\end{theorem}

It is necessary due to $\overset{\cdot}{V}_{g}=0$ and $\left(  4.5\right)  $
that $V_{g}=V_{g0}\geq0$. This is because if $V_{g0}<0$, there will be
$\left(  \overset{\cdot}{R}/R\right)  ^{2}\longrightarrow\eta V_{g0}<0$ when
$R\longrightarrow\infty.$ This is impossible. Thus we discuss only the case
$V_{g0}\geq0$.

When $k\geq0,$ $V_{g}=0$ and $\rho_{g}<0$, there is no solution of $\left(
4.5\right)  -\left(  4.6\right)  .$ In the case $\rho_{g}>0,$ when $k\leq0$ or
$k>0$ but $\rho_{g}\left(  t_{2}\right)  +3p_{g}\left(  t_{2}\right)
-2V_{g}>0,$ or in the case $k\leq0$, $\rho_{g}=0$ and $V_{g}>0,$ the solutions
of $\left(  4.5\right)  -\left(  4.6\right)  $ do not satisfy the boundary condition.

\begin{proof}
If there is such a time $t_{2}$ satisfying $\overset{\cdot}{R}/R=0,$ there
must be $d\left(  \overset{\cdot}{R}/R\right)  /dt>0$ in the period
$t_{1}<t\leq t_{2},$ because $\overset{\cdot}{R}<0$ at the initial time
$t_{1}.$ From $\left(  4.5\right)  -\left(  4.6\right)  $ we have
\begin{equation}
\frac{d}{dt}\left(  \frac{\overset{\cdot}{R}}{R}\right)  =\frac{k}{R^{2}%
}-\frac{3\eta}{2}\left(  \rho_{g}+p_{g}\right)  . \tag{6.14}%
\end{equation}
When $\overset{\cdot}{R}\left(  t_{2}\right)  /R\left(  t_{2}\right)  =0,$
$\left(  6.14\right)  $ is reduced to
\begin{equation}
\frac{d}{dt}\left(  \frac{\overset{\cdot}{R}}{R}\right)  =\frac{\overset
{\cdot\cdot}{R}}{R}=-\frac{\eta}{2}\left(  \rho_{g}+3p_{g}-2V_{g}\right)  .
\tag{6.15}%
\end{equation}
because of $\left(  4.5\right)  .$ In the case with $k\leq0,$ only when
$\rho_{g}<0,$ $d\left(  \overset{\cdot}{R}/R\right)  /dt>0$ is possible
because of $\left(  6.14\right)  $ and $p_{g}=\kappa\rho_{g}$ ($\kappa\geq
0$)$.$ On the other hand, when $\rho_{g}<0$, $d\left(  \overset{\cdot}%
{R}/R\right)  /dt>0$ must be satisfied due to $\left(  6.15\right)  $, and the
contracting process ($\overset{\cdot}{R}\left(  t_{1}\right)  <0$) can come
into being due to $\left(  4.5\right)  $\ provided $k<0$ or $k=0$ but
$V_{g0}>0$.

In the case with $k\leq0$ and $\rho_{g}>0$ or $V_{g}>0,$ there are the
solutions of $\left(  4.5\right)  -\left(  4.6\right)  $ and the contracting
process. But the solutions cannot satisfy the boundary condition because
$\left(  6.14\right)  $ cannot be satisfied.

In the case with $k>0$ and $\rho_{g}<0,$ It is obvious that $\left(
6.14\right)  $ must be satisfied. But there is no solution of $\left(
4.5\right)  -\left(  4.6\right)  $ when $V_{g0}=0$ in the case. When $V_{g}%
>0$, there are the solutions of $\left(  4.5\right)  -\left(  4.6\right)  $,
and the solutions satisfy the boundary condition.

In the case with $k>0$ and $\rho_{g}>0$ or $V_{g}>0,$ there are the solutions
of $\left(  4.5\right)  -\left(  4.6\right)  $ and the contracting process.
But only when there is such a time $t_{2}$\ satisfying $-k/R\left(
t_{2}\right)  +\eta\left(  \rho_{g}\left(  t_{2}\right)  +V_{g0}\right)  =0$
and $\rho_{g}\left(  t_{2}\right)  +3p_{g}\left(  t_{2}\right)  -2V_{g0}%
\leq0,$ the solutions satisfy the boundary condition due $\left(  4.5\right)
$ and $\left(  6.15\right)  .$ Otherwise, the solutions cannot satisfy the
boundary condition,\ e.g. it is obvious due to $\left(  6.15\right)  $ that
the boundary condition cannot be satisfied when $k>0,$ $\rho_{g}>0$ and
$V_{g0}=0.$

In the discussion above, $\left(  4.12\right)  $ is considered. Without loss
of generality, for simplicity, from $\left(  4.12\right)  $ we write $\rho
_{g}$ as
\[
\rho_{g}=\frac{C_{g}}{R^{3\left(  1+\kappa\right)  }},\text{ \ }0\leq
\kappa\leq1/3.
\]
It is possible that $C_{g}>0$, $=0$ or $C_{g}<0$. Thus, $\mid\rho_{g}\mid$
rises monotonously and faster as $R$ decrease than $\mid k/R^{2}\mid$. Proof ends.

The boundary condition $\left(  13\right)  $ is just the condition without
singularity. From the discussion above, we see $\rho_{g}<0$ to be necessary in
order to eliminate the singularity.
\end{proof}

$R_{\min}>0$ so that there is no singularity of space-time

Space contracts with a deceleration $\overset{\cdot\cdot}{R}>0$, because
$\rho_{g}\left(  t_{cr}\right)  <0$ and $\left(  6.5\right)  .$ As mentioned
before, $V_{g,eff}=V_{0}$ and $\overset{\cdot}{V}_{g,eff}=0$ due
$\langle\omega_{s}\rangle=\langle\omega_{v}\rangle=0$ when $t>t_{cr}.$ Thus
$\rho_{g}\left(  t\right)  =-C_{g}/R^{4}.$ Substituting $\rho_{g}\left(
t\right)  =-C_{g}/R^{4}$ into $\left(  6.4\right)  ,$ taking $k=-1,$ we see
that there must be such a $R_{\min}$ satisfying
\begin{align}
\left(  \frac{\overset{\cdot}{R}_{\min}}{R_{\min}}\right)  ^{2}  &  =\frac
{1}{R_{\min}^{2}}+\eta\left[  -\frac{C_{g}}{R_{\min}^{4}}+V_{0}\right]
=0,\tag{6.16}\\
\frac{\overset{\cdot\cdot}{R}_{\min}}{R_{\min}}  &  =\eta\left[  V_{0}%
+\frac{C_{g}}{R_{\min}^{4}}\right]  >0,\text{ }R_{\min}>0. \tag{6.17}%
\end{align}
It is seen that there is the least scale $R_{\min}$ in the present model.

Let $R=R_{\min}$ when $t=t_{FI}$, then $t_{FI}$ is the final time of the
$s-world$ and the initial time of the $v-world$. It is obvious that
$T_{v}\left(  t_{FI}\right)  $ is the highest temperature $T_{v\max},$
$T_{s}\left(  t_{FI}\right)  =T_{s\max}<T_{v\max},$ $\rho_{s}\left(
t_{FI}\right)  =\rho_{s\max},$ $\rho_{v}\left(  t_{FI}\right)  =\rho_{v\max
}>\rho_{s\max},$ $\rho_{s}\left(  t_{FI}\right)  +\rho_{v}\left(
t_{FI}\right)  =\rho_{\max}$ is the largest energy density $\rho_{\max}$. All
the physics quantities $R_{\min},$ $\rho_{s\max},$ $\rho_{v\max}$, $T_{s\max}$
and $T_{v\max}$ must be finite. Because of the cosmological principle, all
$\rho_{s},$ $\rho_{v},$ $\rho_{g},$ $V_{s}$, $V_{v},$ $V_{g}$ and
$p_{g}\leqslant\rho_{g}/3$ are finite when $t<t_{FI}$. Consequently
$T_{Ss\mu\nu}$, $T_{Sv\mu\nu}$ and $T_{Sg\mu\nu}$ must be finite. On the other
hand, because of the cosmological principle, it is obvious that when space is
not in contracting process or does not contract to $R_{\min},$ the physical
quantities must be finite as well. Substituting the finite $T_{Sg\mu\nu}$ or
$T_{Vg\mu\nu}$ into the Einstein field equation$,$ we see that $R_{\mu\nu}$
and $g_{\mu\nu}$ must be finite.

Consequently, there is no singularity of space-time in the present model.

\subsection{Expansion and inflation of space}

It is seen from $\left(  6.16\right)  -\left(  6.17\right)  $ that space will
expands with an acceleration, because $\overset{\cdot}{R}_{\min}=0$ and
$\overset{\cdot\cdot}{R}_{\min}>0$. The expanding process is different from
the contracting process in essence. $\left(  6.14\right)  $ is no longer the
boundary condition for the expanding process. $\left(  6.16\right)  -\left(
6.17\right)  $ or $\left(  6.14\right)  $ is the initial condition. In the
initial stage, $\langle\omega_{v}\rangle=\langle\omega_{v}\rangle=m_{s}%
=m_{v}=0$ so that $\rho_{s}$ and $\rho_{v}$ can transform from one to another.
Hence there must be such a time $\widetilde{t}_{cr}$ at which
\begin{align}
\rho_{g}  &  =\rho_{v}-\rho_{s}=0,\text{ \ }T_{s}=T_{v}\text{, }\nonumber\\
V_{g,eff}  &  =V_{v,eff}+V_{0}-V_{s,eff}=V_{0}. \tag{6.18}%
\end{align}
The transformation of $\rho_{v}$ and $\rho_{s}$ from one to another is very
fast because the coupling constant $\Lambda$ is very large and $m_{s}=m_{v}%
=0$. The universe is in the most symmetric state when $t\sim\widetilde{t}%
_{cr}$. Thus, $\left(  4.5\right)  $ and $\left(  4.6\right)  $ are reduced
to
\begin{align}
\overset{\cdot}{R}^{2}  &  =1+\eta V_{0}R^{2}=0,\tag{6.19}\\
\overset{\cdot\cdot}{R}  &  =\eta V_{0}R>0. \tag{6.20}%
\end{align}
From $\left(  6.19\right)  -\left(  6.20\right)  $ we see that space inflation
must occur when $t>\widetilde{t}_{cr}$.%
\begin{align}
R\left(  t\right)   &  =\sqrt{\frac{1}{\eta V_{0}}}\sinh H\left(
t-\widetilde{t}_{cr}+\tau\right)  ,\text{ \ }H\equiv\sqrt{\eta V_{0}%
},\tag{6.21}\\
&  \sim\frac{1}{2}\exp H\left(  t-\widetilde{t}_{cr}+\tau\right)  ,\text{ when
}H\left(  t-\widetilde{t}_{cr}+\tau\right)  \gg1,\tag{6.22}\\
R\left(  \widetilde{t}_{cr}\right)   &  =\sqrt{\frac{1}{\eta V_{0}}}\sinh
H\tau. \tag{6.23}%
\end{align}

\subsection{The result above is not contradictory to the singularity theorems}

We first intuitively explain the reasons that there is no space-time
singularity. It has been proved that there is space-time singularity under
certain conditions$^{[1]}$. These conditions fall into three categories.
First, there is the requirement that gravity shall be attractive. Secondly,
there is the requirement that there is enough matter present in some region to
prevent anything escaping from that region. The third requirement is that
there should be no causality violations.

Hawking considers it is a reasonable conjecture that $\rho_{g}=\rho>0$ and
$p_{g}\geq0^{[1]}$. But this conjecture is not valid in the present model,
because $\rho_{g}=\rho_{s}-\rho_{v}>0,$ $=0$ or $<0$ are all possible.

As mentioned above, there must be $\rho_{g}\leq0$ when $T\gtrsim T_{cr}$. It
is seen that $\rho_{g}$ does not only stop increasing, but also decreases from
$\rho_{g}>0$ (when $t<t_{cr}$) to $\rho_{g}\left(  t_{cr}\right)  =0$ and
$\rho_{g}\left(  t\right)  <0$ when $t>t_{cr}.$ Hence the second condition of
the singularity theorem is violated.

The key of non-singularity is $\rho_{sg}=-\rho_{vg}$ when $\rho_{s}=\rho_{v}$
and $\rho_{s}$ and $\rho_{v}$ can transform from one to another when $T\gtrsim
T_{cr}$

We explain the reasons that there is no space-time singularity from the
Hawking theorem as follows. S.W. Hawking has proven the following
theorem$^{[1]}$.

The following three conditions cannot all hold:

$(a)$ every inextendible non-spacelike geodesic contains a pair of conjugate point;

$(b)$ the chronology condition holds on ${\Huge \mu;}$

$(c)$ there is an achronal set $\mathfrak{T}$\ such that $E^{+}(\mathfrak{T}%
)$\ or $E^{-}(\mathfrak{T})$\ \ is compact.\ 

The alternative version of the theorem can obtained by the following two propositions.

Proposition $1^{[1]}$:

If $R_{ab}V^{a}V^{b}\geq0$ and if at some point $p=\gamma(s_{1})$ the tidal
force $R_{abcd}V^{c}V^{d}$ is non-zero, there will be values $s_{0}$\ and
$s_{2}$\ \ such that $q=\gamma\left(  s_{0}\right)  $\ and $r=\gamma\left(
s_{2}\right)  $\ will be conjugate along $\gamma\left(  s\right)  $, providing
that $\gamma\left(  s\right)  $\ can be\ extended to these values.

Proposition $2^{[1]}$:

If $R_{ab}V^{a}V^{b}\geq0$ everywhere and if at $p=\gamma(v_{1}),$ $K^{a}%
K^{b}K_{[a}R_{b]cd[e}K_{f]}$\ is non-zero, there will be $v_{0}$\ and $v_{2}%
$\ \ such that $q=\gamma\left(  v_{0}\right)  $\ and $r=\gamma\left(
v_{2}\right)  $\ will be conjugate along $\gamma\left(  v\right)  $ provided
that $\gamma\left(  v\right)  $\ can be\ extended to these values.

An alternative version of the above theorem is as following.

Space-time (${\Huge \mu,g}$) is not timelike and null geodesically complete if:

$(1)$ $R^{ab}K_{a}K_{b}\geq0$ for every non-spacelike vector $\mathbf{K.}$

$(2)$ The generic condition is satisfied, i.e. every non-spacelike geodesic
contains a point at which $K_{[a}R_{b]cd[e}K_{f]}K^{c}K^{d}\neq0$, where
$\mathbf{K}\mathcal{\ }$is the tangent vector to the geodesic.

$(3)$ The chronology condition holds on ${\Huge \mu}$ (i.e. there are no
closed timelike curves).

$(4)$ There exists at least one of the following:

$(A)$ a compact achronal set without edge,

$(B)$ a closed trapped surface,

$(C)$ a point $p$ such that on every past (or every future) null geodesic from
$p$\ the divergence $\widehat{\vartheta}$\ of the null geodesics from $p $
becomes negative (i.e. the null geodesics from $p$ are focussed by the matter
or curvature and start to reconverge).

In fact, $R_{ab}$ is determined by the gravitational energy-momentum tensor
$T_{gab}$. According to the conventional theory, $T_{gab}=T_{ab}$ so that the
above theorem holds$.$

In contrast with the conventional theory, according to conjecture $1$,
\begin{align}
S_{g\mu\nu}  &  \equiv T_{g\mu\nu}-\frac{1}{2}g_{\mu\nu}T_{g}\nonumber\\
&  =\left(  T_{v\mu\nu}-T_{s\mu\nu}\right)  -\frac{1}{2}g_{\mu\nu}\left(
T_{v}-T_{s}\right)  , \tag{6.24}%
\end{align}
$S_{g00}>0$, $=0$ and $<0$ are all possible. Thus, although the strong energy
condition\ still holds, i.e.
\begin{equation}
\left[  \left(  T_{s}^{ab}+T_{v}^{ab}\right)  -\frac{1}{2}g^{ab}\left(
T_{s}+T_{v}\right)  \right]  K_{a}K_{b}\geq0, \tag{6.25}%
\end{equation}
the conditions of propositions $1$ and $2$ and condition $\left(  1\right)  $
no longer hold, because the gravitational mass density $\rho_{g}$ determines
$R_{\mu\nu}$ and $\rho_{g}=\rho_{v}-\rho_{s}\neq\rho_{v}+\rho_{s}=\rho.$ Hence
$\left(  a\right)  $ and $\left(  c\right)  $ do not hold, but $\left(
b\right)  $ still holds, and ${\Huge \mu}$ is timelike and null geodesically complete.

\section{Evolving process of space after inflation}

\subsection{The reheating process}

After inflation, the temperature must sharply descend. In this case, it is
easily seen that the most symmetric state with $\langle\omega_{s}%
\rangle=\langle\omega_{v}\rangle=0$ is no longer stable and must decay into
such a state with $V_{\min}.$ This is the reheating process. Either of the
$S-breaking$ and the $V-breaking$ can come into being, because $s-matter$ and
$v-matter$ are completely symmetric at $T\sim T_{cr}$. Letting the
$V-breaking$ comes into being$,$ then $v-SU(5)\longrightarrow v-SU(3)\times
U(1)$ and $s-SU(5)$ symmetry is still kept. After the phase transition, we
have%
\begin{equation}
\varpi_{v}=\varpi_{v0},\text{ }\varpi_{s}=0,\text{ }V_{v}=-V_{0}\text{, }%
V_{s}=V_{g}=0. \tag{7.1}%
\end{equation}
After transition, $V_{v}\left(  T_{cr}\right)  -V_{v}\left(  T_{v}%
\sim0\right)  =V_{0}$ must first transform into $v-particles$ by the $v-Higgs$
fields. On the other hand, because of the coupling $(2.10),$ the $v-Higgs$
bosons can transform into $s-particles$ by the $s-Higgs$ fields as well$.$
Letting $\alpha V_{0}$ transform the $v-energy$, then $(1-\alpha)V_{0}$
transforms the $s-energy.$ It is necessary $\alpha>(1-\alpha)$. There must be
$\rho_{v}^{\prime}=\rho_{s}^{\prime}$ before the transition because of
$\left(  6.17\right)  $. Thus, after transition, we have
\begin{equation}
\rho_{v}=\rho_{v}^{\prime}+\alpha V_{0}>\rho_{s}=\rho_{s}^{\prime}%
+(1-\alpha)V_{0}. \tag{7.2}%
\end{equation}
After reheating process, all particles must exist in the form of plasma in the
initial stage. After temperature descends further, $v-particles$ will exist in
the forms of nucleons, leptons and photons, and $s-particles$ will form
$s-SU(5)$ color single states whose masses are all non-zero so that
$\rho_{s\gamma}=0$. Consequently, $\rho_{v}=\rho_{vm}+\rho_{v\gamma}$ and
$\rho_{s}=\rho_{sm},$ here $\rho_{vm}\equiv\rho_{vM}+\rho_{vl}$ and $\rho
_{sm}\equiv\rho_{sM}+\rho_{sl}$. Thus, letting $\rho_{sm}>\rho_{vm}$ and the
reheating process ends at $t_{2}$, from $\left(  7.2\right)  $ we have
\begin{equation}
\rho_{vm}\left(  t\right)  +\rho_{v\gamma}\left(  t\right)  >\rho_{sm}\left(
t\right)  >\rho_{vm}\left(  t\right)  ,\text{ when }t>t_{2}. \tag{7.3}%
\end{equation}

After reheating, $dV_{g}/dt\sim0$ and the temperature effects may be
neglected. Ignoring $p_{sm}$ and $p_{vm},$ considering $p_{v\gamma}%
=\rho_{v\gamma}/3$ and $\rho_{s\gamma}=0$ in the $V-breaking,$ from $\left(
4.13\right)  $ and $\left(  7.3\right)  $ we have
\begin{align}
\rho_{g}R^{3}  &  =\left(  \rho_{vm}-\rho_{sm}\right)  R^{3}=-C_{mg}=\rho
_{g0}R_{0}^{3}<0,\tag{7.4}\\
\rho_{s\gamma g}R^{4}  &  =\rho_{s\gamma}R^{4}=C_{\gamma g}=\rho_{s\gamma
0}R_{0}^{4}>0, \tag{7.5}%
\end{align}
where both $C_{mg}$ and $C_{\gamma g}$ are constants. From $\left(
4.8\right)  $ and $\left(  7.4\right)  -\left(  7.5\right)  ,$ $\left(
4.5\right)  -\left(  4.6\right)  $ is reduced to%

\begin{equation}
\overset{\cdot}{R}^{2}=1+\eta\left(  -\frac{C_{mg}}{R}+\frac{C_{\gamma g}%
}{R^{2}}\right)  , \tag{7.6}%
\end{equation}

\begin{equation}
\overset{\cdot\cdot}{R}=\frac{\eta}{2}\left(  \frac{C_{mg}}{R^{2}}%
-2\frac{C_{\gamma g}}{R^{3}}\right)  . \tag{7.7}%
\end{equation}

We discuss $\left(  7.6\right)  -\left(  7.7\right)  $ as follows. $1-\eta
C_{mg}^{2}/4C_{\gamma g}$ is important for the function $\overset{\cdot}%
{R}^{2}$ of $R^{-1}.$

$1.$ If $1-\eta C_{mg}^{2}/4C_{\gamma g}>0,$ there is not $\overset{\cdot}%
{R}=0.$ When $R<R_{1}\equiv2C_{\gamma g}/C_{mg}$, $\overset{\cdot\cdot}{R}<0$
and $\overset{\cdot}{R}>0$, i.e. space expands with a deceleration; when
$R=R_{1}$, $\overset{\cdot\cdot}{R}=0$ and $\overset{\cdot}{R}=\overset{\cdot
}{R}_{\min}=\left(  1-\eta C_{mg}^{2}/4C_{\gamma g}\right)  ^{1/2}>0$; when
$R>R_{1},$ $\overset{\cdot\cdot}{R}>0$ and $\overset{\cdot}{R}>0$ i.e. space
expands with an acceleration. In the process, $\overset{\cdot\cdot}{R}$
increases from $\overset{\cdot\cdot}{R}=0$ to $\overset{\cdot\cdot}{R}_{\max
}=\eta C_{mg}^{3}/54D_{\gamma g}^{2}$ when $R=3D_{\gamma g}/C_{mg},$ then
$\overset{\cdot\cdot}{R}$ decreases from $\overset{\cdot\cdot}{R}_{\max}$ to
$\overset{\cdot\cdot}{R}_{0}>0.$

$2.$ If $1-\eta C_{mg}^{2}/4C_{\gamma g}=0,$ when $R<R_{1}\equiv2C_{\gamma
g}/C_{mg}$, $\overset{\cdot\cdot}{R}<0$ and $\overset{\cdot}{R}>0$; when
$R=R_{1}$, $\overset{\cdot\cdot}{R}=\overset{\cdot}{R}=0$; In the case, space
can be static.

$3.$ If $1-\eta C_{mg}^{2}/4C_{\gamma g}<0,$ when $R<R_{2}\equiv\left(  \eta
C_{mg}/2\right)  \left[  1-\sqrt{1-4C_{\gamma g}/\eta C_{mg}^{2}}\right]  $,
$\overset{\cdot\cdot}{R}<0$ and $\overset{\cdot}{R}>0$; when $R=R_{2}$,
$\overset{\cdot\cdot}{R}<0$ and $\overset{\cdot}{R}=0$. In the case, space
will begin to contract.

The first case is consistent with observations. A computation in detail is the
same as that of Ref. $[9]$.

Even $\chi_{s}$ and $\chi_{v}$ are considered$,$ the above conclusions still
hold qualitatively.

\subsection{The process of space inflation}

Supposing $\lambda\sim g^{4}$ and $g^{2}\sim4\pi/45$ for $SU(5),$ and
considering $m(\Omega_{s})=\sqrt{2}\mu$, from $(5.17)$ we can estimate
$T_{cr}$ ,
\begin{equation}
T_{cr}=\frac{2\mu}{\sqrt{\lambda}}\sim\frac{2\mu}{g^{2}}\sim\frac{\sqrt
{2}m(\Omega_{s})}{4\pi/45}=5m(\Omega_{s}). \tag{7.8}%
\end{equation}
The temperature will strikingly decrease in the process of inflation, but the
potential energy \newline$V\left(  \varpi_{s}\sim\varpi_{v}\sim0\right)  \sim
V_{0}$ cannot decrease to $V_{\min}\left(  T_{v}\right)  $ at once, because
this is a super-cooling process.\textbf{\ }

We can get the expecting results by suitably choosing the parameters in
$(2.8)-(2.10).$ In order to estimate $H=\sqrt{\eta V_{0}},$ taking $V_{0}%
\sim\mu^{4}/4\lambda,$ from $(7.8)$ we have
\begin{equation}
H=aT_{cr}^{2},\text{ \ \ }a\equiv\sqrt{\eta\lambda}/8\sim g^{2}\sqrt{\eta}/8.
\tag{7.9}%
\end{equation}
We can take $T_{cr}$ to be the temperature corresponding to $GUT$ because the
$SU(5)$ symmetry strictly holds at $T_{cr}$.

Taking $T_{cr}\sim5m\left(  \Omega_{s}\right)  \sim5\times10^{15}Gev$ and
$\sqrt{\lambda}/8\sim g^{2}\sim0.035,$ we have $H^{-1}=10^{-35}s.$ If the
duration of the super-cooling state\textbf{\ }is $10^{-33}s\sim\left(
10^{8}Gev\right)  ^{-1},$ $R_{cr}$ will increase $e^{100}\sim10^{43}$ times.
The result is consistent with the Guth's inflation model$^{[15]}$.

If there is no $v-matter,$ because of contraction by gravitation, the world
would become a thermal-equilibrating singular point, i.e., the world would be
in the hot death state. As seen, it is necessary that there are both
$s-matter$ and $v-matter$ and both the $S-breaking$ and the $V-breaking$.

\subsection{To determine $a\left(  t\right)  $}

Letting $a=R/R_{0},$ $\overset{\cdot}{a}_{0}^{2}=H_{0}^{2}\equiv\eta\rho
_{gc},$ $\Omega_{gm0}=\left(  \rho_{sm0}-\rho_{vm0}\right)  /\rho_{gc},$
$\Omega_{v\gamma0}=\rho_{v\gamma0}/\rho_{gc}$ and $\Omega_{g0}=\Omega
_{gm0}-\Omega_{v\gamma0},$ and considering%
\begin{align}
\rho_{sm}-\rho_{vm}-\rho_{v\gamma}  &  =\rho_{gc}\left[  \Omega_{gm0}%
/a^{3}-\Omega_{v\gamma0}/a^{4}\right]  ,\tag{7.10}\\
H_{0}^{2}\left(  1+\Omega g_{0}\right)   &  =1/R_{0}^{2},\text{ }k=-1,
\tag{7.11}%
\end{align}
we rewrite $\left(  7.6\right)  $ as%
\begin{align}
\overset{\cdot}{a}^{2}  &  =H_{0}^{2}\left(  1+\Omega_{g0}\right) \nonumber\\
&  \cdot\left[  1-\frac{1}{\left(  1+\Omega_{g0}\right)  }\left(  \frac
{\Omega_{gm0}}{a}-\frac{\Omega_{v\gamma0}}{a^{2}}\right)  \right]  .
\tag{7.12}%
\end{align}
From $(7.12)$ we have%

\begin{align}
t  &  =t_{0}-\frac{1}{H_{0}\sqrt{1+\Omega_{g0}}}\{\sqrt{1-M+\Gamma}%
-\sqrt{a^{2}-Ma+\Gamma}\nonumber\\
&  +\frac{M}{2}\ln\frac{2-M+2\sqrt{1-M+\Gamma}}{2a-M+2\sqrt{a^{2}-Ma+\Gamma}%
}\}, \tag{7.13}%
\end{align}
If $t_{0}$ is taken as
\begin{align}
t_{0}  &  =\frac{1}{H_{0}\sqrt{1+\Omega_{g0}}}\{1-M+\Gamma\nonumber\\
&  +\frac{M}{2}\ln\left[  1-\frac{2}{M}-\frac{2\sqrt{1-M+\Gamma}}{M}\right]
\}, \tag{7.14}%
\end{align}
then%
\begin{align}
t  &  =\frac{1}{H_{0}\sqrt{1+\Omega_{g0}}}\{\sqrt{a^{2}-Ma+\Gamma}\nonumber\\
&  +\frac{M}{2}\ln\left[  1-\frac{2a}{M}-\frac{2\sqrt{a^{2}-Ma+\Gamma}}%
{M}\right]  \}, \tag{7.15}%
\end{align}
where $M=\Omega_{gm0}/\left(  1+\Omega_{g0}\right)  ,$ and $\Gamma
=\Omega_{v\gamma0}/\left(  1+\Omega_{g0}\right)  .$

Taking $\Omega_{v\gamma0}=0.001,$ $\Omega_{gm0}=0.3\Omega_{v\gamma0}%
+2\sqrt{\Omega_{v\gamma0}},$ $H_{0}^{-1}=9.7776\times10^{9}h^{-1}yr^{[8]}$.
and $h=0.8,$ we get $a\left(  t\right)  .$ $a\left(  t\right)  $ is shown by
the curve $B$ in the figure 1 and describes evolution of the universe from
$14\times10^{9}yr$ ago to now. Taking $\Omega_{v\gamma0}=0.05,$ $\Omega
_{gm0}=2\sqrt{\Omega_{v\gamma0}},$ we get the $a\left(  t\right)  $ which is
shown by the curve $A$ in the figure 1 and describes evolution of the cosmos
from $13.7\times10^{9}yr$ ago to now. Provided $2\left(  \Omega_{v\gamma
0}+\sqrt{\Omega_{v\gamma0}}\right)  >\Omega_{gm0}$ which is equivalent to
$1-\eta C_{mg}^{2}/4C_{\gamma g}>0$, we can get a curve of $a\left(  t\right)
$ which describes evolution of the cosmological scale.

Figure 1

From the two curves we see that the cosmos must undergo a period in which
space expands with a deceleration in the past, and undergo the present period
in which space expands with an acceleration.

It should be noted that $\rho_{g0}=\rho_{v0}-\rho_{s0}$ in the $V-breaking$,
but here $\Omega_{gm0}=\left(  \rho_{s0}-\rho_{v0}\right)  /\rho_{c}%
=-\rho_{g0}/\rho_{c}$. Ignoring $\Omega_{v\gamma0},$ taking $k=-1$ and%

\begin{align}
\sqrt{1-M}  &  =\frac{1}{\sqrt{1+\Omega_{mg}}},\text{ \ }M=\frac{\Omega_{mg}%
}{1+\Omega_{mg}},\tag{7.16}\\
\sqrt{a^{2}-Ma}  &  =\frac{\sqrt{1-z\Omega_{mg}}}{\left(  1+z\right)
\sqrt{1+\Omega_{mg}}},\text{ }a=\frac{1}{1+z}, \tag{7.17}%
\end{align}
and taking
\begin{equation}
t_{0}=\frac{1}{H_{0}\left(  1+\Omega_{mg}\right)  }\left[  1+\frac{\Omega
_{mg}}{2\left(  1+\Omega_{mg}\right)  ^{1/2}}\ln\frac{2+\Omega_{mg}%
+2\sqrt{\left(  1+\Omega_{mg}\right)  }}{\left(  -\Omega_{mg}\right)
}\right]  , \tag{7.18}%
\end{equation}
we can reduce $\left(  7.15\right)  $ to
\begin{align}
t  &  =\frac{1}{H_{0}\left(  1+\Omega_{mg}\right)  }\{\frac{\sqrt
{1-z\Omega_{mg}}}{\left(  1+z\right)  }+\frac{\Omega_{mg}}{2\sqrt
{1+\Omega_{mg}}}\cdot\nonumber\\
&  \ln\frac{2+\left(  1-z\right)  \Omega_{mg}+2\sqrt{\left(  1+\Omega
_{mg}\right)  \left(  1-z\Omega_{mg}\right)  }}{\left(  1+z\right)  \left(
-\Omega_{mg}\right)  }\}. \tag{7.19}%
\end{align}
Replacing $\Omega_{gm0}$ by $\left(  -\Omega_{gm0}\right)  $ because
$\Omega_{gm0}=\left(  \rho_{s0}-\rho_{v0}\right)  /\rho_{c}=-\rho_{g0}%
/\rho_{c}$, we see $\left(  7.19\right)  $ to be the same as $\left(
3.44\right)  $ in Ref. $[8]$.

\subsection{The relationship between luminosity distance and its redshift}

From $(7.12)$ and the RW metric we have%

\begin{equation}
\int\nolimits_{a}^{1}\frac{cda}{R\overset{\cdot}{a}}=-\int\nolimits_{r}%
^{0}\frac{dr}{\sqrt{1+r^{2}}}, \tag{7.20}%
\end{equation}%
\begin{align}
H_{0}d_{L}  &  =H_{0}R_{0}r(1+z)=\frac{2c}{\left(  \Omega_{gm0}-2\Omega
_{v\gamma0}\right)  ^{2}-4\Omega_{v\gamma0}}\cdot\nonumber\\
&  \{2\left(  1+\Omega_{g0}\right)  -\left(  1+z\right)  \Omega_{gm0}-\left[
2\left(  1+\Omega_{g0}\right)  -\Omega_{gm0}\right]  \cdot\nonumber\\
&  \sqrt{1-\left(  \Omega_{gm0}-2\Omega_{v\gamma0}\right)  z+\Omega_{v\gamma
0}^{2}z^{2}}\}, \tag{7.21}%
\end{align}
where $z=\left(  1/a\right)  -1$ is the redshift caused by $R$ increasing.

Considering $\Omega_{gm0}$ in $\left(  7.21\right)  $ corresponds to
$-\Omega_{m0}$ in $\left(  3.81\right)  $ in Ref. $[8]$ and $\Omega
_{g0}=\Omega_{gm0}-\Omega_{v\gamma0},$ we see that $(7.21)$ is consistent with
$\left(  3.81\right)  $.

Ignoring $\Omega_{v\gamma0},$ taking $\Omega_{gm0}\longrightarrow-\Omega
_{gm0}$, we reduce $(7.21)$ to
\begin{align}
H_{0}d_{L}  &  =\frac{2c}{\Omega_{gm0}^{2}}\cdot\nonumber\\
&  \left\{  2+\Omega_{gm0}\left(  1-z\right)  -\left[  2+\Omega_{gm0}\right]
\sqrt{1-\Omega_{gm0}z}\right\}  , \tag{7.22}%
\end{align}
which is consistent with $\left(  3.78\right)  $ in Ref. $[8]$. Approximating
to $\Omega_{m0}^{1}$ and $z^{2}$, we obtain
\begin{equation}
H_{0}d_{L}=z+\frac{1}{2}z^{2}\left(  1+\frac{1}{2}\Omega_{gm0}\right)  .
\tag{7.23}%
\end{equation}

Taking $\Omega_{v\gamma0}=0.001,$ $\Omega_{gm0}=0.3\Omega_{v\gamma0}%
+2\sqrt{\Omega_{v\gamma0}}$ and $H_{0}^{-1}=9.7776\times10^{9}h^{-1}yr^{[8]}$
and $h=0.8,$ from $\left(  7.22\right)  $ we get the $d_{L}-z$ relation which
is shown by the curve $A$ in the figure 2; Taking $\Omega_{v\gamma0}=0.05,$
$\Omega_{gm0}=2\sqrt{\Omega_{v\gamma0}}$ we get the $d_{L}-z$ relation which
is shown by the curve $B$ in the figure 2.

Figure 2

\section{Existing forms and distribution forms of $s-SU(5)$ color single
states}

In the $V-breaking$, all $s-gauge$ particles and $s-fermions$ are massless.
When the temperature $T_{s}$ is high enough, all $s-particles$ must exist in
plasma form. When $T_{s}$ is low, all $s-particles$ will exist in $s-SU(5)$
color-single state form (conjecture 2). Let $A$, $B$, $C$, $D$, $E$ denote the
5 sorts of colors. A component of \underline{$10$} representation carries
color $\alpha\beta$, $\alpha,$ $\beta=A$, $B$, $C$, $D$, $E,$ $\alpha\neq
\beta.$ A component of \underline{$5$} representation carries color $\alpha.$
A gauge boson carries colors $\alpha\beta^{\ast}.$\ There are the following
sorts of the $s-SU(5)$ color-single states.

2-fermion states: \underline{$\alpha$}\underline{$\alpha$}$^{\ast}$ or
$\left(  \underline{\alpha\beta}\right)  \left(  \underline{\alpha\beta
}\right)  ^{\ast}$, $\alpha\neq\beta.$ 3-fermion states: $\left(
\underline{AB}\right)  (\underline{CD})\underline{E}$ or $\left(
\underline{AB}\right)  \underline{A}^{\ast}\underline{B}^{\ast}.$ 4-fermion
states: $\left(  \underline{AB}\right)  \underline{C}\underline{D}%
\underline{E}$. 5-fermion states:\ $\underline{A}\underline{B}\underline
{C}\underline{D}\underline{E}$ or $\left(  \underline{AB}\right)
(\underline{BC})\left(  \underline{CD}\right)  (\underline{DE})\left(
\underline{EA}\right)  .$ Gauge boson single-states:\ $\left(  \underline
{\alpha\beta^{\ast}}\right)  \left(  \underline{\alpha^{\ast}\beta}\right)  $
or $\left(  \underline{\alpha\beta^{\ast}}\right)  \left(  \underline
{\beta\gamma^{\ast}}\right)  \left(  \underline{\gamma\alpha^{\ast}}\right)  $
etc.$,$ $\alpha\neq\beta\neq\gamma.$ Fermion-gauge boson single-states:
\underline{$\alpha^{\ast}$}$\left(  \underline{\alpha\beta^{\ast}}\right)
$\underline{$\beta$}, \underline{$\alpha^{\ast}$}$\left(  \underline
{\alpha\gamma^{\ast})(\gamma\beta^{\ast}}\right)  $\underline{$\beta$} etc..

The masses of all color single states are non-zero, hence $\rho_{s\gamma}=0$.
The fermions with the spin $s=1/2$ and the least mass and the bosons with the
spin $s=0$ and the least mass are stable, because there is no the electroweak
interaction among $SU(5)$ color single states. Of course, there are the
$s-antiparticles$ corresponding to the $s-colour$ single states above as well.

There are interaction among the $SU(5)$ color single states by exchanging the
$s-SU(5)$ color single states. The interaction radius must be very small
because the masses of all $s-SU(5)$ color single states are non-zero. Thus we
can approximately regard the $s-SU(5)$ color single states as ideal gas
without collision. Consequently the $s-SU(5)$ color single states cannot form
clusters and must distribute loosely in space, but it is possible they form
$s-superclusterings$ as the neutrinos.

\section{New predictions and an inference}

\subsection{New predictions}

\subsubsection{It is possible that huge voids are not empty and are equivalent
to huge concave lenses. The density of hydrogen inside the huge voids is more
less than that predicted by the conventional theory.}

There is no interaction except the gravitation among the $s-SU(5)$ singlets
(the interaction by exchanging $SU(5)$ singlets between two $SU(5)$ singlets
may be neglected), because $SU(5)$ is a simple group. Hence the $s-SU(5)$
singlets cannot form massive $S-objects$, the $s-SU(5)$ color single states
can be regarded as ideal gas without collision, the decoupling temperature of
the $s-SU(5)$ singlets must be very high, and the velocities of the $s-SU(5)$
singlets must be very large and invariant.

The ideal gas has the effect of free flux damping for clustering, i.e. the
$s-SU(5)$ singlets with very high velocities prevents clustering to form. Thus
the $s-SU(5)$ singlets must loosely distribute in space or form
$s-superclusters$ (similar to neutrino superclusters) which are the huge
$v-voids$ for $v-observers$.

Consequently, huge $v-voids$ in the $V-breaking$ are, in fact, superclusters
of $s-SU(5)$ singlets. The huge $v-voids$ are not empty. There must be
$s-matter$ inside them, and $\rho_{s}^{^{\prime}}\gg\rho_{v}^{^{\prime}}$,
$\rho_{s}^{^{\prime}}>\rho_{s}$, and $\rho_{v}^{^{\prime}}<\rho_{v}$ because
of the repulsion between $v-matter$ and $s-matter$. Here $\rho_{s}^{^{\prime}%
}$and $\rho_{v}^{^{\prime}}$ denote the densities inside the huge $v-voids$,
and $\rho_{s}$ and $\rho_{v}$ denote the densities outside the huge $v-voids$.
Because there is the repulsion between $s-matter$ and $v-matter$ and there is
the gravitation among $s-particles$, a huge $v-void$ can form. The characters
of such a huge $v-void$ are as follows:

$A$. A $v-void$ must be huge, because there is no other interaction among the
$s-SU(5)$ color single states except the gravitation and the masses of the
$s-SU(5)$ singlets are very small.

$B$. When $v-photons$ pass through such a huge $v-void$, the $v-photons$ must
suffer repulsion from $\rho_{s}^{^{\prime}}$ and are scattered by the $v-void$
as they pass through a huge concave lens. Consequently, the galaxies behind
the huge $v-void$ seem to be darker and more remote. Hence the huge voids are
equivalent to huge concave lenses.

$C$. Both density of matter and density of dark matter in huge voids must be
more lower than those predicted by the conventional theory. Consequently, the
density of hydrogen and the density of helium inside the huge voids must be
more less than that predicted by the conventional theory.

The predict can be confirmed or negated by the observation of hydrogen distribution.

This is a decisive prediction which distinguishes the present model from other models.

\subsubsection{The gravitation between two galaxies with distant long enough
will be less than that predicted by the conventional theory.}

There must be $s-matter$ between two $v-galaxies$ with distance long enough,
hence the gravitation between the two $v-galaxies$ must be less than that
predicted by the conventional theory due to the repulsion between $s-matter$
and $v-matter$. When the distance between two $v-galaxies$ is small, the
gravitation is not influenced by $s-matter,$ because $\rho_{s}$ must be small
when $\rho_{v}$ is big.

\subsubsection{A black hole with its mass and density big enough will
transform into a white hole}

Letting there be a $v-black$ hole with its mass and density to be so big that
its temperature can arrive at $T_{v}\gtrsim T_{cr}=2\mu/\sqrt{\lambda} $ since
the black hole contracts by its self-gravitation, then the expectation values
of the Higgs fields inside the $v-black$ hole will change from $\varpi
_{v}=\varpi_{v0}$ and $\varpi_{s}=0$ into $\varpi_{v}=\varpi_{s}=0$.
Consequently, inflation must occur. After inflation, the most symmetric state
will transit into the $V-breaking$. Thus, the energy of the black hole must
transform into both $v-energy$ and $s-energy$. Thus, a $v-observer$ will find
that the black hole disappears and a white hole appears.

In the process, part of $v-energy$ transforms into $s-energy.$ A $v-observer$
will consider the energy not to be conservational because he cannot detect
$s-matter$ except by repulsion.\textbf{\ }The\textbf{\ }%
transformation\textbf{\ }of black holes is different from the Hawking
radiation. This is the transformation of the vacuum expectation values of the
Higgs fields. There is no contradiction between the transformation and the
Hawking radiation or another quantum effect, because both describe different
processes and based on different conditions. According to the present model,
there still are the Hawking radiation or other quantum effects of black holes.
In fact, the universe is just a huge black hold. The universe can transform
from the $S-breaking$ into the $V-breaking$ because of its contraction. This
transformation is not quantum effects.

\subsection{An inference : $\lambda_{eff}=\lambda=0,$ although $\rho_{vac}%
\neq0$}

The effective cosmological constant $\lambda_{eff}=\lambda+\rho_{g,vac}.$ The
conventional theory can explain evolution with a small $\lambda_{eff},$ but
$\rho_{g,vac}=\rho_{vac}\ggg\lambda_{eff},$ Consequently, the issue of the
cosmological constant appears.

$\rho_{vac}=0$ can be obtained by some supersymmetric model, but it is not a
necessary result. On the other hand, the particles predicted by the
supersymmetric theory have not been found, although their masses are not large.

$\rho_{vac}=0$ is a necessary result of our quantum field theory without
divergence$^{[6]}$ . In this theory, $\rho_{vac}=0$ is naturally obtained
without normal order of operators, there is no divergence of loop corrections,
all known results are obtained, and dark matter which can form dark galaxies
is predicted$^{[7]}$.

As mention above, this model can explain evolution of the universe without
$\lambda_{eff},$ hence%
\begin{equation}
\lambda_{eff}=0. \tag{9.1}%
\end{equation}

Applying the conventional quantum field theory to the present model, we have
$\rho_{vac}=\rho_{s,vac}+\rho_{v,vac},$ here $\rho_{vac}$ is the energy
density of the vacuum state. According to the conjecture 1, $s-particles$ and
$v-particles$ are symmetric. Hence both ground states must be symmetric as
well. Hence
\begin{equation}
\rho_{s,vac}=\rho_{v,vac}=\rho_{vac}/2. \tag{9.2}%
\end{equation}
According to conjecture 1, $\rho_{gs}=-\rho_{gv}$ when $\rho_{s}=\rho_{v}$.
Consequently, although%
\begin{equation}
\rho_{vac}=\rho_{s,vac}+\rho_{v,vac}=2\rho_{s,vac}\gg0 \tag{9.3}%
\end{equation}
we have still
\begin{align}
\rho_{g,vac}  &  =\rho_{sg,vac}+\rho_{vg,vac}=\rho_{s,vac}-\rho_{v,vac}%
=0,\tag{9.4}\\
\lambda_{eff}  &  =\lambda+\rho_{g,vac}=\lambda=0. \tag{9.5}%
\end{align}
Here $\lambda$ is the Einstein cosmological constant. This is a direct
inference of the present model, and independent of a quantum field theory.

Because of $\left(  9.4\right)  ,$ for the vacuum state in the $S-breaking$ or
the $V-breaking$, the Einstein field equation is reduced to
\begin{align}
R_{\mu\nu}-\frac{1}{2}g_{\mu\nu}R  &  =-8\pi G\left(  T_{s,vac,\mu\nu
}-T_{v,vac,\mu\nu}\right) \nonumber\\
&  =-8\pi G\left(  T_{v,vac,\mu\nu}-T_{s,vac,\mu\nu}\right)  =0. \tag{9.6}%
\end{align}
This is a reasonable result.

\section{Discussion and conclusions}

\subsection{Discussion}

The problem of total energy conservation in the general relativity is unsolved
up to now, because tensors at different points cannot summed up. Based on the
same reason, the problem of total gravitation mass conservation in the general
relativity is unsolved as well. Hence I can only discuss $T_{g;\nu}^{\mu\nu
}=0$ and $T_{,\nu}^{\mu\nu}$.

From $\left(  2.22\right)  $ and $\left(  4.9\right)  $ we see that there are
the following corresponding relations between the conventional theory and the
present model.%
\begin{equation}
T_{;\nu}^{\mu\nu}=0\longrightarrow T_{g;\nu}^{\mu\nu}=0, \tag{10.1}%
\end{equation}%
\begin{align}
\overset{\cdot}{\rho}+3\left(  \rho+p\right)  \frac{\overset{\cdot}{R}}{R}  &
=-\overset{\cdot}{V}\nonumber\\
&  \longrightarrow\overset{\cdot}{\rho}_{g}+3\left(  \rho_{g}+p_{g}\right)
\frac{\overset{\cdot}{R}}{R}=-\overset{\cdot}{V}_{g}. \tag{10.2}%
\end{align}
$\left(  10.2\right)  $ is consistent with $\left(  10.1\right)  $. For
$T_{g}^{\mu\nu}=T_{s}^{\mu\nu}-T_{v}^{\mu\nu},$ there is no such a condition
similar to the dominant energy condition. Hence, there is no such a theorem
similar to the positive energy theorem for $T_{g}^{\mu\nu}$. All $\rho_{g}>0,$
$=0$ or $\rho_{g}<0$ are possible.

If $\overset{\cdot}{V}_{g}=0$, $\rho_{s}$ cannot transform into $\rho_{v}$.
For example, when temperature is low or so high that $\langle\omega_{s}%
\rangle=\langle\omega_{v}\rangle=V_{s}=V_{v}=0$ and $V_{g}=V_{0},$ the
transformation of $\rho_{s}$ and $\rho_{v}$ from one to another may be
neglected. In the case, $\rho_{g}$ is the same as $\rho$ on the left of
$\left(  10.2\right)  $ and $\left(  10.1\right)  $ and $\left(  10.2\right)
$ obviously hold. It i.e. seen that in general, $\left(  10.1\right)  $ and
$\left(  10.2\right)  $ can hold.

When $\overset{\cdot}{V}_{g}\neq0,$ $\rho_{s}$ and $\rho_{v}$ can transform
from one to another. Although $\rho_{s}$ transforms to $\rho_{v}$ so that
$\overset{\cdot}{\rho}_{g}<0$, $\left(  10.1\right)  $ and $\left(
10.2\right)  $ still hold. For example, all $V_{s},$ $\rho_{s}$ and $\rho_{v}$
will increase when space contracts in the $S-breaking,$ but $\left(
10.1\right)  $ and $\left(  10.2\right)  $ still hold. Such a process is not
different from a equilibrium or stable process.

It is possible that the differential law of energy-momentum conservation
holds,
\begin{equation}
T_{S,\nu}^{\mu\nu}=T_{V,\nu}^{\mu\nu}=T_{,\nu}^{\mu\nu}=0, \tag{10.3}%
\end{equation}
where $T^{\mu\nu}$ may contain the contribution of gravitational field as
well. This is because there is no restriction for $T^{\mu\nu}$.

\subsection{Conclusions of the model based on the RW metric}

A new conjecture is proposed that there are $s-matter$ and $v-matter$ which
are symmetric, whose gravitational masses are opposite to each other, although
whose masses are all positive. Both can transform from one to another when
temperature $T\gtrsim T_{cr}$. Consequently there is no singularity in the
model. The cosmological constant $\lambda_{eff}=\lambda=0$ is determined
although the energy density of the vacuum state is still very large. A theorem
related to singularity is presented.

The conjecture are not in contradiction with all given experiments and
astronomical observations up to now, although the conjecture corrects the
equivalence principle. This is because $SU(5)$ singlets which violate the
equivalence principle cannot be observed and can only loosely distribute in
space as so-called dark energy.

There are two sorts of breaking modes, i.e. the $S-breaking$ and the
$V-breaking$. In the $V-breaking$ $v-SU(5)$ is broken into $v-SU(3)\times
U\left(  1\right)  $ and $s-SU(5)$ is still kept$.$ Consequently,
$v-particles$ get their masses and form $v-atoms$, $v-observers$ and
$v-galaxies$ etc., while $s-gauge$ bosons and $s-fermions$ are still massless
and must form $s-SU(5)$ color-single states after reheating. There is no
interaction among the $s-SU(5)$ color-single states except the gravitation,
because $SU(5)$ group is a simple group. Hence they must distribute loosely in
space, cannot be observed and can cause space to expand with an acceleration.
Thus, $v-matter$ is identified with conventional matter (include dark matter)
and $s-matter$ is similar to the dark energy. But in contrast with the dark
energy, the gravitational mass of $s-matter$ is negative in the $V-breaking$.

Based on the present model$,$ the space evolving process is as follows.
Firstly, in the $S-breaking$, $\rho_{g}=\rho_{s}-\rho_{v}>0,$ hence space
contracts and $T_{s}$ rises$.$ When $T_{s}$ arrives the critical temperature
$T_{cr},$ $\langle\omega_{s}\rangle=\langle\omega_{v}\rangle=0,$ the $\left[
v-SU(5)\right]  \times\left[  s-SU(5)\right]  $ symmetry is realized, and the
masses of all particles originating from the couplings with the Higgs fields
are zero so that $\rho_{s}$ and $\rho_{v}$ can transform from one into
another. As a consequence, $\rho_{g}=\rho_{s}-\rho_{v}<0,$ and $V_{g}=V_{0}$.
When space contracts further, $R$ arrives the least scale $R_{\min}>0$ and
$T_{v}$ arrives the highest temperature $T_{\max}$. Then space expands and
inflation must occur. After the inflation, the phase transition of the vacuum
(the reheating process) occurs. After the reheating process, this state with
the highest symmetry transits to the state with the $V-breaking$. Space in the
$V-breaking$ have two evolving stages. Space firstly expands with a
deceleration because $\rho_{g}=\rho_{v}-\rho_{s}>0$; then expands with an
acceleration up to now because $\rho_{g}<0$ and $k=-1$. The results above is
still valid when $V\rightleftarrows S$ and $v\rightleftarrows s.$ It is seen
that the world in the $S-breaking$ and the world in the $V-breaking$ can
transform from one into another.

There are the critical temperature $T_{cr}$, the highest temperature $T_{\max
}$, the least scale $R_{\min}$ and the largest energy density $\rho_{\max}$ in
the universe. $V_{0}$ and $T_{cr}$ are two new important constants.

A formula is derived which well describes the relation between a luminosity
distance and the redshift corresponding to it.

Three new predicts have been given.

Huge $v-voids$ in the $V-breaking$ are not empty, but are superclusters of
$s-particles$. The huge voids are equivalent to huge concave lens. The density
of hydrogen and the density of helium in the huge voids predicted by the
present model must be much less than that predicted by the conventional theory.

The gravitation between two galaxies with distant long enough will be less
than that predicted by the conventional theory.

It is possible that a $v-black$ hole with its big enough mass and density can
transform into a huge white hole by its self-gravitation.

It is possible that the differential law of energy-momentum conservation
holds, $T_{,\nu}^{\mu\nu}=0,$ where $T^{\mu\nu}$ may contain the contribution
of gravitational field as well. This is because there is no restriction for
$T^{\mu\nu}$.

\part{An explanation for some phenomena based on the first model}

\section{Introduction to explanation for some phenomena}

The evolution of the universe has been explained and the cosmological constant
issue has been solved based on the first model. There are two ways to explain
the primordial nucleosynthesis and the cosmic microwave background radiation
$(CMBR)$ by suitably choosing parameters in this model. The first way is based
on this model and the $F-W$ dark matter model$^{\left[  7\right]  }$ or mirror
dark matter model$^{\left[  16\right]  }$; The second way is based on this
model and the cold-dark matter model. Generalizing the equations governing
nonrelativistic fluid motion$^{\left[  8\right]  }$ to the present model, we
have derived the equations of structure formation. According to the equations,
galaxies can form earlier than that according to the Friedmann model; Galaxies
distribution is not uniform in radial direction. There are two sphere layers
in which the density of galaxies is relatively larger. $SU(5)$ color single
states must loosely distribute in space or form superclusters which are
equivalent to huge voids for observers.

Only the expanding process after reheating is considered in the present part.
According to the model, after reheating, the density of the gravitational
potential energy $V_{g}=V_{v}+V_{0}-V_{s}\sim0$ and $\overset{\cdot}{V}%
_{g}\sim0$. Thus the evolution equations $\left(  4.5\right)  -\left(
4.6\right)  $ is reduced to
\begin{equation}
\overset{\cdot}{R}^{2}+k=\eta\rho_{g}R^{2},\text{ }k=-1,\text{\ }\eta
\equiv8\pi G/3, \tag{11.1}%
\end{equation}

\begin{equation}
\overset{\cdot\cdot}{R}=-\frac{1}{2}\eta\left(  \rho_{g}+3p_{g}\right)  R.
\tag{11.2}%
\end{equation}%
\begin{equation}
\rho_{g}=\rho_{v}-\rho_{s},\text{ }p_{g}=p_{v}-p_{s},\text{\ in the
}V-breaking. \tag{11.3}%
\end{equation}

\subsection{The $F-W$ dark matter model}

The $F-W$ dark matter model$^{[7]}$ is a necessary inference of the quantum
field theory without divergence$^{[6]}$. The $F-W$ dark matter model is
similar with the mirror dark matter model.

According to the mirror dark matter model, it is impossible that the density
of matter is equal to that of mirror matter in order to explain the primordial
nucleosynthesis and $CMBR$. This is too difficult to understanding, because
matter and mirror matter are symmetric and both can transform from one into
another when temperature is high enough.

In contrast with the mirror dark matter model, according to the $F-W$ model,
$F-matter$ and $W-matter$ are completely symmetric in both properties and
density, there is no interaction except the gravitation and Higgs couplings
between $F-matter$ and $W-matter$. Thus, $W-matter$ is dark matter for a
$F-observer,$ and vice versa. Both $F-matter$ and $W-matter$ belong to
$v-matter$. We regard $F-matter$ as conventional matter, and $W-matter$ as
dark matter or mirror matter.

In order to analyze the primordial nucleosynthesis, we divide $F-matter$ into
three sorts. The first sort is called $F-massive$ particles whose masses are
denoted by $M_{vF},$ and $M_{vF}>1MeV$. It is composed of $F-nucleons$ and
unknown $F-particles$. The unknown $F-particles$ are invisible for a time so
that they belong to cold dark matter. The second sort is called $F-light$
particles whose masses are denoted by $m_{vFl},$ $1MeV>m_{vFl}\gg1eV$, e.g.
$F-electrons$. The third sort is called $F-photon-like$ particles whose masses
are zero, e.g. $F-photons$ and $F-neutrinos$ whose masses are approximately
regarded as zero.

Let the densities of the three sorts of particles are $\rho_{vFM},$
$\rho_{vFl},$ and $\rho_{vF\gamma},$ respectively, then
\begin{equation}
\rho_{vF}=\rho_{vFM}+\rho_{vFl}+\rho_{vF\gamma}. \tag{11.4}%
\end{equation}
Considering $F-matter$ and $W-matter$ to be completely symmetric, we have
\begin{align}
\rho_{vFM}  &  =\rho_{vWM},\text{ }\rho_{vFl}=\rho_{vWl},\text{ }%
\rho_{vF\gamma}=\rho_{vW\gamma},\tag{11.5}\\
\rho_{vW}  &  =\rho_{vWM}+\rho_{vWl}+\rho_{vW\gamma}=\rho_{vF}.\tag{11.6}\\
\rho_{v}  &  =\rho_{vF}+\rho_{vW}=2\rho_{vF}=\rho_{vM}+\rho_{vl}+\rho
_{v\gamma}.\tag{11.7}\\
\rho_{vM}  &  =2\rho_{vFM},\text{ }\rho_{vl}=2\rho_{vFl},\text{ }\rho
_{v\gamma}=2\rho_{vF\gamma}.\tag{11.8}\\
\rho_{vFM}  &  =\rho_{vFB}+\rho_{vUF},\text{ }\rho_{vUF}=\rho_{vUW},\text{
}\rho_{vFB}=\rho_{vWB}, \tag{11.9}%
\end{align}
where $\rho_{vFB}$ is the density of F-baryons and $\rho_{vUF}$ is the density
of unknown F-particles. Both $W-matter$ and unknown $F-particles$ are
invisible for a $F-observer$, hence the density $\rho_{dm}$ of dark matter is
\begin{align}
\rho_{dm}  &  =\rho_{vW}+\rho_{vUF}\equiv\rho_{cd}+\rho_{hd},\tag{11.10}\\
\rho_{cd}  &  \equiv\rho_{vWB}+\rho_{vUW}+\rho_{vUF},\tag{11.11}\\
\rho_{hd}  &  \equiv\rho_{vWl}+\rho_{vW\gamma}, \tag{11.12}%
\end{align}
where $\rho_{cd}$ and $\rho_{hd}$ are the density of cold-dark matter and the
density of hot-dark matter, respectively. Observations show $\rho_{hd0}\ll
\rho_{cd0}.$ It is seen from $\left(  11.6\right)  $ and $\left(  11.9\right)
-\left(  11.10\right)  $ that dark matter is composed of three parts. The
first part is the $W-baryons$ which can form galaxies (dark galaxies) as the
$F-baryons.$ The second part is unknown $F-matter$ which is invisible for a
time. The third part is composed of unknown $W-matter,$ $W-photons$ and light
$W-particles$ which are possibly always invisible and have only cosmological effects.

The density of visible matter is%
\[
\rho_{vis}=\rho_{vFB}+\rho_{vFl}+\rho_{vF\gamma}.
\]

$\rho_{vF}$ and $\rho_{vW}$ can transform from one into another when
temperature is high enough by the couplings of $F-Higgs$ and $W-Higgs$ fields.

In $V-breaking$, all $s-particles$ must form $s-SU(5)$ color single states
whose masses are not zero, i.e. there is no $s-photon$ so that $\rho_{s\gamma
}=0$. Thus, s-particles may be divided into two sorts. The first sort is
composed of massive s-particles whose masses $M_{s}>1MeV$ and density is
$\rho_{sM};$ The second sort is composed of light s-particles with their
masses $m_{sl}$ and density $\rho_{sl}$, $1MeV>m_{sl}\gg1eV$. Thus,
\begin{equation}
\rho_{sm}=\rho_{sM}+\rho_{sl}. \tag{11.13}%
\end{equation}
$\rho_{sl}/\rho_{sM}$ is a undetermined parameter.

When temperature $T<1MeV,$ $p_{sM}\ll\rho_{sM}$ and $p_{vM}\ll\rho_{vM}$\ so
that $p_{sM}$ and $p_{vM}$ may be neglected.\ Considering $V_{g}%
=\overset{\cdot}{V}_{g}=0$ after particles decouple, and $p_{v\gamma}%
=\rho_{v\gamma}/3,$ letting $p_{sl}=\kappa\rho_{sl}$ and $p_{vl}=\kappa
\rho_{vl}$, from $\left(  11.1\right)  -\left(  11.2\right)  $ we have%
\[
\frac{d\left(  \rho_{Mg}R^{3}\right)  }{dt}+\frac{d\left(  \rho_{lg}%
R^{3\left(  1+\kappa\right)  }\right)  }{R^{3\kappa}dt}+\frac{d\left(
\rho_{v\gamma g}R^{4}\right)  }{Rdt}=0,
\]%
\begin{align}
\rho_{Mg}R^{3}  &  =\left(  \rho_{vM}-\rho_{sM}\right)  R^{3}=\rho_{vM}\left(
1-S_{M}\right)  R^{3}\sim C_{Mg},\nonumber\\
\rho_{lg}R^{3\left(  1+\kappa\right)  }  &  =\left(  \rho_{vl}-\rho
_{sl}\right)  R^{3\left(  1+\kappa\right)  }\sim C_{lg},\text{ \ }%
\rho_{v\gamma g}R^{4}\sim C_{v\gamma g}, \tag{11.14}%
\end{align}
where all $C_{mg},$ $C_{lg}$ and $C_{v\gamma g}$ are parameters, and
$S_{M}\equiv\rho_{sM}/\rho_{vM}$ is a undetermined parameter. When temperature
is high enough, e.g. $T\gtrsim T_{cr}$, all masses may be regarded photon-like
particles. $\kappa$ is a function of temperature,%
\begin{align}
\kappa &  \lesssim1/3,\text{ \ when }T\gtrsim1MeV,\text{ }\nonumber\\
0  &  <\kappa<1/3\text{ when }1MeV>T\gtrsim1eV\nonumber\\
0  &  \lesssim\kappa,\text{ \ when }T\lesssim1eV. \tag{11.15}%
\end{align}

\subsection{The cold-dark matter model}

Both matter and dark matter belong to v-matter according to this cosmological
model. Cold-dark matter is composed of such particles which have large masses
and very weak interaction. The present observations show that there are dark
energy and dark matter$.$ Visible matter is ordinary baryon matter. Let the
density of dark energy, the density of visible matter, the density of
cold-dark matter, the density of hot-dark matter be $\rho_{de},$ $\rho_{B},$
$\rho_{cd},$ and $\rho_{hd},$ respectively, then $\rho_{dm}=\rho_{cd}%
+\rho_{hd}$ and the total density is $\rho_{t}=\rho_{de}+\rho_{dm}+\rho
_{B}+\rho_{l}+\rho_{\gamma}$. Observations show
\begin{align}
\frac{\rho_{de0}}{\rho_{t0}}  &  =0.73,\text{ \ }\frac{\rho_{dm0}}{\rho_{t0}%
}=0.23,\text{ \ }\frac{\rho_{B0}}{\rho_{t0}}=0.04,\nonumber\\
\rho_{dm0}/\rho_{B0}  &  \sim6,\text{ \ }\rho_{cd0}\gg\rho_{hd0}\sim
\rho_{\gamma0}\text{, \ }\rho_{dm0}\sim\rho_{cd0}. \tag{11.16}%
\end{align}
After nucleons decouple, $\rho_{dm}\sim R^{-3}$ and $\rho_{cd}\sim R^{-3}.$
From $\left(  11.16\right)  $ we have
\begin{equation}
\rho_{dm}\sim\rho_{cd}\sim6\rho_{B}. \tag{11.17}%
\end{equation}

According to this cosmological model, there is no dark energy, but there are
s-matter and v-matter. In the V-breaking, s-matter corresponds the dark
energy, but $\rho_{sg}=-\rho_{s}.$ For primordial nucleosynthesis and $CMBR,$
$\left(  11.1\right)  -\left(  11.3\right)  $ is applicable. The hot-dark
matter is composed of neutrinos mainly and $\rho_{hd}\sim\rho_{\gamma}$. For
simplicity, we replace $\rho_{hd}+\rho_{\gamma}$ by $\rho_{\gamma}$. Combining
this cosmological model and the the cold-dark matter model, we have
\begin{equation}
\rho_{v}=\rho_{B}+\rho_{cd}+\rho_{l}+\rho_{\gamma},\text{ \ }\rho_{hd}%
+\rho_{\gamma}\longrightarrow\rho_{\gamma} \tag{11.18}%
\end{equation}
Thus $\left(  11.1\right)  $ can be write as
\begin{align}
H^{2}  &  =\left(  \overset{\cdot}{R}/R\right)  ^{2}=-k/R^{2}+\eta\{\left(
\rho_{B}+\rho_{cd}+\rho_{l}+\rho_{\gamma}\right) \nonumber\\
&  -\left(  \rho_{sM}+\rho_{sl}\right)  \}. \tag{11.19}%
\end{align}

In section 12, primordial nucleosynthesis is explained; In section 13, $CMBR$
is explained; In section 14, dynamics of structure formation and the
distributive form of the $SU(5)$ color single states;\ Section 15 is the
conclusion of the part.

\section{Primordial nucleosynthesis}

\subsection{Explanation for primordial nucleosynthesis the by the Friedmann
model}

The primordial helium abundance $Y_{4}$ is determined by $n_{n}/n_{p}^{[17]}%
$,
\begin{align}
Y_{4}  &  =2/\left[  1+\left(  n_{n}/n_{p}\right)  ^{-1}\right]  ,\tag{12.1}\\
n_{n}/n_{p}  &  =exp\left(  -\bigtriangleup m/kT_{d}\right)  ,\text{
}\bigtriangleup m=m_{n}-m_{p},\nonumber
\end{align}
where $n_{n}/n_{p}$ is the neutron-proton ratio in the unit comoving volume at
the freeze-out temperature $T_{d}.$ $T_{d}$ is determined by $\Gamma
=\Gamma\left(  T\right)  $ and $H\left(  T\right)  =\overset{\cdot}{R}/R,$
e.g.
\begin{equation}
\Gamma\left(  T\right)  =H\left(  T\right)  , \tag{12.2}%
\end{equation}
here $\Gamma$ is the interaction rate experienced by a nucleon. According to
the conventional theory
\begin{equation}
\Gamma\sim G_{F}^{2}T^{5}, \tag{12.3}%
\end{equation}
where $G_{F}$ is the weak interaction Fermi constant.

Considering the cold-dark matter model and the Friedmann model with $k=0$ and
the effective cosmological constant $\lambda_{eff}$, we have
\begin{equation}
\widetilde{H}^{2}=\eta\left(  \rho_{B}+\rho_{cd}+\rho_{l}+\rho_{\gamma
}+\lambda_{eff}\right)  \tag{12.4}%
\end{equation}

The freezing-out temperature is $T_{d}\sim0.8MeV$ at which $\Gamma\left(
Td\right)  \sim H\left(  T_{d}\right)  .$ Consequently $\rho_{B}$ may be
neglected. The reasons are as follows.%
\begin{align}
\rho_{B}  &  =\rho_{B0}\left(  \frac{R_{0}}{R}\right)  ^{3},\text{ \ }%
\rho_{\gamma}=\rho_{\gamma0}\left(  \frac{R_{0}}{R}\right)  ^{4},\nonumber\\
\frac{\rho_{B}}{\rho_{\gamma}}  &  =\frac{\rho_{B0}}{\rho_{\gamma0}}\frac
{R}{R_{0}},\text{ \ }\frac{R}{R_{0}}\sim\frac{T_{\gamma0}}{T_{\gamma}}.
\tag{12.5}%
\end{align}
$T_{\gamma}$ is the effective temperature of photons. $T_{\gamma}=T$ before
photons decouple. $T_{\gamma0}=0.235\times10^{-4}eV.$ Thus $T_{d}/T_{0}%
\sim10^{10}.$\ Substituting $T_{d}/T_{0}$ and $\rho_{B0}/\rho_{\gamma0}%
\sim6000^{[5]}$ into $\left(  12.4\right)  $, we have $\rho_{B}\left(
T_{d}\right)  \ll\rho_{\gamma}\left(  T_{d}\right)  $ so that $\rho_{B}\left(
T_{d}\right)  $ may be neglected. According to the cold-dark matter model,
$\rho_{dm}\sim\rho_{cd}\sim6\rho_{B}\sim R^{-3}$ when $T\sim T_{d},$ and
$\lambda_{eff}$ is very small and is invariant. Hence $\rho_{cd}$ and
$\lambda_{eff}$ may be neglected when $T\sim T_{d}$ (see $\left(
11.14\right)  ,$ $\left(  11.17\right)  -\left(  11.18\right)  $).

When $T\sim T_{d},$ electrons and neutrinos may be regarded as photon-like
particles. Considering three generations of neutrinos and electrons and their
antiparticles, we have%
\[
g^{\ast}=2+\left(  7/8\right)  \times\left(  2\times2+3\times2\right)  =43/4
\]
Thus, from the Friedmann model we have
\begin{equation}
\widetilde{H}^{2}\left(  T_{d}\right)  \sim\eta\left(  \rho_{l}+\rho_{\gamma
}\right)  =\eta\frac{\pi^{2}}{30}g^{\ast}T_{d}^{4}. \tag{12.6}%
\end{equation}
Taking $\left(  12.2\right)  $ as a rough approximation, from $\left(
12.3\right)  $ and $\left(  12.6\right)  $ we get $T_{d}\sim0.8MeV.$
Substituting $T_{d}\sim0.8MeV$ into $\left(  12.1\right)  ,$ we get
\begin{equation}
n_{n}/n_{p}\sim1/7,\text{ \ }Y_{4}\sim1/4. \tag{12.7}%
\end{equation}

\subsection{Explanation for primordial nucleosynthesis by this cosmological
model and the F-W dark matter model or the cold dark matter model}

As mentioned before, $F-matter$ is regarded as the conventional matter. Thus
$\rho_{vFB}=\rho_{B},$ $\rho_{vFl}=\rho_{l},$ $\rho_{vF\gamma}=\rho_{\gamma}$
and $T_{v}=T.$

Thus, $\left(  12.3\right)  $ still holds, provided $T$ is replaced by $T_{v}%
$. This is because both $W-particles$ and $s-particles$ do not influence
$F-nuclear$ interactions. Thus, provided $\left(  12.6\right)  $ is derived
from the present model, this model can explain primordial nucleosynthesis.

Observations show $\rho_{dm}/\rho_{B}\sim6.$ As mentioned above, when $T\sim
T_{d}$, $\rho_{B}\ll\rho_{\gamma},$ i.e. $\rho_{vFB}\ll\rho_{vF\gamma}$.
Consequently, $\rho_{vFB}$ and $\rho_{cd}$ may be neglected when $T\sim T_{d}$
due to $\rho_{cd}\lesssim\rho_{dm}.$ Thus $\rho_{vM}$ may be neglected due to
$\left(  11.8\right)  -\left(  11.9\right)  .$

Before photons decouple, $T_{\gamma}=T$. For the stage in which photon-like
particles is dominative,
\begin{equation}
\frac{T_{s}}{T_{s0}}=\frac{R_{0}}{R}=\frac{T_{v}}{T_{v0}} \tag{12.8}%
\end{equation}
we have $T_{s}\sim T_{v}$. Thus, because of $\left(  11.15\right)  -\left(
11.17\right)  $ and $\rho_{sM}\sim\rho_{vM}\ll\rho_{vF\gamma},$ $\rho_{sM}$
may be neglected when $T_{v}\sim T_{d}$. On the other hand, when $T_{v}\sim
T_{d},$
\begin{equation}
k/\eta R^{2}\ll\rho_{vF\gamma}\sim R^{-4}, \tag{12.9}%
\end{equation}
so that $k/\eta R^{2}$ may be neglected when $T_{v}\sim T_{d}.$

Taking%
\begin{equation}
\rho_{vWl}\left(  T\right)  +\rho_{vW\gamma}\left(  T\right)  -\rho
_{sl}\left(  T\right)  \sim0,\text{ when }T\gtrsim T_{d}, \tag{12.10}%
\end{equation}
neglecting $\rho_{vFM}$ and $\rho_{sM}$, when $T_{v}\sim T_{d}$, we can reduce
$\left(  11.1\right)  $ as
\begin{align}
H^{2}  &  =\eta\{2\left(  \rho_{vFM}+\rho_{vFl}+\rho_{vF\gamma}\right)
-\left(  \rho_{sM}+\rho_{sl}\right)  \}\nonumber\\
&  \sim\eta\left(  \rho_{vFl}+\rho_{vF\gamma}\right)  =\eta\frac{\pi^{2}}%
{30}g^{\ast}T_{d}^{4}, \tag{12.11}%
\end{align}
where $g^{\ast}=43/4.$ Thus, $\left(  12.11\right)  $ is the same as $\left(
12.6\right)  .$ It is seen that this model and the $F-W$ model can still
explain the primordial nucleosynthesis and $Y_{4}$, although $\rho_{vF}%
=\rho_{vW}$. This is different from the mirror dark matter model.

It is easily seen from $\left(  11.10\right)  -\left(  11.12\right)  $ that
the cold dark matter model is equivalent to $\rho_{vWl}=\rho_{vW\gamma}=0.$
From $\left(  12.10\right)  $ we see that provided $\rho_{sl}=0$ is taken,
when $T\sim T_{d},$ we can still explain the primordial nucleosynthesis based
this model and the cold dark matter model.

For the cold-dark matter model, $\left(  -k/R^{2}\right)  ,$ $\rho_{B\text{,
}}\rho_{cd}$ and $\rho_{sM}$ may be neglected because of the same reasons as
above. From $\left(  11.18\right)  -\left(  11.19\right)  $ we see that
provided $\rho_{sl}\sim0$ is taken when $T\sim T_{d},$ we can still explain
the primordial nucleosynthesis based this cosmological model and the cold dark
matter model.

\section{Cosmic microwave background radiation}

\subsection{The recombination temperature $T_{rec}$}

It is the same as the conventional theory that there are the inflation and big
bang processes in the present model. Hence there must be $CMBR.$

As mentioned before, $v-F-matter$ is regarded as the conventional matter.
Thus, $T=T_{v}.$ The recombination mechanism and temperature of this model is
the same as those of the conventional theory, because there is no interaction
between the $v-F-particles$ and the $v-W-particles$ except the gravitation and
there is no interaction between the $v-F-particles$ and the $s-particles$
except the repulsion as well. From the Saha formulas$^{[17]}$ we can determine
the recombination temperature $T_{vrec},$%
\begin{align}
\frac{1-\chi}{\chi^{2}}  &  =1.1\times10^{-8}\xi T_{v}^{3/2}\exp\frac
{13.6}{T_{v}}\nonumber\\
&  =3.96\times10^{-14}\xi A^{3/2}\exp\frac{57872}{A},\tag{13.1}\\
\chi &  =n_{p}/\left(  n_{p}+n_{H}\right)  ,\text{ }\nonumber\\
A  &  \equiv T_{\gamma v}/T_{v\gamma0}=T_{v}/T_{v\gamma0},\nonumber
\end{align}
where $n_{p}=n_{e}$ and $n_{H}$ are the number densities of protons and
hydrogens, respectively, $T_{v\gamma0}=2.35\times10^{-4}ev$, and $13.6ev$ is
the ionization potential energy of hydrogen. Before $v-F-photons$ decouple,
$T_{\gamma v}=T_{v}.$

Taking $\xi\sim5\times10^{-10}$ and $\chi=0.1,$ we obtain$^{[5]}$.%
\begin{equation}
T_{vrec}=0.295ev=T_{rec},\text{ }\left(  1+z_{rec}\right)  =T_{vrec}%
/T_{v\gamma0}=1255. \tag{13.2}%
\end{equation}
$T_{vrec}$ is just the result of the conventional theory.

\subsection{The temperature of matter-radiation equality}

Let the temperature of matter-radiation equality be $T_{eq},$ then $T_{eq}\sim
T_{vrec}\ll m_{e}\sim m_{vFl}.$ Thus $p_{vFl}$ may be neglected so that
$\rho_{vFl}\propto R^{-3}$. When $T_{v}\leq T_{eq}$, because of $\left(
11.8\right)  $ and $\left(  11.14\right)  $, the density of $v-matter$ can be
written as%
\begin{equation}
\rho_{vm}=2\rho_{vFm}=2\left(  \rho_{vFM}+\rho_{vFl}\right)  =\rho_{vm0}%
R_{0}^{3}/R^{3}. \tag{13.3}%
\end{equation}
In contrast with the conventional theory, according to the $F-W$ model, not
only there are $F-photon$ (ordinary photons), but also $W-photons$ (dark
photons), and $\rho_{vF\gamma}=\rho_{vW\gamma}.$ From this we can estimate
$T_{eq}$ as follows.

Because $\left(  11.14\right)  $ and $\left(  11.8\right)  $ or $\left(
11.5\right)  ,$ the density of photon-like particles is
\begin{equation}
\rho_{v\gamma}\equiv2\rho_{vF\gamma}=2\rho_{vF\gamma0}R_{0}^{4}/R^{4}%
=2\times\frac{\pi^{2}}{30}g_{\gamma}^{\ast}T_{v\gamma}^{4}, \tag{13.4}%
\end{equation}%
\begin{equation}
g_{\gamma}^{\ast}=2+\frac{7}{8}\times6\times\left(  \frac{4}{11}\right)
^{4/3}=3.36. \tag{13.5}%
\end{equation}
Here the photons and the three species of neutrinos are considered$^{[18]}$.

Letting $\rho_{vm}\left(  T_{eq}\right)  =\rho_{v\gamma}\left(  T_{eq}\right)
$, Considering $\left(  R_{0}/R\right)  =\left(  T_{v\gamma}/T_{v\gamma
0}\right)  $, from $\left(  13.3\right)  -\left(  13.4\right)  $ we have%
\begin{equation}
\frac{\rho_{vm0}}{2\rho_{vF\gamma0}}=\frac{1}{2}\cdot\frac{\rho_{m0}}%
{\rho_{\gamma0}}=\frac{R_{0}}{R}=\frac{T_{v\gamma}}{T_{v\gamma0}} \tag{13.6}%
\end{equation}
Considering $T_{vF\gamma0}=T_{0}=2.35\times10^{-4}ev,$ from $\left(
13.4\right)  -\left(  13.5\right)  $ we get%
\begin{equation}
\rho_{v\gamma0}=2\rho_{vF\gamma0}=6.7425\times10^{-51}Gev^{4}. \tag{13.7}%
\end{equation}
Observations show that the total density of matter and dark matter is
$\rho_{0}=\Omega_{0}\rho_{c}=0.27\rho_{c}$, $H_{0}^{2}\equiv\eta\rho_{c}.$
According to the $F-W$ model, this implies%
\begin{align}
\rho_{0}  &  =\rho_{v0}=2\left(  \rho_{vFM0}+\rho_{vFl0}+\rho_{vF\gamma
0}\right) \nonumber\\
&  \simeq2\left(  \rho_{vFM0}+\rho_{vFl0}\right)  =\rho_{vm0}=\Omega_{0}%
\rho_{c}\nonumber\\
&  =1.8789\times10^{-26}h^{2}\Omega_{0}\cdot kg\cdot m^{-3}\nonumber\\
&  =9.238\times10^{-48}Gev^{4},\text{ \ when }h=0.65. \tag{13.8}%
\end{align}
where $h=0.5-0.8.$ Here $\rho_{\gamma0}$ is neglected due to $\rho_{B0}%
/\rho_{\gamma0}\sim6000^{[5]}.$ From $\left(  13.6\right)  -\left(
13.8\right)  $ we obtain%
\begin{equation}
T_{veq}=T_{v\gamma eq}=0.32ev,\text{ when }h=0.65. \tag{13.9}%
\end{equation}
$\rho_{\gamma0}=\rho_{v\gamma0}/2$ and $\rho_{m0}=\rho_{vm0}$ in $\left(
13.6\right)  $ are the densities of radiation and matter, respectively, in the
conventional theory. Let $T_{eq}^{\prime}$ be is the temperature of
matter-radiation equality in the conventional theory, from $\left(
13.6\right)  $ we see
\begin{equation}
T_{eq}^{\prime}=2T_{veq}=0.64ev=2T_{veq}. \tag{13.10}%
\end{equation}

\subsection{Decoupling temperature}

The decoupling temperature of photons $T_{dec}$ is determined by
$\Gamma\left(  T_{dec}\right)  $ and $H\left(  T_{dec}\right)  .$ According to
the conventional theory$^{[17]}$,%

\begin{align}
\Gamma &  =n_{e}\sigma_{Th},\text{ }n_{e}=n_{p}\equiv\chi\xi n_{\gamma
}\nonumber\\
\left(  n_{p}+n_{H}\right)   &  =\xi n_{\gamma}=\xi\left(  \frac{2.4}{\pi^{2}%
}T^{3}\right)  ,\nonumber\\
\xi &  \sim5\times10^{-10},\text{ }\sigma_{Th}=1.71\times10^{3}GeV^{-2}%
.\nonumber\\
\Gamma &  =5.4\times!0^{-36}\chi\xi A^{3}GeV, \tag{13.11}%
\end{align}
where $A\equiv T_{\gamma}/T_{\gamma0}=T_{v}\gamma/T_{v\gamma0}$.

From the present model and the $F-W$ model $\left[  7\right]  $ we have the
same result as $\left(  13.11\right)  $, because F-matter is regarded as
conventional matter. In order to determine the decoupling temperature
$T_{dec},$ only it is necessary to determine $H\left(  T_{v}\right)  .$

When $T_{v}=T_{vdec},$ $\rho_{vm}$ is the same as $\left(  13.3\right)  $
because $T_{dec}\sim T_{veq}<1ev$. Let when $T_{v}=T_{veq},$ $T_{s}=T_{sq}$
and $\rho_{sm}=\rho_{sM}+\rho_{sl}\equiv s_{mq}\rho_{vm}\left(  T_{veq}%
\right)  .$ Considering
\begin{equation}
\frac{T_{sq}}{T_{s}}=\frac{R}{R_{eq}}=\frac{T_{veq}}{T_{v}} \tag{13.12}%
\end{equation}
and $\rho_{v\gamma,eq}=\rho_{vm,eq}$, we can rewrite $\rho_{vm},$
$\rho_{v\gamma}$ and $\rho_{sm}$ as follows.
\begin{align}
\rho_{vm}  &  =\frac{\rho_{vm}}{\rho_{vm0}}\frac{\rho_{vm0}}{\rho_{c}}\rho
_{c}=\frac{R_{0}^{3}}{R^{3}}\Omega_{m0}\rho_{c}\nonumber\\
&  =A^{3}\Omega_{m0}\rho_{c},\text{ }A\equiv\frac{T_{v\gamma}}{T_{v\gamma0}%
}=\frac{R_{0}}{R},\tag{13.13}\\
\rho_{v\gamma}  &  =\frac{\rho_{v\gamma}}{\rho_{v\gamma,eq}}\frac
{\rho_{v\gamma,eq}}{\rho_{vm,eq}}\frac{\rho_{vm,eq}}{\rho_{vm0}}\frac
{\rho_{vm0}}{\rho_{c}}\rho_{c}\nonumber\\
&  =\left(  \frac{T_{v\gamma0}}{T_{v\gamma,eq}}\right)  A^{4}\Omega_{m0}%
\rho_{c},\tag{13.14}\\
\rho_{sm}  &  =\frac{\rho_{sm}}{\rho_{sm,eq}}\frac{\rho_{sm,eq}}{\rho_{vm,eq}%
}\frac{\rho_{vm,eq}}{\rho_{vm0}}\frac{\rho_{vm0}}{\rho_{c}}\rho_{c}\nonumber\\
&  =s_{mq}A^{3}\Omega_{m0}\rho_{c},\tag{13.15}\\
s_{mq}  &  \equiv\frac{\rho_{sm,eq}}{\rho_{vm,eq}}=\frac{\rho_{sm0}}%
{\rho_{vm0}},\text{ e.g. }s_{mq}=1.5. \tag{13.16}%
\end{align}
$\left(  11.14\right)  $ is considered in $\left(  13.16\right)  $. $k=-1$ may
be neglected because $R_{0}/R_{vdec}\sim T_{vdec}/T_{0}\sim10^{3}$ and%
\begin{align}
\left(  -k/\eta R_{0}^{2}\right)  \left(  R_{0}/R_{vdec}\right)  ^{2}  &
\ll\rho_{g0}\left(  R_{0}/R_{vdec}\right)  ^{3},\tag{13.17}\\
R_{vdec}  &  =R_{0}T_{v\gamma0}/T_{vdec},\text{\ }T_{vdec}=T_{v\gamma
dec}.\nonumber
\end{align}
Considering\textbf{\ }$H_{0}=\sqrt{\eta\rho_{c}}=65km\cdot\left(  s\cdot
Mpc\right)  ^{-1}=1.4\times10^{-42}Gev,$\textbf{\ }$\Omega_{vm0}%
=0.27,$\textbf{\ }and\textbf{\ }$T_{v0}/Tveq=\left(  2.35\times10^{-4}\right)
/0.32=0.734\times10^{-3},$ from $\left(  13.13\right)  -\left(  13.16\right)
$ we get%
\begin{align}
H^{2}  &  =\eta\rho_{g}=\eta\left(  \rho_{vm}+\rho_{v\gamma}-\rho_{sm}\right)
\nonumber\\
&  =0.27\times1.4^{2}\times10^{-82}A^{3}\cdot\nonumber\\
&  \left(  1-s_{mq}+0.734\times10^{-3}A\right)  \left(  Gev\right)  ^{2}.
\tag{13.18}%
\end{align}
This is the only difference between this model and the Friedmann model in
order to determine the decoupling temperature.

Taking the crude approximation%
\begin{equation}
\Gamma\left(  T_{v}\right)  =H\left(  T_{v}\right)  , \tag{13.19}%
\end{equation}
from $\left(  13.11\right)  $ and $\left(  13.18\right)  -\left(
13.19\right)  $ we can represent $\chi$ and $\chi A^{3/2}$ by $A$. Thereby
from $\left(  13.1\right)  $, $\left(  13.11\right)  $ and $\left(
13.18\right)  -\left(  13.19\right)  $ we can determine $A$ and $\chi.$
\begin{equation}
\chi A^{3/2}=\xi^{-1}1.347\times10^{-7}\left(  1-t_{mq}+0.734\times
10^{-3}A\right)  ^{1/2} \tag{13.20}%
\end{equation}%
\begin{align}
A^{3/2}  &  =\xi^{-1}1.347\times10^{-7}\left(  1-t_{mq}+0.734\times
10^{-3}A\right)  ^{1/2}\nonumber\\
&  +\xi^{-1}7.185\times10^{-28}\nonumber\\
&  \times\left(  1-t_{mq}+0.734\times10^{-3}A\right)  \times\exp\frac
{57872}{A} \tag{13.21}%
\end{align}
Taking $s_{mq}=1.5$ and $\xi=5\times10^{-10},$ we have%

\begin{align}
A  &  =1+z_{dec}=1096.5,\tag{13.22}\\
T_{vdec}  &  =AT_{\gamma0}=0.257eV,\text{ \ }\chi=0.004. \tag{13.23}%
\end{align}
This result is consistent with that of the Friedmann model $T_{dec}=0.25eV$
and $\chi=0.004^{\left[  17\right]  }$. It can be proved that $z_{dec}$ is not
susceptible for change of $s_{mq}$ in the scope $1.1-1.7.$

For the cold-dark matter model, $\left(  13.11\right)  $ still holds. Thus,
provided $\rho_{sl}=0$ and $s_{mq}=1.1$ in $\left(  13.18\right)  $ and
$\left(  13.21\right)  ,$ considering $\rho_{vm}=\rho_{B}+\rho_{cd}$,
$\rho_{\gamma}=\rho_{v\gamma}$ and $\rho_{sm}=\rho_{sM},$ from $\left(
11.19\right)  $ we can still obtain the result $\left(  13.22\right)  -\left(
13.23\right)  .$ In fact, in the case, there is only one sort of photon-like
particles, i.e. ordinary photons and neutrinos. Hence $Tveq\longrightarrow
T^{\prime}veq=2Tveq=0.64eV$ so that $0.734\times10^{-3}=\left(  T_{v0}%
/Tveq\right)  =\left(  2.35\times10^{-4}\right)  /0.32$ becomes $\left(
T_{v0}/T^{\prime}veq\right)  =\left(  2.35\times10^{-4}\right)  /0.64.$ In
order to get the result $\left(  13.22\right)  -\left(  13.23\right)  ,$
provided $s_{mq}=1.5\longrightarrow1.1$. Thus, we can still obtain the results above.

Sum up, we see that the primordial nucleosynthesis and $CMBR$ can be explained
based on this model and the $F-W$ dark matter model or the cold dark matter model.

\subsection{Space is open}

From $\left(  13.2\right)  $ we see $T_{vrec}=T_{rec}$. On the other hand, as
mentioned before, $F-matter$ is same as the conventional matter and there are
the same reheating process so that $T_{vreh}=T_{reh}$.

The interaction of $s-Higgs$ fields and $v-Higgs$ fields $\left(  2.10\right)
$ may be neglected, because $m\left(  \Omega_{s}\right)  $ and $m\left(
\Omega_{v}\right)  $ are very large after reheating. Thus, there is no
interaction except the repulsion or the gravitation among $F-matter$,
$W-matter$ and $s-matter$. The speed in $F-matter$ is not influenced by
$W-matter$ and $s-matter$. On the other hand, $F-matter$ and $W-matter$ are
symmetric and $F-matter$ is identified as conventional matter. Hence we have
\begin{align}
c_{vFs}  &  =c_{vWs}\equiv c_{v}=c_{s},\tag{13.24}\\
T_{vrec}  &  =T_{rec},\text{ \ }T_{vreh}=T_{reh} \tag{13.25}%
\end{align}
Here $c_{vFs}=\partial p_{vF}/\partial\rho_{vF}$ and $c_{vWs}=\partial
p_{vW}/\partial\rho_{vW}$ are the sound speeds in the $F-matter$ or in the
$W-matter,$ respectively.

After reheating, temperature is so high that there are the $F-plasma$, the
$W-plasma$ and the $S-plasma$. The $F-plasma$ and the $W-plasma$ are the same
as the conventional plasma, because $F-matter$ and $W-matter$ are symmetric.

Let $\bigtriangleup t_{vhc}$ be the duration in which $T_{vreh}$ descends into
$T_{vrec}$ according to this model and $\bigtriangleup t_{hc}$ be that
according to the Friedmann model, there must be
\begin{equation}
\bigtriangleup t_{vhc}>\bigtriangleup t_{hc}. \tag{13.26}%
\end{equation}
The reasons are as follows.

$\bigtriangleup t_{hc}$ is determined by the Friedmann model and the cold-dark
matter model in the conventional theory. As mentioned before, according to the
Friedmann model and the cold-dark matter model, $\lambda_{eff}\ll\rho_{B}%
\ll\rho_{\gamma}$ when $T\gtrsim T_{d}$. Thus $\lambda_{eff},$ $\rho_{B}$ and
$\rho_{cd}\sim6\rho_{B}$ may be neglected. Thus, $\left(  12.4\right)  $ is
reduced to
\begin{equation}
\widetilde{H}^{2}\sim\eta\left(  \rho_{l}+\rho_{\gamma}\right)  =\eta\frac
{\pi^{2}}{30}g^{\ast}T^{4}\text{ \ when.}T\gtrsim T_{d}, \tag{13.27}%
\end{equation}%
\begin{equation}
\widetilde{H}^{2}=\eta\left(  \rho_{B}+\rho_{cd}+\rho_{l}+\rho_{\gamma
}\right)  \text{ when }T_{d}>T\gtrsim T_{rec}. \tag{13.28}%
\end{equation}
Thus $\bigtriangleup t_{hc}$ is determined by $\left(  13.27\right)  -\left(
13.28\right)  .$ When $T_{d}>T\gtrsim T_{rec},$ $\rho_{B},$ $\rho_{cd}$ and
$\rho_{sM}$ cannot be neglected because $\rho_{B}$ and $\rho_{sM}\varpropto
R^{-3}$ and $\rho_{\gamma}\varpropto R^{-4}$

$\bigtriangleup t_{vhc}$ is determined by $\left(  11.1\right)  $ and $\left(
12.10\right)  $. As mentioned before, $k$ may be neglected when $T_{v}\gtrsim
T_{rec}\sim1eV.$ Thus we can rewrite $\left(  11.1\right)  $ as%
\begin{equation}
H^{2}\left(  T_{v}\right)  =\eta\lbrack2\left(  \rho_{vFM}+\rho_{vFl}%
+\rho_{vF\gamma}\right)  -\left(  \rho_{sM}+\rho_{sl}\right)  ]. \tag{13.29}%
\end{equation}
Considering $\rho_{B}=\rho_{vFB},$ $\rho_{l}=\rho_{vFl\text{, }}\rho_{\gamma
}=\rho_{vF\gamma}$ $\left(  12.10\right)  $ and $\left(  11.11\right)  ,$
neglecting $\rho_{sM}$ and $\rho_{vFM},$ we again get%
\begin{equation}
H^{2}=\eta\left(  \rho_{vFl}+\rho_{vF\gamma}\right)  =\eta\frac{\pi^{2}}%
{30}g^{\ast}T_{v}^{4}\text{ when }T_{v}\gtrsim T_{d}. \tag{13.30}%
\end{equation}
Considering $T_{v}=T$, we have
\begin{equation}
H=\widetilde{H}\text{ \ \ when \ }T\gtrsim T_{d}. \tag{13.31}%
\end{equation}

From $\left(  11.8\right)  -\left(  11.9\right)  ,$ $\left(  11.11\right)  $
and $\left(  13.16\right)  $ we have
\begin{align}
2\rho_{vFM}  &  =\rho_{B}+\rho_{cd},\text{ }\rho_{vFl}+\rho_{vF\gamma}%
=\rho_{l}+\rho_{\gamma}.\nonumber\\
\rho_{sM}  &  \sim1.5\times2\rho_{vFM}. \tag{13.32}%
\end{align}
When $T_{d}>T\gtrsim T_{rec},$ from $\left(  11.14\right)  -\left(
11.15\right)  $ and $\left(  12.10\right)  $ we have
\begin{equation}
\rho_{vWl}\left(  T\right)  +\rho_{vW\gamma}\left(  T\right)  -\rho
_{sl}\left(  T\right)  \lesssim0. \tag{13.33}%
\end{equation}
Considering $\left(  13.32\right)  -\left(  13.33\right)  $ and comparing
$\left(  13.28\right)  $ and $\left(  13.29\right)  $, we find%
\begin{equation}
H<\widetilde{H}\text{ \ \ when \ }T_{d}>T\gtrsim T_{rec}. \tag{13.34}%
\end{equation}
Consequently, it is necessary due to $\left(  13.31\right)  $ and $\left(
13.34\right)  $ that $T_{reh}=T_{vreh}$ can descend faster to $T_{rec}%
=T_{vrec}$ according to the Friedmann model than that according to the present
model and the $F-W$ model. Thus, $\left(  13.26\right)  $ must holds$.$
Considering $c_{vs}=c_{s}$, we have
\begin{equation}
c_{vs}\bigtriangleup t_{vhc}=c_{s}\bigtriangleup t_{vhc}>c_{s}\bigtriangleup
t_{hc}. \tag{13.35}%
\end{equation}
Based on $\bigtriangleup t_{hc}=3.8\times10^{5}yr,$ space is regarded to be
flat according to the conventional theory. But based on $\bigtriangleup
t_{vhc}>\bigtriangleup t_{hc}$ and $c_{vs}=c_{s},$ space is open $\left(
k<0\right)  $ according to the this model. This shows this model to be self consistent.

Combining the cold dark matter model and this model, we can still get the
result similar to $\left(  13.35\right)  $. The cold dark matter model is
equivalent to $\rho_{vm}=\rho_{B}+\rho_{cd},$ $\rho_{vl}+\rho_{v\gamma}%
=\rho_{l}+\rho_{\gamma}$, $\rho_{sM}=\rho_{sm}$, $\rho_{vWl}+\rho_{vW\gamma
}=0$ so that $\rho_{sl}=0,$ and $s_{sq}=\rho_{sm}/\rho_{vm}=1.1$. When
$T\gtrsim T_{d},$ $\left(  13.30\right)  $ still holds because $\rho_{sm},$
$\rho_{B}+\rho_{cd}$ may be neglected. When $T_{d}>T\gtrsim T_{rec},$ we have%
\begin{equation}
H^{2}=\eta\left(  \rho_{B}+\rho_{cd}+\rho_{vl}+\rho_{v\gamma}-\rho
_{sm}\right)  . \tag{13.36}%
\end{equation}
It is seen from $\left(  13.28\right)  $ and $\left(  13.36\right)  $ that
$\left(  13.34\right)  $ still holds because $\rho_{sm}\sim1.1\rho_{vm}$.

\section{Dynamics of structure formation and the distributive form of the
$SU(5)$ singlets}

\subsection{The equations of structure formation}

Generalizing the equations governing nonrelativistic fluid motion$^{[8]}$ to
the model without singularity, in the $V-breaking,$ considering $\left(
1\right)  -\left(  3\right)  ,$ we have%

\begin{equation}
\left(  \frac{\partial}{\partial t}+\mathbf{v}_{v}\mathbf{\cdot\nabla}\right)
\mathbf{v}_{v}=-\frac{\mathbf{\nabla}p_{v}}{\rho_{v}}-\mathbf{\nabla}\Phi,
\tag{14.1}%
\end{equation}%
\begin{equation}
\frac{\partial}{\partial t}\rho_{v}+\mathbf{\nabla}\cdot\left(  \rho
_{v}\mathbf{v}_{v}\right)  =0 \tag{14.2}%
\end{equation}%
\begin{equation}
\nabla^{2}\Phi=4\pi G\left(  \rho_{v}-\rho_{s}\right)  , \tag{14.3}%
\end{equation}%
\begin{equation}
\left(  \frac{\partial}{\partial t}+\mathbf{v}_{s}\mathbf{\cdot\nabla}\right)
\mathbf{v}_{s}=-\frac{\mathbf{\nabla}p_{s}}{\rho_{s}}+\mathbf{\nabla}\Phi,
\tag{14.4}%
\end{equation}%
\begin{equation}
\frac{\partial}{\partial t}\rho_{s}+\mathbf{\nabla}\cdot\left(  \rho
_{s}\mathbf{v}_{s}\right)  =0, \tag{14.5}%
\end{equation}
where $\partial/\partial t+\mathbf{v}_{v}\mathbf{\cdot\nabla}$ is call the
convective derivative$^{[8]}$. We can produce the linearized equations of
motion by collecting terms of first order in perturbations about a homogeneous
background $\rho_{v}=\rho_{v0}+\delta\rho_{v}$ etc.. Letting%

\begin{align}
\mathbf{v}_{v}  &  =\mathbf{v}_{v0}+\delta\mathbf{v}_{v},\quad
\ \ \ \ \mathbf{v}_{s}=\mathbf{v}_{s0}+\delta\mathbf{v}_{s}\tag{14.6}\\
\rho_{v}  &  =\rho_{v0}+\delta\rho_{v},\quad\ \ \ \ \rho_{s}=\rho_{s0}%
+\delta\rho_{s},\tag{14.7}\\
\delta_{v}  &  =\frac{\delta\rho_{v}}{\rho_{v0}},\quad\ \ \ \ \delta_{s}%
=\frac{\delta\rho_{s}}{\rho_{s0}}, \tag{14.8}%
\end{align}
where $\mathbf{v}_{0}=H\mathbf{x=}\left(  \overset{\cdot}{a}/a\right)
\mathbf{x}$ is the Hubble expansion. Now, for sufficiently small
perturbations, we get%

\begin{equation}
\left(  \frac{\partial}{\partial t}+\mathbf{v}_{v0}\mathbf{\cdot\nabla
}\right)  \delta\mathbf{v}_{v}=-\frac{\nabla\delta p_{v}}{\rho_{v0}}%
-\nabla\delta\Phi-\left(  \delta\mathbf{v}_{v}\cdot\nabla\right)
\mathbf{v}_{v0}, \tag{14.9}%
\end{equation}%
\begin{equation}
\left(  \frac{\partial}{\partial t}+\mathbf{v}_{v0}\cdot\nabla\right)
\delta_{v}=-\mathbf{\nabla\cdot}\delta\mathbf{v}_{v}, \tag{14.10}%
\end{equation}%
\begin{equation}
\nabla^{2}\Phi=4\pi G\left(  \rho_{v0}\delta_{v}-\rho_{s0}\delta_{s}\right)  ,
\tag{14.11}%
\end{equation}%
\begin{equation}
\left(  \frac{\partial}{\partial t}+\mathbf{v}_{0}\mathbf{\cdot\nabla}\right)
\delta\mathbf{v}_{s}=-\frac{\mathbf{\nabla}\delta p_{s}}{\rho_{s0}%
}+\mathbf{\nabla}\delta\Phi-\left(  \delta\mathbf{v}_{s}\cdot\mathbf{\nabla
}\right)  \mathbf{v}_{0}, \tag{14.12}%
\end{equation}%
\begin{equation}
\left(  \frac{\partial}{\partial t}+\mathbf{v}_{0}\cdot\mathbf{\nabla}\right)
\delta_{s}=-\mathbf{\nabla\cdot}\delta\mathbf{v}_{s}. \tag{14.13}%
\end{equation}
Defining the comoving spatial coordinates%
\begin{equation}
\mathbf{x}(t)=a\left(  t\right)  \mathbf{r}\left(  t\right)  ,\quad
\delta\mathbf{v}\left(  t\right)  =a\left(  t\right)  \mathbf{u}\left(
t\right)  ,\text{ }\mathbf{\nabla}_{x}=\frac{\mathbf{\nabla}_{r}}{a}.
\tag{14.14}%
\end{equation}
$\left(  14.9\right)  -\left(  14.10\right)  $ and $\left(  14.12\right)
-\left(  14.13\right)  $ can be rewritten as
\begin{equation}
\overset{\cdot}{\mathbf{u}}_{v}+2\frac{\overset{\cdot}{a}}{a}\mathbf{u}%
_{v}\mathbf{=}\frac{\mathbf{\nabla}\delta\Phi}{a^{2}}-\frac{\mathbf{\nabla
}\delta p_{v}}{\rho_{v0}} \tag{14.15}%
\end{equation}%
\begin{equation}
\overset{\cdot}{\delta}_{v}=-\mathbf{\nabla\cdot}\delta\mathbf{u}_{v},
\tag{14.16}%
\end{equation}%
\begin{equation}
\overset{\cdot}{\mathbf{u}}_{s}+2\frac{\overset{\cdot}{a}}{a}\mathbf{u}%
_{s}\mathbf{=-}\frac{\mathbf{\nabla}\delta\Phi}{a^{2}}-\frac{\mathbf{\nabla
}\delta p_{s}}{\rho_{s0}} \tag{14.17}%
\end{equation}%
\begin{equation}
\overset{\cdot}{\delta}_{s}=-\mathbf{\nabla\cdot}\delta\mathbf{u}_{s},
\tag{14.18}%
\end{equation}
Letting
\begin{align}
\delta_{vk}  &  =\delta_{vk}\left(  t\right)  \exp\left(  -i\mathbf{k}%
_{v}\cdot\mathbf{r}\right)  ,\text{ \ }c_{v}^{2}=\frac{\partial p_{v}%
}{\partial\rho_{v}},\tag{14.19}\\
\delta_{sk}  &  =\delta_{sk}\left(  t\right)  \exp\left(  -i\mathbf{k}%
_{s}\cdot\mathbf{r}\right)  ,\ \ c_{s}^{2}=\frac{\partial p_{s}}{\partial
\rho_{s}}, \tag{14.20}%
\end{align}
from $\left(  14.15\right)  -\left(  14.20\right)  $ and $\left(
14.11\right)  $ we can get%

\begin{equation}
\overset{\cdot\cdot}{\delta}_{vk}+2\frac{\overset{\cdot}{a}}{a}\overset{\cdot
}{\delta}_{vk}=4\pi G\left(  \rho_{v0}\delta_{vk}-\rho_{s0}\delta_{sk}\right)
-\frac{c_{v}^{2}k_{v}^{2}}{a^{2}}\delta_{vk}, \tag{14.21}%
\end{equation}%
\begin{equation}
\overset{\cdot\cdot}{\delta}_{sk}+2\frac{\overset{\cdot}{a}}{a}\overset{\cdot
}{\delta}_{sk}=4\pi G\left(  \rho_{s0}\delta_{sk}-\rho_{v0}\delta_{vk}\right)
-\frac{c_{s}^{2}k_{s}^{2}}{a^{2}}\delta_{sk}, \tag{14.22}%
\end{equation}
Where dots stand for $d/dt=\left(  \partial/\partial t+\mathbf{v}%
_{0}\mathbf{\cdot\bigtriangledown}_{r}\right)  .$ It is necessary that
$\delta_{sk}<0$ when $\delta_{vk}>0$ and that $\delta_{sk}>0$ when
$\delta_{vk}<0,$ because there is only repulsion between $s-matter$ and
$v-matter.$ Consequently,
\begin{align}
\rho_{v0}\delta_{vk}-\rho_{s0}\delta_{sk}  &  =\rho_{v0}\delta_{vk}+\rho
_{s0}\mid\delta_{sk}\mid,\tag{14.23}\\
\rho_{s0}\delta_{sk}-\rho_{v0}\delta_{vk}  &  =\rho_{s0}\delta_{sk}+\rho
_{v0}\mid\delta_{vk}\mid. \tag{14.24}%
\end{align}

\subsection{Three predictions.}

1. Galaxies can form faster and earlier according to the present model than
that according to the Friedmann model.

From $\left(  14.21\right)  $ and $\left(  14.23\right)  $ we see that
$\delta_{vk}\left(  t\right)  $ can grow faster than that determined by the
Friedmann model because $\rho_{v0}\delta_{vk}+\rho_{s0}\mid\delta_{sk}%
\mid>\rho_{v0}\delta_{vk}.$ The origin is the repulsion between $s-matter$ and
$v-matter$.

2. Galaxies distribution is not uniform in radial direction. There are two
sphere layers in which the density of galaxies is relatively larger.

According to the present model, there is $\overset{\cdot}{a}_{\min}$. If
$\overset{\cdot}{a}_{\min}$ is so small that it may be neglected, $v_{0}$ may
be neglected as well. For simplicity, $\rho_{s0}\delta_{sk}$ in $\left(
14.21\right)  $ is neglected, because $\rho_{v0}\delta_{vk}+\rho_{s0}%
\mid\delta_{sk}\mid>\rho_{v0}\delta_{vk}.$ Thus, in the period in which
$\overset{\cdot}{a}_{\min}$ is may be neglected, for long-wavelength, i.e.
$\left(  4\pi G\rho_{v0}-c_{v}^{2}k_{v}^{\prime2}\right)  >0$, from $\left(
14.21\right)  $ and $\left(  14.23\right)  $ we see that $\delta_{vk}\left(
t\right)  $ can grow exponentially,%

\begin{equation}
\overset{\cdot\cdot}{\delta}_{vk}\left(  t\right)  =\left(  4\pi G\rho
_{v0}-c_{v}^{2}k_{v}^{\prime2}\right)  \delta_{vk},\quad\ \ \ \delta
_{vk}\left(  t\right)  =\exp\left(  t/\tau\right)  , \tag{14.25}%
\end{equation}
where $\tau=1/\sqrt{4\pi G\rho_{v0}-c_{v}^{2}k_{v}^{\prime2}},$ $k_{v}%
^{\prime}=k_{v}/a.$ For observation, this implies that the density of galaxies
and the brightness will be relatively larger in the sphere layer corresponding
to $\overset{\cdot}{a}_{\min}\sim0$ than those in near other sphere layer. Of
course, the density of galaxies on the sphere-surface is still uniform.

If $\left(  \mid\delta_{sk}\mid/\delta_{vk}\right)  $ in $\left(
14.23\right)  $ is regarded as invariant approximately, we have%
\begin{equation}
\tau=\left[  4\pi G\left(  \rho_{v0}+\rho_{s0}\mid\delta_{sk}\mid/\delta
_{vk}\right)  -c_{v}^{2}k_{v}^{\prime2}\right]  ^{-1/2}. \tag{14.26}%
\end{equation}

In the $V-breaking$, $v-particles$ can form particles with large masses as
$v-atoms$ and $v-molecules$ by the electroweak interaction. The density
$\rho_{v\gamma}$ of photon-like particles may be neglected and matter is
dominative in low temperatures. In this stage, the velocities $u_{v}^{\prime
}s$ of the particles with large masses may be neglected so that $\left(
\overset{\cdot}{a}/a\right)  \cdot\mathbf{u}_{v}$ in $\left(  14.15\right)  $
may be neglected as well. Thus, in the period in which $\rho_{vm}$ is
relatively larger and $\left(  \overset{\cdot}{a}/a\right)  \cdot
\mathbf{u}_{v}$ may be neglected, for long-wavelength, from $\left(
14.15\right)  $ and $\left(  14.21\right)  $ we get again the result similar
$\left(  14.25\right)  .$

Thus, galaxies distribution is not uniform in radial direction. There are two
sphere layers in which the density of galaxies is relatively larger.

3. The $s-SU(5)$ color single states must loosely distribute in space or form
$s-superclustering$, i.e. huge $v-voids.$

There is no interaction except the gravitation among the $s-SU(5)$ color
single states, because $SU(5)$ is a simple group and $SU(5)$ symmetry holds in
the V-breaking. There is no interaction except the repulsion between
$v-matter$ and $s-matter$. Consequently, the $s-SU(5)$ color single states
have decoupled after reheating, and can be regarded as ideal gas without
collision. The ideal gas has the effect of free flux damping for clustering,
i.e. the $s-SU(5)$ color single states with very high velocities prevents
$\delta_{sk}\left(  t\right)  $ to increase. The velocities $u_{s}$ of the
$s-SU(5)$ color single states must be very large and invariant, because their
decoupling temperature, i.e. the temperature after the reheating, is very
high, and there is no interaction except the repulsion or gravitation. Thus
$\overset{\cdot}{\delta}_{s}=-\mathbf{\nabla\cdot u}_{s}$ cannot be small.
Hence when $\overset{\cdot}{a}/a$ cannot be neglected, $\left(  \overset
{\cdot}{a}\overset{\cdot}{\delta}_{sk}/a\right)  $ in $\left(  14.22\right)  $
cannot be neglected as well. The damping term $\left(  \overset{\cdot}%
{a}\overset{\cdot}{\delta}_{sk}/a\right)  $ reduces $\delta_{sk}\left(
t\right)  $ to grow. When $\left(  \overset{\cdot}{a}/a\right)  $ may be
neglected, the perturbation $\delta_{sk}$ will slowly grow in power rules.
Thus, the $s-SU(5)$ color single states must loosely distribute in space or
form $s-superclusters$ which are the huge $v-voids$ for $v-observers$. In the
huge $v-voids,$ $\rho_{s}\gg\rho_{v},$ because of the repulsion between
$v-matter$ and $s-matter$. The features of the huge v-voids have been
discussed before.

\section{Conclusions of the explanation for some phenomena}

Based on the first model without singularity and the $F-W$ dark matter model
or cold dark matter model, the primordial nucleosynthesis and $CMBR$ are
explained. Based on the analysis of the baryon-elementary wave which began at
reheating and ended at recombination, we see that space is open. This shows
this model to be self-consistent. Generalizing equations governing
nonrelativistic fluid motion to the present model, we have derived the
equations of structure formation. Based on the equations, the following three
predictions are obtained. Galaxies can form faster and earlier according to
the present model than according to the Friedmann model; Galaxies distribution
is not uniform in radial direction. There are two sphere layers in which the
density of galaxies is relatively larger. $SU(5)$ singlets must loosely
distribute in space or form superclusters which are equivalent to huge voids
for observers.

\part{A cosmological model without singularity based on a new metric}

\section{Introduction to the model based on a new metric}

As mentioned in section 2, the existing probability of the $S-breaking$ and
the $V-breaking$ must be equal, because the $s-Higgs$ fields and the $v-Higgs$
fields are symmetric. This equality can be realized by the following two sorts
of modes.

$(1)$ The whole universe is in the same breaking (e.g. the $S-breaking$), but
this sort of breaking (the $S-breaking$) can transform to the other (the
$V-breaking$) as space contracts to the least scale $R_{\min}.$

In this case, a hypersurface of simultaneity must be complete as the same as
the Friedmann model. The $3-$dimensional volume of a complete hypersphere is
finite, and the 3-dimensional volume of a complete hyperplane or
hyper-hyperboloid is infinite. Consequently, if the density of matter is
finite and the cosmological principle holds, $k$ cannot be changeable, because
a complete hypersphere cannot transform into a complete hyperplane or
hyper-hyperboloid in the case. This mode has been discussed in the first part.

$(2)$ There are the two sorts of breaking in the universe simultaneously.
Consequently, the universe must be composed of infinite $s-cosmic$ islands
with the $S-breaking$ and $v-cosmic$ islands with $V-breaking$. Every cosmic
island must be finite. Thus, the supersurface of a cosmic island must be
incomplete and the radial coordinate of a cosmic island is finite. For
example, we may take the maximum of the radial coordinate to be $r_{\max
}\lesssim1$ as the same as this case $k=1$ in the Friedmann model. This
implies that not only is the maximum of the radial coordinate of a supersphere
finite, but also the maximum of the radial coordinate of a superplane or a
hyper-hyperboloid is finite and may be taken as $1$. We discuss the first sort
of modes in the present part.

In this case, not that we describe the evolution of the whole universe, but
that we describe approximately the evolution of a cosmic island. In fact, in
contrast with the conventional theory, the universe as a whole does not change
in essence according to the present model, although every cosmic island is
changing (expands, contracts or inflates)\textbf{.}

We will see that the probability that particles get out of a cosmic island is
very little. Consequently a cosmic island seems to be the whole cosmos for an
observer inside it.

In this case, the cosmological principle does no longer hold strictly, the RW
metric is no longer applicable, and a new metric is necessary for the
incomplete hypersurfaces.

The inferences of the first model can still be obtained by the second model.
The differences between the two model are follows.

$1.$ In addition to conjectures $1$ and $2$ in the first model, the second
model increases a new conjecture about incomplete supersurfaces.

$2.$ There are new inferences.

$A.$ The universe must be composed of infinite $s-cosmic$ islands, $v-cosmic$
islands and transitional regions.

$B.$ Some huge redshifts (e.g. the big redshifts of quasi-stellar objects)
possibly are `the mass redshifts' which is caused by less mass $m_{eT}$ of an
electron than given $m_{e}$.

$C$. It is possible that a $v-black$ hole with its big enough mass and density
can transform into a huge white hole in the $V-breaking$ or the $S-breaking.$

In section $17$, the conjecture about incomplete supersurfaces are presented;
in section $18$, in section $6$, contraction of space, the highest temperature
and inflation of space are discussed; in section $7$, some new predictions are
given; section $8$ is the conclusion of the part.

\section{Conjectures}

\subsection{Conjectures}

In addition to conjectures 1 and 2, the following conjecture is necessary for
the present model. In order to describe the conjecture, we first define
incomplete hypersurfaces of simultaneity.

\begin{definition}
Such hypersurfaces of simultaneity are called incomplete, if a hypersphere has
one hole, and a hyperplane or a hyper-hyperboloid is finite.
\end{definition}

\begin{description}
\item
\begin{conjecture}
\textbf{\ }The metric describing incomplete hypersurfaces can be obtained by
the replacement of the curvature factor $k$ in the RW metric by $K=K\left(
\underline{\rho}_{g}\right)  $ which is a monotone and finite function of
$\underline{\rho}_{g}$ and $dK/d\underline{\rho}_{g}>0$. Here $\underline
{\rho}_{g}$ is the gravitational mass density in the comoving coordinates.
\end{conjecture}
\end{description}

In this case, $K$ does no longer describe the quality of an incomplete
supersurface in whole and only describe differential quality of the incomplete
supersurface. $K$ corresponding to a cosmic island can change from $K>0$ to
$K=0$ or $K<0,$ because a cosmic island is finite and its gravitational mass
density $\rho_{g}$ can change from $\rho_{g}>0$ to $\rho_{g}=0$ or $<0.$

\subsection{The new metric is applicable to incomplete hypersurfaces of
simultaneity}

The only difference between the new metric and the RW metric is that $K$ is
changeable in the new metric.

A few explanations for the conjecture as follows:

$\left(  \mathbf{1}\right)  $ The basic promise of a hypersurface of
simultaneity to be complete is that the universe evolves as a whole. If the
universe is composed of infinite $s-cosmic$ islands and $v-cosmic$ islands
which evolve individually, the hypersurface of every cosmic island must be incomplete.

$k$ in the RW metric cannot change from $k=1$ to $0$ and $-1$ for complete
hypersurfaces because of the relative topological theorems. But the
topological theorems are not applicable to the incomplete supersurfaces. In
fact, there is no the same breaking, no unified evolution, and no the same
features for the cosmic islands. Hence the metrices of the cosmic islands must
be different from each other.

$k$ can describe the topology feature for complete hypersurfaces. But
$K\left(  \underline{\rho}_{g}\right)  $ describes only the differential
feature for the incomplete hypersurfaces.

The radial coordinate $r$ in the spherical coordinates is finite for an
incomplete hypersurface of simultaneity and $r_{\max}$ satisfies $r_{\max}%
^{2}\lesssim1/K_{\max}$ $\left(  K_{\max}>0\text{ in this model}\right)  $,
because a cosmic island must be finite.

$\left(  \mathbf{2}\right)  $ In the Friedmann model the equation of
$\overset{\cdot}{R}$ is regarded as a first integral of the equation of
$\overset{\cdot\cdot}{R}$ because of the matter conservation equation
$\overset{\cdot}{\rho}=-3\left(  \rho+p\right)  \overset{\cdot}{R}/R.$ In this
model, the equation of $\overset{\cdot}{\rho}$ becomes the equation of
$\overset{\cdot}{\rho}_{g}$ $\left(  18.50\right)  $ or $\left(  18.30\right)
$ which is more complex, because $\overset{\cdot}{K}\neq0$. The equation
$\left(  18.50\right)  $ or $\left(  18.30\right)  $ does not contradict any
known experiment and observation, because $\rho_{g}$ is the gravitational mass
density, rather than mass density. The gravitational mass may be not
conservational, although energy is still conservational. In other words,
$\overset{\cdot}{K}\neq0$ and energy conservation are compossible. On the
other hand, from $\left(  18.49\right)  ,$ $\left(  18.44\right)  $ and
$\left(  18.41\right)  $ we see that the the conservation equation of
$\rho_{g}$ holds approximately when $\overset{\cdot}{K}\sim\partial K/\partial
r\sim0$.

$\left(  \mathbf{3}\right)  $ Let $h_{ab}\left(  t,x^{i}\right)  $ be the
induced metric of the space-time metric $g_{ab}$ on a homogeneous
hypersurface. Only when the uniqueness of the homogeneous hypersurface clan
holds, the variables $t$ and $x^{i}$ of $h_{ab}\left(  t,x^{i}\right)  $ can
be separated, i.e. $h_{ab}\left(  t,x^{i}\right)  =a^{2}\left(  t\right)
\widehat{h}_{ab}\left(  x^{i}\right)  ,$ so that the metric can be written as
the $RW$ metric$^{\left[  19\right]  }$. When $\rho_{g}=0$ or a hypersurfaces
is incomplete, the uniqueness of the homogeneous hypersurface clan cannot
hold. Thus, $t$ and $x^{i}$ in $h_{ab}\left(  t,x^{i}\right)  $ cannot be
separated completely and the new metric $\left(  K=K\left(  t,x^{i}\right)
\right)  $ is necessary.

In fact, only when $\rho_{g}\sim0,$ $\overset{\cdot}{K}\neq0$. When
$\underline{\rho}_{g}\neq0$ so that $\overset{\cdot}{K}\sim0$, the evolution
equations of the present model are consistent with those of the Friedmann
model in form.

\section{Evolution equations of space}

The only difference between the two models is the curvature factors in them to
be different from each other. Hence $\left(  2.1\right)  -\left(  3.12\right)
$ are still applicable to the second model, because they are independent of
the curvature factor.

\subsection{The evolution equations in a new metric}

Based on the new metric described by conjecture 3, we have
\begin{align}
(ds)^{2}  &  =-\left(  dt\right)  ^{2}\nonumber\\
&  +R^{2}\left(  t\right)  \left\{  \frac{\left(  dr\right)  ^{2}%
}{1-K(\underline{\rho}_{g})r^{2}}+\left(  rd\theta\right)  ^{2}+\left(
r\sin\theta d\varphi\right)  ^{2}\right\}  . \tag{18.1}%
\end{align}
$\left(  18.1\right)  $ implies $g_{00}=-1,$ $g_{11}=R^{2}\left(  t\right)
/\left(  1-K(\underline{\rho}_{g})r^{2}\right)  $, $g_{22}=R^{2}\left(
t\right)  r^{2}$, $g_{33}=R^{2}\left(  t\right)  r^{2}\sin^{2}\theta$ and the
others to be zero. Form this we can determine $\Gamma_{\mu\nu}^{\lambda}$ and
$R_{\mu\nu}$,%
\begin{align}
R_{00}  &  =g_{00}\left\{  \frac{3\overset{\cdot\cdot}{R}}{R}+\frac{r^{2}%
}{2\left(  1-Kr^{2}\right)  }\left[  \frac{2\overset{\cdot}{R}\overset{\cdot
}{K}}{R}+\overset{\cdot\cdot}{K}+\frac{3r^{2}\overset{\cdot}{K}^{2}}{2\left(
1-Kr^{2}\right)  }\right]  \right\}  ,\tag{18.2}\\
R_{11}  &  =g_{11}\{\frac{2\overset{\cdot}{R}^{2}}{R^{2}}+\frac{2K}{R^{2}%
}+\frac{\overset{\cdot\cdot}{R}}{R}\nonumber\\
&  +\frac{r^{2}}{2\left(  1-Kr^{2}\right)  }\left[  \frac{4\overset{\cdot}%
{R}\overset{\cdot}{K}}{R}+\overset{\cdot\cdot}{K}+\frac{3r^{2}\overset{\cdot
}{K}^{2}}{2\left(  1-Kr^{2}\right)  }\right]  +\frac{r\partial K}%
{R^{2}\partial r}\},\tag{18.3}\\
R_{22}  &  =g_{22}\left[  \frac{2\overset{\cdot}{R}^{2}}{R^{2}}+\frac
{2K}{R^{2}}+\frac{\overset{\cdot\cdot}{R}}{R}+\frac{r^{2}}{2\left(
1-Kr^{2}\right)  }\frac{\overset{\cdot}{R}\overset{\cdot}{K}}{R}%
+\frac{r\partial K}{2R^{2}\partial r}\right]  ,\tag{18.4}\\
R_{33}  &  =g_{33}\left[  \frac{2\overset{\cdot}{R}^{2}}{R^{2}}+\frac
{2K}{R^{2}}+\frac{\overset{\cdot\cdot}{R}}{R}+\frac{r^{2}}{2\left(
1-Kr^{2}\right)  }\frac{\overset{\cdot}{R}\overset{\cdot}{K}}{R}%
+\frac{r\partial K}{2R^{2}\partial r}\right]  ,\tag{18.5}\\
R_{01}  &  =R_{10}=\frac{r\overset{\cdot}{K}}{\left(  1-Kr^{2}\right)
},\text{ \ the others are zero}. \tag{18.6}%
\end{align}

Matter in the universe may be regarded approximately as ideal gas distributed
evenly in the whole space. The energy-momentum tensor densities of the ideal
gas are
\begin{equation}
T_{aM\mu\nu}=\left(  \rho_{a}+p_{a}\right)  U_{a\mu}U_{a\nu}+p_{a}g_{\mu\nu
},\text{ }a=s,\text{ }v\text{.} \tag{18.7}%
\end{equation}
Considering the potential energy densities in $(2.14)-\left(  2.15\right)  $,
we can rewrite $\left(  18.7\right)  $ as
\begin{equation}
\widetilde{T}_{a\mu\nu}=\left[  \widetilde{\rho}_{a}+\widetilde{p}_{a}\right]
U_{\mu}U_{\nu}+\widetilde{p}_{a}g_{\mu\nu}, \tag{18.8}%
\end{equation}%
\begin{equation}
\widetilde{\rho}_{a}=\rho_{a}+\widetilde{V}_{a}\left(  \varpi_{a}\right)
,\ \ \ \widetilde{p}_{a}=p_{a}-\widetilde{V}_{a}(\varpi_{a}). \tag{18.9}%
\end{equation}
From $\left(  2.17\right)  -\left(  2.18\right)  $ we have
\begin{align}
\widetilde{V}_{s}\left(  \varpi_{s}\right)   &  =V_{s}\left(  \varpi
_{s}\right)  +V_{0},\text{ \ }\widetilde{V}_{v}\left(  \varpi_{v}\right)
=V_{v}\left(  \varpi_{v}\right)  ,\tag{18.10}\\
T_{Sg\mu\nu}  &  =\widetilde{T}_{s\mu\nu}-\widetilde{T}_{v\mu\nu}. \tag{18.11}%
\end{align}
in the $S-breaking$, and
\begin{align}
\widetilde{V}_{s}\left(  \varpi_{s}\right)   &  =V_{s}\left(  \varpi
_{s}\right)  ,\text{ \ }\widetilde{V}_{v}\left(  \varpi_{v}\right)
=V_{v}\left(  \varpi_{v}\right)  +V_{0},\tag{18.11}\\
T_{Vg\mu\nu}  &  =\widetilde{T}_{v\mu\nu}-\widetilde{T}_{s\mu\nu}, \tag{18.13}%
\end{align}
in the $V-breaking$. In fact, $V_{v}=0$ because $\langle\omega_{v}\rangle=0$
in the $S-breaking$, hence $\widetilde{\rho}_{v}=\rho_{v}$ and $\widetilde
{p}_{v}=p_{v}.$ Similarly, $V_{s}=0,$ $\widetilde{\rho}_{s}=\rho_{s}$ and
$\widetilde{p}_{s}=p_{s}$ in the $V-breaking$. In the present model, the
4-velocities are
\begin{align}
U^{\mu}  &  =\frac{dx^{\mu}}{d\tau}\text{, }x^{1}=r\text{, }x^{2}%
=\theta,\text{ }x^{3}=\varphi,\text{ }\tag{18.14}\\
U^{1}  &  =\frac{dr}{d\tau}=\frac{q}{\sqrt{1-v^{2}}},\text{ }U^{2}%
=U^{3}=0,\tag{18.15}\\
q  &  =\frac{dr}{dt},\text{ \ }v=q\sqrt{\frac{R^{2}}{1-Kr^{2}}},\tag{18.16}\\
U_{0}  &  =g_{0\nu}U^{\nu}=g_{00}U^{0}=-U^{0}=\frac{-1}{\sqrt{1-v^{2}}%
},\tag{18.17}\\
U_{1}  &  =g_{1\nu}U^{\nu}=\sqrt{\frac{R^{2}}{1-Kr^{2}}}\cdot\frac{v}%
{\sqrt{1-v^{2}}}. \tag{18.18}%
\end{align}
Let
\begin{align}
S_{g\mu\nu}  &  \equiv T_{g\mu\nu}-\frac{1}{2}g_{\mu\nu}T_{g},\text{ }%
T_{g}=T_{g\lambda}^{\lambda}=-\widetilde{\rho}_{g}+3\widetilde{p}%
_{g},\tag{18.19}\\
S_{g00}  &  =\frac{1}{2}\left(  \widetilde{\rho}_{g}+3\widetilde{p}%
_{g}\right)  +\frac{v^{2}}{1-v^{2}}\left(  \widetilde{\rho}_{g}+\widetilde
{p}_{g}\right)  ,\tag{18.20}\\
S_{g11}  &  =g_{11}\left[  \frac{1}{2}\left(  \widetilde{\rho}_{g}%
-\widetilde{p}_{g}\right)  +\frac{v^{2}}{1-v^{2}}\left(  \widetilde{\rho}%
_{g}+\widetilde{p}_{g}\right)  \right]  ,\tag{18.21}\\
S_{g22}  &  =g_{22}\frac{1}{2}\left(  \widetilde{\rho}_{g}-\widetilde{p}%
_{g}\right)  ,\text{ \ }S_{33}=g_{33}\frac{1}{2}\left(  \widetilde{\rho}%
_{g}-\widetilde{p}_{g}\right)  ,\tag{18.22}\\
S_{g01}  &  =S_{g10}=\sqrt{\frac{R^{2}}{1-Kr^{2}}}\cdot\frac{-v}{1-v^{2}%
}\left(  \widetilde{\rho}_{g}+\widetilde{p}_{g}\right)  . \tag{18.23}%
\end{align}
the others are all zero. Substituting $\left(  18.2\right)  -\left(
18.23\right)  $ into $\left(  2.16\right)  ,$ we obtain%
\begin{align}
&  \dot{R}^{2}+K+\frac{r^{2}}{3\left(  1-Kr^{2}\right)  }R\overset{\cdot}%
{R}\overset{\cdot}{K}+\frac{r\partial K}{3\partial r}\nonumber\\
&  =\eta\left[  \rho_{g}+V_{g}+\frac{v^{2}}{1-v^{2}}\left(  \rho_{g}%
+p_{g}\right)  \right]  R^{2},\text{ }\eta\equiv8\pi G/3, \tag{18.24}%
\end{align}%
\begin{align}
&  \ddot{R}-\frac{r^{2}}{6\left(  1-Kr^{2}\right)  }\overset{\cdot}{R}%
\overset{\cdot}{K}-\frac{r\partial K}{6R\partial r}\nonumber\\
&  =-\frac{\eta}{2}\left[  \left(  \rho_{g}+3p_{g}\right)  -2V_{g}%
+\frac{4v^{2}}{1-v^{2}}\left(  \rho_{g}+p_{g}\right)  \right]  R. \tag{18.25}%
\end{align}%
\begin{align}
&  \frac{R^{2}r^{2}}{\left(  1-Kr^{2}\right)  }\left[  \frac{3\overset{\cdot
}{R}\overset{\cdot}{K}}{R}+\overset{\cdot\cdot}{K}+\frac{3r^{2}\overset{\cdot
}{K}^{2}}{2\left(  1-Kr^{2}\right)  }\right]  +\frac{r\partial K}{\partial
r}\nonumber\\
&  =6\eta\frac{v^{2}}{1-v^{2}}\left(  \rho_{g}+p_{g}\right)  R^{2},
\tag{18.26}%
\end{align}%
\begin{equation}
\frac{r\overset{\cdot}{K}}{1-Kr^{2}}=3\eta\frac{\left(  -v\right)  }{1-v^{2}%
}\sqrt{\frac{R^{2}}{1-Kr^{2}}}\left(  \rho_{g}+p_{g}\right)  , \tag{18.27}%
\end{equation}
In the $S-breaking$,
\begin{equation}
\rho_{g}=\rho_{s}-\rho_{v},\text{ \ }V_{g}=V_{s}+V_{0}-V_{v},\text{ }%
p_{g}\equiv p_{s}-p_{v}. \tag{18.28}%
\end{equation}
In the $V-breaking$,%
\begin{equation}
\rho_{g}=\rho_{v}-\rho_{s},\text{ \ }V_{g}=V_{v}+V_{0}-V_{s},\text{ }%
p_{g}\equiv p_{v}-p_{s}. \tag{18.29}%
\end{equation}
From $\left(  18.24\right)  -\left(  18.27\right)  $ we get%
\begin{align}
&  \overset{\cdot}{K}\{1+\frac{r^{2}}{1-Kr^{2}}\cdot\nonumber\\
&  \left[  \frac{K}{3}+\frac{r}{6}\frac{\partial K}{\partial r}-\frac{\eta}%
{2}R^{2}\left(  1+\frac{2v^{2}}{1-v^{2}}\right)  \left(  \rho_{g}%
+p_{g}\right)  \right]  \}\nonumber\\
&  =\eta\{R^{2}\frac{d}{dt}\left[  \rho_{g}+V_{g}+\frac{v^{2}}{1-v^{2}}\left(
\rho_{g}+p_{g}\right)  \right] \nonumber\\
&  +\left(  3+\frac{4v^{2}}{\left(  1-v^{2}\right)  }\right)  \left(  \rho
_{g}+p_{g}\right)  R\overset{\cdot}{R}\} \tag{18.30}%
\end{align}

\subsection{The asymptotic solution of the equations $\left(  18.24\right)
-\left(  18.27\right)  $}

\subsubsection{The asymptotic solution of $K\left(  t,r\right)  $}

As mentioned in the section above, the radial coordinate $r$ must be finite
and $r_{\max}^{2}<1/K_{\max}.$ $K\left(  t,r\right)  $ in $\left(
18.24\right)  -\left(  18.27\right)  $ has such an asymptotic solution in the
following form,
\begin{align}
K  &  =\frac{D-D_{0}}{1+S\left(  r,D\right)  },\tag{18.31}\\
S\left(  r,D\right)   &  \equiv\left(  r^{2}-1\right)  \left[  1-\exp\left(
-Ar^{2}\right)  \right]  ,\text{ }A\equiv\frac{1-D^{2}}{D^{2}\left(  t\right)
+\delta^{2}}\tag{18.32}\\
D\left(  \underline{\rho}_{g}\right)   &  \equiv\frac{2}{\pi}\arctan Q\left(
\underline{\rho}_{g}\right)  ,\text{ }\underline{\rho}_{g}=V_{g}+\frac
{\rho_{g}R^{3}}{R_{1}^{3}}, \tag{18.33}%
\end{align}
where $\underline{\rho}_{g}$ is the gravitational mass density in the comoving
coordinates, $Q\left(  \underline{\rho}_{g}\right)  $ is a monotonously
increasing function of $\underline{\rho}_{g}$, and $0\leq D_{0}<1,$ $\delta
$\ is a small positive parameter. $Q\left(  \underline{\rho}_{g}\right)  $ can
be so chosen that
\begin{align}
Q\left(  0\right)   &  =0,\text{ \ }dQ/d\underline{\rho}_{g}>0,\nonumber\\
Q\left(  \underline{\rho}_{g}\right)   &  \longrightarrow\infty\text{ when
}\underline{\rho}_{g}\longrightarrow\infty,\nonumber\\
Q\left(  \underline{\rho}_{g}\right)   &  \longrightarrow-\infty\text{ when
}\underline{\rho}_{g}\longrightarrow-\underline{\rho}_{g0}/A_{\rho},
\tag{18.34}%
\end{align}
e.g.
\begin{equation}
Q\left(  \underline{\rho}_{g}\right)  =\ln\frac{A_{\rho}\underline{\rho}%
_{g}+\underline{\rho}_{g0}}{\underline{\rho}_{g0}}, \tag{18.35}%
\end{equation}
where $\underline{\rho}_{gc}$\ and $A_{\rho}$ are all positive parameters.\ In
the present model, $\underline{\rho}_{g}$ must be finite in any case. We will
see $V_{0}\geq\underline{\rho}_{g}>-\underline{\rho}_{g0}/A_{\rho}$.

In general, $\underline{\rho}_{g}=\underline{\rho}_{g}\left(  r,t\right)  $ so
that $D=D\left(  r,t\right)  $. In fact, we will see that $\underline{\rho
}_{g}\simeq\underline{\rho}_{g}\left(  t\right)  $ in many cases, because the
cosmological principle holds approximately. Thus, we take $\underline{\rho
}_{g}=\underline{\rho}_{g}\left(  t\right)  $ and $D=D\left(  t\right)  $ as
an approximate solution in the following.

$D\left(  \underline{\rho}_{g}\right)  $ is a monotone and finite function of
$\underline{\rho}_{g}$ due to $\left(  18.31\right)  -\left(  18.35\right)
,$
\begin{align}
D  &  \sim0\text{ provided }\underline{\rho}_{g}\sim0,\nonumber\\
K  &  \sim-\frac{D_{0}}{r^{2}}\text{ only when }r\text{ is large and
}\underline{\rho}_{g}\sim0,\tag{18.36}\\
K  &  \sim-D_{0},\text{ when }r\sim0\text{ and }\underline{\rho}_{g}%
\sim0\U{ff0c}\tag{18.37}\\
K  &  \longrightarrow\pm1-D_{0},\text{ when }Q\left(  \underline{\rho}%
_{g}\right)  \longrightarrow\pm\infty.\text{ } \tag{18.38}%
\end{align}
It is considered in $\left(  18.36\right)  -\left(  18.37\right)  $ that
$V_{g}=0$ when $\underline{\rho}_{g}\sim0$, because $V_{g}\sim0$ after reheating.

It is seen from $\left(  18.36\right)  -\left(  18.38\right)  $\ that
$K\left(  r,t\right)  $ is dependent on $r$ only when $r$ is very large and
$\underline{\rho}_{g}\sim0$. In this case, $K\sim\left(  r\partial K/\partial
r\right)  \sim-D_{0}/r^{2}\sim0$ and $\left(  r\overset{\cdot}{K}\right)
\sim\overset{\cdot}{D}/r\sim0$ may still be neglected, because $r$ is very
large. Thus $\left(  18.24\right)  -\left(  18.27\right)  $ is still
approximately independent of $r$.

\subsubsection{Discussion of $K$}

$1.$ $\overset{\cdot}{K}\sim0$ when $Q^{2}\longrightarrow\infty$.

For monotonously expanding space, $\underline{\rho}_{g}\left(  t\right)  $ is
a monotone decreasing function of $t,$ hence $\overset{\cdot}{\underline{\rho
}}_{g}\lesssim0.$ Considering $D\left(  \underline{\rho}_{g}\right)  $ and $K$
are all monotonously increasing functions of $\underline{\rho}_{g}$ since
$dQ/d\underline{\rho}_{g}=\left(  2/\pi\right)  a/\left(  \underline{\rho}%
_{g}+\underline{\rho}_{g0}/A_{\rho}\right)  >0,$ from $\left(  18.31\right)
-\left(  18.35\right)  $ we have%
\begin{equation}
\overset{\cdot}{D}=\frac{2}{\pi}\frac{dQ/d\underline{\rho}_{g}}{1+Q^{2}%
}\overset{\cdot}{\underline{\rho}}_{g}\lesssim0, \tag{18.39}%
\end{equation}%
\begin{align}
\overset{\cdot}{K}  &  =\frac{\overset{\cdot}{D}}{1+S}\left[  1+r^{2}\left(
r^{2}-1\right)  \frac{K}{1+S}\frac{2D\left(  1+\delta^{2}\right)  }{\left(
D^{2}+\delta^{2}\right)  ^{2}}\exp\left(  -Ar^{2}\right)  \right]
,\nonumber\\
\overset{\cdot}{K}  &  \sim\frac{\overset{\cdot}{D}}{1+S},\text{ when
}D,\text{ }K\text{, or }r\sim0,\text{ or }Ar^{2}\text{ is large} \tag{18.40}%
\end{align}
Considering $K,$ $S$ and $D$ are all finite, from $\left(  18.31\right)
-\left(  18.32\right)  $ we have
\begin{equation}
\overset{\cdot}{K}\sim0,\text{ provided }\overset{\cdot}{D}\sim0\text{ or
}S\text{ is large,} \tag{18.41}%
\end{equation}%
\begin{align}
\overset{\cdot\cdot}{K}  &  \sim0,\text{ }\overset{\cdot\cdot}{D}\sim0,\text{
provided }\overset{\cdot}{\underline{\rho}}_{g}\sim0\nonumber\\
\text{and }\underline{\overset{\cdot\cdot}{\rho}}_{g}  &  \sim0\text{ or
}Q^{2}\longrightarrow\infty. \tag{18.42}%
\end{align}
$\overset{\cdot}{D}$, $\overset{\cdot\cdot}{D}$, $\overset{\cdot}{K}$ and
$\overset{\cdot\cdot}{K}$ must be finite$.$

$2.$ $r\partial K/\partial r\sim0,$ provided $A\sim0,$ or $r\sim0,$ or
$K\sim0$.

From $\left(  18.31\right)  -\left(  18.32\right)  $ we have
\begin{equation}
\frac{r\partial K}{\partial r}=-\frac{2Kr^{2}}{\left(  1+S\right)
}\{1-\left[  1+A\left(  1-r^{2}\right)  \right]  \exp\left(  -Ar^{2}\right)
\}. \tag{18.43}%
\end{equation}
From $\left(  18.43\right)  $, and $\left(  18.26\right)  -\left(
18.27\right)  $ we
\begin{align}
\frac{r\partial K}{\partial r}  &  \sim0,\text{ provided }A\sim0,\text{ or
}r\sim0,\text{ or }K\sim0\text{,}\nonumber\\
\text{or}\overset{\cdot}{K}  &  \sim\overset{\cdot\cdot}{K}\sim0. \tag{18.44}%
\end{align}
When $r$ is very large and $\underline{\rho}_{g}\sim0$%
\begin{equation}
\frac{r\partial K}{\partial r}\sim-\frac{2D_{0}}{r^{2}}. \tag{18.45}%
\end{equation}
In the case, $r\partial K/\partial r$ may still be neglected, when
$D_{0}/r^{2}\ll1$. It is seen that $\left(  18.31\right)  $ is indeed an
asymptotic solution of $K\left(  t,r\right)  $.

Sum up, it is seen that we may take $r\partial K/\partial r\sim0$ so that
$K\left(  r,t\right)  \sim K\left(  t\right)  $ approximately.

$3.$ Reduction of $\left(  18.24\right)  -\left(  18.27\right)  $ and $\left(
18.30\right)  $ when $Q^{2}$ is very large.

As mentioned above, $K\sim\pm1-D_{0}$ and $r\partial K/\partial r\sim
\overset{\cdot}{K}\sim\overset{\cdot\cdot}{K}\sim0$ provided $Q^{2}%
\longrightarrow\infty$, and $K\sim D\left(  t\right)  -D_{0}$ when $r\sim0$.
In the two cases, , from $\left(  18.26\right)  -\left(  18.27\right)  $ we
get%
\begin{equation}
\frac{\left(  -v\right)  }{1-v^{2}}\left(  \rho+p\right)  \sim0,\text{ }%
\frac{r\partial K}{\partial r}\sim0. \tag{18.46}%
\end{equation}
and $\left(  18.24\right)  -\left(  18.25\right)  $ is reduced to
\begin{equation}
\dot{R}^{2}+K=\eta\left[  \rho_{g}+V_{g}\right]  R^{2}, \tag{18.47}%
\end{equation}%
\begin{equation}
\ddot{R}=-\frac{\eta}{2}\left[  \left(  \rho_{g}+3p_{g}\right)  -2V_{g}%
\right]  R. \tag{18.48}%
\end{equation}
When $Q^{2}\longrightarrow\infty$, $\left(  18.30\right)  $ is reduced to
\begin{equation}
R^{2}\frac{d}{dt}\left[  \rho_{g}+V_{g}\right]  +3\left(  \rho_{g}%
+p_{g}\right)  R\overset{\cdot}{R}=0. \tag{18.49}%
\end{equation}
The equations are independent of $r$.

When $r\sim0$, $\left(  18.30\right)  $ is reduced to%
\begin{equation}
\overset{\cdot}{D}\sim\overset{\cdot}{K}=\eta\{R^{2}\frac{d}{dt}\left(
\rho_{g}+V_{g}\right)  +3\left(  \rho_{g}+p_{g}\right)  R\overset{\cdot}%
{R}\}<0. \tag{18.50}%
\end{equation}
When $\underline{\rho}_{g}\sim0$ and $r\sim0$, considering $V_{g}=0$ after
reheating and $p_{g}=\rho_{\gamma g}/3>0$ when $\rho_{g}=0$ , from $\left(
18.24\right)  -\left(  18.25\right)  $ and $\left(  18.31\right)  -\left(
18.33\right)  $ we have
\begin{equation}
\overset{\cdot}{R}^{2}=D_{0}>0\text{, }\ddot{R}/R=-\left(  3\eta/2\right)
p_{g}<0. \tag{18.51}%
\end{equation}

$4.$ Reduction of $\left(  18.24\right)  -\left(  18.27\right)  $ and $\left(
18.30\right)  $ when $\underline{\rho}_{g}\sim0$.

The results $\left(  18.46\right)  -\left(  18.49\right)  $ still hold for
very large $r$, provided $Q^{2}$ is large enough. But the results do no long
hold for very large $r$ and $\underline{\rho}_{g}\sim0$ due to $Q\left(
0\right)  =0.$ When $\underline{\rho}_{g}\sim0$ and $r$ is very large$,$
$D\left(  \underline{\rho}_{g}\right)  \sim0$ and $K\sim-D_{0}/r^{2}$ due to
$\left(  18.37\right)  $. From $\left(  18.39\right)  $ we have $\overset
{\cdot}{D}=\left(  2/\pi\right)  \left(  A_{\rho}/\rho_{g0}\right)
\underline{\overset{\cdot}{\rho}}$ and $\overset{\cdot}{K}=\overset{\cdot}%
{D}/r^{2}$ when $\underline{\rho}_{g}\sim0$. Thus, $r\partial K/\partial r\sim
K\sim-D_{0}/r^{2}\sim0$\ and $\left(  r\overset{\cdot}{K}\right)  \sim
\overset{\cdot}{D}/r\sim0$ may be neglected when $\underline{\rho}_{g}\sim0$
and $r$ is very large. Considering $V_{g}=0$ after reheating, $\left(
18.24\right)  -\left(  18.27\right)  $ is reduced to%
\begin{equation}
\left(  \frac{\dot{R}}{R}\right)  ^{2}\simeq-\frac{\overset{\cdot}{D}%
}{3\left(  1+D_{0}\right)  }\frac{\dot{R}}{R}+\eta\rho_{g}, \tag{18.52}%
\end{equation}%
\begin{equation}
\frac{\ddot{R}}{R}\simeq\frac{\overset{\cdot}{D}\dot{R}}{6\left(
1+D_{0}\right)  R}-\frac{3\eta}{2}p_{g}, \tag{18.53}%
\end{equation}%
\begin{equation}
3\frac{\dot{R}\overset{\cdot}{D}}{R}+\overset{\cdot\cdot}{D}+\frac
{3\overset{\cdot}{D}^{2}}{2\left(  1+D_{0}\right)  }\simeq0, \tag{18.54}%
\end{equation}%
\begin{equation}
0\simeq\frac{\left(  -v\right)  }{1-v^{2}}p_{g}, \tag{18.55}%
\end{equation}
$\dot{R}>0$ because space is expanding. Hence
\begin{equation}
\overset{\cdot}{D}<0,\text{ }\underline{\overset{\cdot}{\rho}}<0,\text{ }%
\ddot{R}<0\text{ when \ }\underline{\rho}_{g}\sim0. \tag{18.56}%
\end{equation}
due to $\left(  18.52\right)  -\left(  18.53\right)  $.

When $\underline{\rho}_{g}\sim0$ and $r\sim0$, we have $D\left(
\underline{\rho}_{g}\right)  \sim0,$ $K\sim-D_{0}$, $\overset{\cdot}%
{K}=\overset{\cdot}{D}=\left(  2/\pi\right)  \left(  A_{\rho}/\rho
_{g0}\right)  \underline{\overset{\cdot}{\rho}}$, and $r\partial K/\partial
r\sim\left(  r\overset{\cdot}{K}\right)  \sim0$ may be neglected. Thus
$\left(  18.24\right)  -\left(  18.25\right)  $ is reduced to%
\begin{equation}
\dot{R}^{2}\simeq D_{0}+\eta\rho_{g}>0,\text{ \ }\ddot{R}\simeq-\frac{3\eta
}{2}p_{g}R\simeq-\frac{\eta}{2}\rho_{\gamma g}R<0. \tag{18.57}%
\end{equation}
In this case, $\left(  18.56\right)  $ still holds. When $\underline{\rho}%
_{g}\sim0$ and $r$ is large,
\[
\overset{\cdot}{\rho}_{g}=-\frac{3\dot{R}}{2R}p_{g}=\frac{\overset{\cdot}{D}%
}{2\left(  1+D_{0}\right)  }p_{g}<0.
\]
When $\underline{\rho}_{g}\sim0$ and $r\sim0$,
\[
\overset{\cdot}{\rho}_{g}=-3p_{g}\overset{\cdot}{R}/R+\overset{\cdot}{D}/\eta
R^{2}<0.
\]

$5.$ It is possible that $\ddot{R}>0$ when $\overset{\cdot}{R}=0$.

As mentioned above, when $\underline{\rho}_{g}\sim0,$ we have $\overset
{\cdot\cdot}{R}<0$ and $\dot{R}>0$ so that $\overset{\cdot}{\rho}_{g}<0$.
Hence space will continue to expand with a deceleration and $\underline{\rho
}_{g}$ will become negative.

Let $t=t_{2}$ when $\underline{\rho}_{g}\left(  t_{2}\right)  =0.$ Because of
$\left(  18.56\right)  ,$\ $\overset{\cdot}{\underline{\rho}}_{g}<0$\ so that
$\underline{\rho}_{g}\left(  t\right)  =\underline{\rho}_{g}\left(
t_{2}\right)  +\overset{\cdot}{\underline{\rho}}_{g}\left(  t-t_{2}\right)
<0$ when $t>t_{2}.$\ Thus $\overset{\cdot}{R}\left(  t\right)  =0$ is possible
when $t=t_{1}>t_{2},$ because $\overset{\cdot\cdot}{R}<0$ when $t=t_{2}$ and
$\underline{\rho}_{g}\left(  t\right)  <0.$

When $\overset{\cdot}{R}\left(  t_{1}\right)  =0,$ from $\left(  18.24\right)
-\left(  18.25\right)  $ we have%
\begin{equation}
\ddot{R}=-\frac{K}{2R}-\frac{3\eta}{2}\left[  p_{g}+\frac{v^{2}}{1-v^{2}%
}\left(  \rho_{g}+p_{g}\right)  \right]  R. \tag{18.58}%
\end{equation}
From $\left(  18.31\right)  -\left(  18.35\right)  ,$ $\left(  18.40\right)  $
and $\left(  18.43\right)  $ we see that when $A_{\rho}\underline{\rho}%
_{g}/\underline{\rho}_{g0}\sim-1,$ $K\sim-1-D_{0},$ $Q^{2}\longrightarrow
\infty,$ $\overset{\cdot}{K}\sim0$ and $r\partial K/\partial r\sim0.$ Thus,
$r\overset{\cdot}{K}$ and $r\partial K/\partial r$ may be neglected when
$\left(  -A_{\rho}\underline{\rho}_{g}/\underline{\rho}_{g0}\right)  $ is
large. Consequently, $\left(  18.58\right)  $ may be reduced to
\begin{equation}
\ddot{R}=-\frac{K}{2R}-\frac{3\eta}{2}p_{g}R\simeq-\frac{K}{2R}-\frac{\eta}%
{2}\rho_{\gamma g}R. \tag{18.59}%
\end{equation}
It i.e. seen that all $\ddot{R}>0,$ $=0$ and $\ddot{R}<0$ are possible when
$\overset{\cdot}{R}\left(  t_{1}\right)  =0$, because $K\sim-1-D_{0}<0$. If
$\ddot{R}\left(  t_{1}\right)  <0,$ space will contract with a deceleration
when $t>t_{1}.$ If $\ddot{R}\left(  t_{1}\right)  =0,$ space will be static
when $t>t_{1}.$ If $\ddot{R}\left(  t_{1}\right)  >0,$ space will expand with
an acceleration when $t>t_{1}.$ This is because $\underline{\rho}_{g}$ will
decrease as space expands (see $\left(  18.65\right)  $). We will discuss the
case $\ddot{R}\left(  t_{1}\right)  >0$ in the present paper.

$6.$ A conservation equation.

It is impossible that the $s-particles$ transforms into the $v-particles$ when
temperature is low. When temperature is high enough ($T\gtrsim T_{cr},$ see
the following), it is possible that $\langle\omega_{s}\rangle=$ $\langle
\omega_{v}\rangle=0$ so that the masses of all particles are zero. Thus, the
$s-particles$ and the $v-particles$ can transform from one to another by
$\left(  2.10\right)  .$ We discuss $\left(  18.49\right)  $ as follows.

As mentioned above, $\overset{\cdot}{K}\sim0$ when $Q^{2}\longrightarrow
\infty.$ In the case, $\left(  18.49\right)  $ is a conservation equation. Let
$\rho_{g}=\rho_{mg}+\rho_{\lg}+\rho_{\gamma g},$ $p_{mg}\sim0$ and $p_{\lg
}=\kappa\rho_{\lg}$ ($0<\kappa<1/3$) for a given temperature. For photon-like
particles, $p_{\gamma g}=\rho_{\gamma g}/3$. Here $\rho_{m}$ is the mass
density of the particles with so large masses that $\rho_{m}\gg p_{m},$
$\rho_{l}$ is the mass density of the particles whose masses are small, but
cannot be neglected for the given temperature. From $\left(  18.49\right)  $
we have%
\begin{align}
-R^{2}\frac{d}{dt}V_{g}  &  =\frac{1}{R}\frac{d}{dt}\left(  \rho_{mg}%
R^{3}\right) \nonumber\\
&  +\frac{1}{R^{1+3\kappa}}\frac{d}{dt}\left(  \rho_{\lg}R^{3\left(
1+\kappa\right)  }\right)  +\frac{1}{R^{2}}\frac{d}{dt}\left(  \rho_{\gamma
g}R^{4}\right)  . \tag{18.60}%
\end{align}
$V_{g}=dV_{g}/dt=0$ after reheating. $\left(  18.60\right)  $ may be regarded
as the conservation equation of the gravitation mass. The transformation of
the $s-particles$ and the $v-particles$ may be neglected when temperature is
low. After particles decouple, from $\left(  18.60\right)  $ we have
\begin{align}
\rho_{sm}R^{3}  &  =C_{sm},\text{ }\rho_{vm}R^{3}=C_{vm}\text{, }\rho
_{sl}R^{3\left(  1+\kappa\right)  }=C_{sl}\text{, }\nonumber\\
\rho_{vl}R^{3\left(  1+\kappa\right)  }  &  =C_{vl}\text{, }\rho_{s\gamma
}R^{4}=C_{s\gamma}\text{, }\rho_{v\gamma}R^{4}=C_{v\gamma}, \tag{18.61}%
\end{align}
$C_{sm}$ etc. are constants. It is seen that the individual conservation of
$s-energy$ and $v-energy$ holds only approximately in low temperatures. In the
$S-breaking$, $\rho_{v\gamma}=0$ because the masses of all $v-SU(5)$
color-single states are not zero. Similarly, in the $V-breaking$,
$\rho_{s\gamma}=0$.

When $\overset{\cdot}{K}$ cannot be neglected, the conservation equation
$\left(  18.49\right)  $ is no long held. From $\left(  18.50\right)  $ we see
that there must be $\underline{\overset{\cdot}{\rho}}_{g}\neq0$ when
$\overset{\cdot}{K}\neq0.$ This does not contradict any known experiment and
observation, because $\rho_{g}=\rho_{v}-\rho_{s}$ or $=\rho_{s}-\rho_{v}$,
rather than energy density $\rho=\rho_{v}+\rho_{s}$. Although the
gravitational mass is not conservational, the total energy, whose density is
$\rho,$ can still be conservational. Consequently $\overset{\cdot}{K}\neq0$
and energy conservation are compossible.

$7.$ $\overset{\cdot}{K}$ cannot be neglected in the period from $K=0$ to
$\rho_{g}=0.$

Let $D\left(  \underline{\rho}_{g}^{\prime}\right)  -D_{0}\sim0,$ then
$K\sim0$ and $\underline{\rho}_{g}>0$ due to $\left(  18.31\right)  $ and
$\left(  18.33\right)  $, but $\overset{\cdot}{K}\neq0$ so that $v=v\left(
r,t\right)  $ cannot be neglected due to $\left(  18.27\right)  $. In the
case, $\left(  18.24\right)  -\left(  18.27\right)  $ is dependent on $r$.
$\left(  18.24\right)  -\left(  18.27\right)  $ is independent of $r$ when
$\rho_{g}\sim0$ due to $\left(  18.52\right)  -\left(  18.57\right)  $. When
$D\left(  \underline{\rho}_{g}\right)  >D_{0}$ due to $\underline{\rho}%
_{g}>\underline{\rho}_{g}^{\prime},$ $K>0$ and $Q$ is large. In the case, as
mentioned above, $\overset{\cdot}{K}$ and $r\partial K/\partial r$ may be
neglected and $\left(  18.24\right)  -\left(  18.27\right)  $ is reduced to
$\left(  18.46\right)  -\left(  18.49\right)  .$

$8\mathbf{.}$ The definition of $\underline{\rho}_{g}$.

Letting $\left(  \rho_{mg1}+\rho_{\lg1}+\rho_{\gamma g1}\right)  \neq0$
($\rho_{gm1}\equiv\rho_{gm}\left(  t_{1}\right)  $ etc.)$,$ considering the
Higgs potential, we define $\underline{\rho}_{g}$ as
\begin{equation}
\underline{\rho}_{g}=V_{g}+\left(  \rho_{mg}+\rho_{\lg}+\rho_{\gamma
g}\right)  R^{3}/R_{1}^{3}, \tag{18.62}%
\end{equation}
where $\rho_{bg}=\rho_{sb}-\rho_{vb},$ $b=m,$ $l$ or $\gamma,\ \rho_{v\gamma
g}=0$ and $V_{g}=V_{s}+V_{0}-V_{v}$ in the $S-breaking$. When $T\gtrsim
T_{cr}$, then $\rho_{g}=\rho_{s}-\rho_{v}=0$ and $V_{g}=V_{0}$. Thus%
\begin{equation}
\underline{\rho}_{g}=V_{0}>0 \tag{18.63}%
\end{equation}
so that $\overset{\cdot}{K}\sim0$.

$V_{g}=0$ after inflation. Thus, we have%
\begin{equation}
\underline{\rho}_{g}=\left(  \rho_{mg}+\rho_{\lg}+\rho_{\gamma g}\right)
R^{3}/R_{1}^{3}. \tag{18.64}%
\end{equation}
When temperature is so low that $\rho_{l}\gg p_{l},$ we may take $\rho_{l}%
\sim0.$ Thus, considering $\rho_{gm}R^{3}\simeq\rho_{gm1}R_{1}^{3}$ and
$\rho_{g\gamma}R^{4}=\rho_{g\gamma1}R_{1}^{4},$ we can reduce $\left(
18.64\right)  $ to
\begin{equation}
\underline{\rho}_{g}=\rho_{vm1}-\rho_{sm1}+\rho_{g\gamma1}R_{1}/R. \tag{18.65}%
\end{equation}
When $R_{1}/R\ll1,$ we have $\overset{\cdot}{K}\sim\underline{\overset{\cdot
}{\rho}}_{g}\sim0$.

\section{Contraction of space, the highest temperature and inflation of space}

The temperature effects in the second model are the same as those in the first
model. Analogously to the first model, there are $T_{cr},$ $T_{\max}$,
$R_{\min}$ and inflation, and there is no singularity in the second model as well.

\subsection{Contraction of space}

The contracting process in the second model is similar to that in the first
model. The difference is that the boundary condition $\left(  6.13\right)  $
is more obviously satisfied in the second model than in the first model.

In the $S-breakng,$ we consider the space contraction process. In low
temperatures, $V_{g}=0$, $\rho_{sm}\gg\rho_{s\gamma},$ $\underline{\rho}%
_{g}=\underline{\rho}_{sm}-\underline{\rho}_{vm}\gtrsim\underline{\rho}_{g0}.$
In this case, $\overset{\cdot}{K}\sim0$ and $K>0$ could be regarded as a
constant and space will contract because of $\left(  18.47\right)  -\left(
18.48\right)  .$ When $T_{s}\ll T_{cr},$ the transformation $\rho_{s}$ into
$\rho_{v}$ may be ignored. In this case, space will monotonously contract
faster and faster.

$T_{s}$ and $T_{v}$\ must go up high as $R$ decreases. When $T_{s}\sim T_{cr}%
$, the masses all particles tend to zero. Consequently the thermal equilibrium
between $s-matter$ and $v-matter$ comes into being due to $\left(
2.10\right)  $. After thermal equilibrium, the number and the energy of every
sort of particles will satisfy statistical distribution determined by their
spins. Considering $\left(  5.26\right)  $ and the symmetry of $s-particles$
and $v-particles,$ when $T_{s}\gtrsim T_{cr}$, we have%
\begin{align}
T_{s}  &  =T_{v}=T,\text{ \ \ }\rho_{s}\left(  T_{s}\right)  =\rho_{v}\left(
T_{v}\right)  \equiv\rho\left(  T\right)  =\frac{\pi^{2}}{30}g^{\ast}%
T^{4},\tag{19.1}\\
g_{s}^{\ast}  &  =g_{sB}+7g_{sF}/8=g_{v}^{\ast}=g_{vB}+7g_{vF}/8\equiv
g^{\ast}.\nonumber
\end{align}
where $g_{aB}$ $\left(  g_{aF}\right)  $ is the total number of the spin
states of $a-bosons$ $\left(  a-fermions\right)  ,$ and%
\begin{align}
\rho_{a}\left(  T_{a}\right)  R^{4}\left(  T_{a}\right)   &  =D_{a},\text{
\ }\rho_{a}\left(  T_{a}\right)  =\frac{\pi^{2}}{30}g_{a}^{\ast}T_{a}%
^{4},\tag{19.2}\\
TR\left(  T\right)   &  =T_{cr}R\left(  T_{cr}\right)  . \tag{19.3}%
\end{align}
$V_{s}=V_{v}=0$ when $T_{s}\geq T_{cr}$ due to $\langle\omega_{s}%
\rangle=\langle\omega_{v}\rangle=0.$ Considering $\left(  5.26\right)  ,$
$\left(  18.63\right)  ,$\ $\left(  18.47\right)  -\left(  18.48\right)  $ and
$\left(  19.1\right)  ,$\ when $T_{s}\geq T_{cr},$ we have
\begin{align}
\rho_{g}+V_{g}  &  =\rho_{s}\left(  T_{s}\right)  -\rho_{v}\left(
T_{v}\right)  +V_{s}+V_{0}-V_{v}=V_{0},\tag{19.4}\\
K  &  =K\left(  V_{0}\right)  \equiv K_{cr}=1-D_{0}>0, \tag{19.5}%
\end{align}%
\begin{equation}
\dot{R}^{2}=-K_{cr}+\eta V_{0}R^{2}, \tag{19.6}%
\end{equation}%
\begin{equation}
\ddot{R}=\eta V_{0}R>0. \tag{19.7}%
\end{equation}
It is seen that when $T\geq T_{cr},$ $s-matter$ and $v-matter$ are completely
symmetric, and both $s-SU(5)$ and $v-SU(5)$ hold strictly. $\left(
19.4\right)  $ and $\left(  19.6\right)  -\left(  19.7\right)  $ are the
certain results of space contraction and conjecture 1$.$ This is different
from the conventional theory in essence. If there is only one sort of matter
as the conventional theory or $\rho_{s}$ and $\rho_{v}$ cannot transform from
one into other, space will continue to contract and $T_{s}$ and $T_{v}$ will
continue to rise provided $\rho_{g}R^{2}-K>0.$ In fact, in this case, space
will contract to a singular point and $T_{s}$ and $T_{v}$ tend to infinite.

$\langle\omega_{v}\rangle=\langle\omega_{s}\rangle=0$ is the sufficient but is
not the necessary condition for $\left(  19.4\right)  $. In fact, provided the
following conditions are realized, $\left(  19.4\right)  $ can come into being
when $T_{s}<T_{cr}$ as well.

$A$. $m\left(  \omega_{a},T_{v},T_{s}\right)  ,$ $m\left(  f_{a},T_{v}%
,T_{s}\right)  $ and $m\left(  g_{a},T_{v},T_{s}\right)  $ will decrease
because $T_{v}$ and $T_{s}$ arise. Here $f_{a}$ and $g_{a}$ denote fermions
and gauge bosons, respectively. When $m\left(  \omega_{a},T_{v},T_{s}\right)
\sim m\left(  f_{a},T_{v},T_{s}\right)  $ or $m\left(  g_{a},T_{v}%
,T_{s}\right)  ,$ $\Omega_{a}$ or $\varphi_{a}$ and $f_{a}$ or $g_{a}$ can
transform from one into other by the $SU(5)$ couplings.

$B$. When $m\left(  \omega_{s},T_{v},T_{s}\right)  \sim m\left(  \omega
_{v},T_{v},T_{s}\right)  ,$ $\omega_{s}$ and $\omega_{v}$ can transform from
one into other by $\left(  2.10\right)  $.

The two conditions can be realized as well when $T_{s}\sim T_{v}\lesssim
T_{cr}^{\left[  9\right]  }$.

Even when $\chi_{s}$ and $\chi_{v}$ are considered$,$ the above conclusions
still hold qualitatively.

\subsection{There is non-singularity in the second model, the highest
temperature and inflation of space}

It is seen from $\left(  19.6\right)  -\left(  19.7\right)  $ there no
singularity of space-time in the second model.

If space does not contract, it is necessary that there is no space-time
singularity because of the cosmological principle. $\left(  19.6\right)
-\left(  19.7\right)  $ is consistent with the Lemaitre model without
singularity in which $\rho_{g}=0$, $K=1$ and the cosmological constant
$\lambda_{eff}>0^{\left[  20\right]  }.$ It is seen that there is no
singularity in the present model. This can be explained in detail as follows.

Let $R_{cr}=R\left(  T_{cr}\right)  .$ If
\begin{equation}
\overset{\cdot}{R}_{cr}^{2}=-K_{cr}+\eta V_{0}R_{cr}^{2}\geq0,\text{ \ i.e.
\ }R_{cr}\geq\sqrt{K_{cr}/\eta V_{0}}, \tag{19.8}%
\end{equation}
$R$ can continue to decrease with a deceleration or stop contracting. Hence
there must be the least scale $R_{\min}\leq R_{cr},$\ the critical temperature
$T_{cr}$, the highest temperature $T_{\max}$\ and the largest energy density
$\rho_{\max}:$
\begin{align}
0  &  <R_{\min}=\sqrt{K_{cr}/\eta V_{0}}\leq R_{cr},\text{ \ }T_{cr}\equiv
2\mu/\sqrt{\lambda},\tag{19.9}\\
T_{\max}  &  =T\left(  R_{\min}\right)  =T_{cr}R_{cr}/R_{\min}\geq
T_{cr},\tag{19.10}\\
\rho_{\max}  &  =\rho_{s\max}+\rho_{v\max}=2\frac{\pi^{2}}{30}g^{\ast}T_{\max
}^{4}. \tag{19.11}%
\end{align}
$\left(  19.2\right)  -\left(  19.3\right)  $ is considered in $\left(
19.9\right)  -\left(  19.11\right)  .$

It is seen from $\left(  19.9\right)  $ that the boundary condition $\left(
6.13\right)  $ must be satisfied. Consequently, there must be no singularity
in the second model as well, because of the theorem related to singularity and
the discussion in section $6$.

It is seen from $\left(  19.6\right)  -\left(  19.7\right)  $ that when $R$
decreases to $R_{\min}$, space inflation must occur.%
\begin{align}
R  &  =\sqrt{\frac{K_{cr}}{\eta V_{0}}}\cosh\sqrt{\eta V_{0}}\left(
t-t_{FI}\right) \nonumber\\
&  =\sqrt{\frac{K_{cr}}{\eta V_{0}}}\cosh H\left(  t-t_{FI}\right)  ,\text{
\ }\sqrt{\eta V_{0}}\equiv H,\tag{19.12}\\
&  =\sqrt{\frac{K_{cr}}{\eta V_{0}}}=R_{\min}>0\text{ when}\text{ \ }%
t=t_{FI}\tag{19.13}\\
&  \sim\frac{1}{2}\sqrt{\frac{K_{cr}}{\eta V_{0}}}\exp H\left(  t-t_{FI}%
\right)  \text{ when}\text{ \ }H\left(  t-t_{FI}\right)  >>1, \tag{19.14}%
\end{align}
$t_{FI}$ is the final of the s-world and the beginning moment. $\left(
K_{cr}/R_{\min}^{2}\right)  $\ or $V_{0}$\ and $T_{cr}$\ are two new important
constants, and $T_{\max}$\ and $\rho_{\max}$\ are determined by $R_{\min},$
$R_{cr}$ and $T_{cr}$. It is seen from $\left(  19.9\right)  -\left(
19.11\right)  $\ that all $R$, $T$ and $\rho$ must be finite in this case. The
meanings of the parameters are that when $T=T_{cr},$ $\langle\omega_{s}%
\rangle=\langle\omega_{v}\rangle=0$ and $R=R_{cr},$ and when $R=R_{\min}$,
$T=T_{\max}$ or $\rho=\rho_{\max}$ and $\overset{\cdot}{R}=0.$

We know that the duration of inflation $\tau$ must be finite, because
$T$\ will descend extremely fast due to inflation and the inflation process is
a supercooled process. When $T<T_{cr},$ $V=V_{0}$ is the maximum so that the
phase transition (i.e. reheating process) must occur. After $\tau,$ $R$ has a
large enough increase.

$R_{cr}\geq\sqrt{K_{cr}/\eta V_{0}}$ in $\left(  19.8\right)  $\ is the
condition of space inflation. By $\left(  19.1\right)  -\left(  19.3\right)  $
and $\left(  19.9\right)  -\left(  19.11\right)  ,$ considering $V_{0}=\mu
^{4}/4\lambda$ and $K_{cr}>0,$ we can rewrite $R_{cr}\geq\sqrt{K_{cr}/\eta
V_{0}}$ as
\begin{align}
\left(  TR\right)  ^{2}  &  =\left(  T_{cr}R_{cr}\right)  ^{2}\geq\frac
{K_{cr}}{\eta}\left(  \frac{4}{\mu}\right)  ^{2}\text{ \ or }\nonumber\\
D_{s}  &  =\rho_{cr}R_{cr}^{4}\geq g^{\ast}\frac{\left(  \pi K_{cr}\right)
^{2}}{30\eta^{2}}\left(  \frac{4}{\mu}\right)  ^{4}\equiv D_{cr}. \tag{19.15}%
\end{align}
If $R\left(  T_{cr}\right)  <\sqrt{K_{cr}/\eta V_{0}}$ or $D_{s}<D_{cr},$ this
implies that $\overset{\cdot}{R}=0$ already occurs before $R$ contracts to
$R_{cr}$ or $T_{s}$ rises to $T_{cr},$ i.e., $R_{\min}>R_{cr}$ and $T\left(
R_{\min}\right)  =T_{\max}<T_{cr}.$ Consequently $T_{cr}$ and $R_{cr}$ cannot
be reached. there are still $\langle\omega_{s}\left(  T_{s}\right)
\rangle\neq0$ and $\langle\omega_{v}\rangle=0$. In this case, there are still
$\langle\omega_{s}\left(  T_{s}\right)  \rangle\neq0$ and $\langle\omega
_{v}\rangle=0,$ and all $R_{\min}$, $T_{\max},$ $\rho_{g},$ $\rho_{s}$ and
$\rho_{v}$ must still be finite, i.e. there is no space-time singularity. In
this case, it is necessary that
\begin{equation}
\overset{\cdot}{R}=0,\text{ \ \ }\overset{\cdot\cdot}{R}>0,\text{ \ when
}R=R_{\min}>R_{cr}, \tag{19.16}%
\end{equation}
because $R_{\min}$ is the end of contracting $R$. In this case, space will
expand still in the $S-breaking$, but space inflation cannot occur.

We see from mentioned above that in any case of the contracting process, there
must be $R_{\min}>0$ and the finite $T_{\max}.$ Because of the cosmological
principle, all $\rho_{s},$ $\rho_{v},$ $\widetilde{\rho}_{Sg}=\widetilde{\rho
}_{s}-\widetilde{\rho}_{v},$ $\widetilde{p}_{Sg}=\widetilde{p}_{s}%
-\widetilde{p}_{v}$ and $p_{a}\leqslant\rho_{a}/3$ are finite. Hence all
$T_{s\mu\nu},$ $T_{v\mu\nu}$ and $T_{g\mu\nu}=T_{s\mu\nu}-T_{v\mu\nu}$ are
finite. Substituting the finite $T_{g\mu\nu}$ into the Einstein field equation
$\left(  2.13\right)  ,$ we see that $R_{\mu\nu}$ and $g_{\mu\nu}$ must be finite.

According to the present model, the universe is composed of infinite s-cosmic
islands and v-cosmic islands. The cosmic islands are different from each
other. Hence every possibility can be realized. There must be no singularity
in any case.

Thus, we have proved that there is no singularity in present model.

The process of space inflation is the same as that in the first model (see
section $7.2$).

\section{Evolving process of space after inflation}

We discuss the expanding process after reheating as follows. After reheating
process, expansion of space can be divide into three stages. The three stages
are the early stage, the stage from $\rho_{g}\sim0$ to $\overset{\cdot}{R}%
\sim0$ and the final stage in which space expands with an acceleration.

\subsection{The three stage of the universe evolution}

1. The reheating process and the early stage in which space expands with a deceleration

As the same as the first model, there is the reheating process after
inflation. $\left(  7.1\right)  -\left(  7.3\right)  $ is still applicable for
the second model.

After the reheating, $V_{g}=0.$ $K,$ $\overset{\cdot}{K}$ and $r\partial
K/\partial r$ may be neglected because $\underline{\rho}_{g}=\underline{\rho
}_{v}-\underline{\rho}_{s}\gg\underline{\rho}_{g}/A_{\rho}$ in the early
stage. Thus, $\left(  18.24\right)  -\left(  18.25\right)  $ and $\left(
18.30\right)  $ are reduced to $\left(  18.47\right)  -\left(  18.49\right)  $
in which $V_{g}=0$. Thus, space must expand with a deceleration in the early
stage after reheating.

2. The stage from $\rho_{g}\left(  t_{2}\right)  \sim0$ to $\overset{\cdot}%
{R}\left(  t_{1}\right)  \sim0.$

After temperature descends sharply, $\rho_{g}$ will decrease to $\rho
_{g}\left(  t_{2}\right)  =0$ and finally $\rho_{g}<0,$ because $\rho
_{mg}=\rho_{vm}-\rho_{sm}\sim R^{-3},$ $\rho_{\lg}=\rho_{vl}-\rho_{sl}\sim
R^{-3\left(  1+\kappa\right)  }$ and $\rho_{\gamma g}=\rho_{v\gamma}\sim
R^{-4}$. When temperature is so low that $p_{l}\ll\rho_{l},$ we may rewrite
$\rho_{v}=\rho_{vm}+\rho_{v\gamma}$ and $\rho_{s}=\rho_{sm}$. The stage from
$\rho_{g}\left(  t_{2}\right)  \sim0$ to $\overset{\cdot}{R}\left(
t_{1}\right)  \sim0$ ($t_{1}>t_{2}$) has been discussed in section $18.2$ (see
$\left(  18.52\right)  -\left(  18.59\right)  $). In contrast with the
conventional theory, both $\rho_{g}\left(  t_{1}\right)  <0$ and
$\overset{\cdot\cdot}{R}\left(  t_{1}\right)  >0$ are possible when
$\overset{\cdot}{R}\left(  t_{1}\right)  =0$ so that space will expand with an
acceleration up to now.

3. The final stage in which space expands with an acceleration

Space will continue to expand because $\ddot{R}\left(  t_{1}\right)  >0$ after
$\overset{\cdot}{R}\left(  t_{1}\right)  =0.$ Thus, $\rho_{g}\left(  t\right)
$ will continue to decrease from $\rho_{g}\left(  t_{1}\right)  <0$ as $R$
increases. Temperature will decrease as $R$ increases. When temperature is so
low that $p_{\alpha}\ll\rho_{\alpha},$ we may rewrite $\rho_{v}=\rho_{vm}%
+\rho_{v\gamma}$ and $\rho_{s}=\rho_{sm}$. Considering $\rho_{mg}\sim R^{-3}$
and $\rho_{\gamma g}\sim R^{-4}$, we obtain $\underline{\rho}_{g}$ determined
by $\left(  18.65\right)  $. Thus $\underline{\rho}_{g}\simeq\underline{\rho
}_{mg1}$ when $R_{1}/R\ll1$ due to $\left(  18.65\right)  $ and $p_{g}\sim
\rho_{\gamma g}/3$ may be neglected. Consequently, $\overset{\cdot}{K}\sim0,$
$\overset{\cdot\cdot}{K}\sim0,$ $v\left(  \rho_{g}+p_{g}\right)  \sim0$ and
$r\partial K/\partial r\sim0,$ and $\left(  18.24\right)  -\left(
18.25\right)  $ and $\left(  18.30\right)  $ are reduced to
\begin{equation}
\dot{R}^{2}=-K+\eta\rho_{g}R^{2},\text{ }\ddot{R}=-\frac{\eta}{2}\rho
_{g}R>0,\text{ }K\sim-1-D_{0} \tag{20.1}%
\end{equation}%
\begin{equation}
\overset{\cdot}{\rho}_{g}+3\rho_{g}\overset{\cdot}{R}/R=0. \tag{20.2}%
\end{equation}
Thus, space will expand with an acceleration when $t>t_{1}$.

Sum up, we see that after reheating, space first expands with a deceleration,
then comes to static (in fact, it is $\overset{\cdot}{R}/R=\overset{\cdot}%
{(R}/R)_{\min}\geq0$), and finally expands with an acceleration up to now.

\subsection{Determination of $a\left(  t\right)  $ in the second model}

The exact solutions of $K\left(  t,r\right)  ,$ $v\left(  t,r\right)  $,
$R\left(  t,r\right)  $, $\rho_{g}\left(  t,r\right)  $ and $p_{g}\left(
t,r\right)  $ are hardly determined. Hence we have to discuss the asymptotic
solutions. The asymptotic solution of $K\left(  t,r\right)  $ had been
discussed in section $18.2$. We discuss the asymptotic solutions of $R\left(
t,r\right)  $ and $\rho_{g}\left(  t,r\right)  $ as follows.

The evolution equations have been derived in section $18.2$. From $\left(
18.38\right)  $ we see that $K\sim1-D_{0}>0$ when $A_{\rho}\underline{\rho
}_{g}\gg\underline{\rho}_{g0}$ and $K\sim-1-D_{0}<0$ when $A_{\rho}%
\underline{\rho}_{g}\longrightarrow-\underline{\rho}_{g0}$. In the two cases,
$\left(  18.24\right)  -\left(  18.27\right)  $ is reduced to $\left(
18.46\right)  -\left(  18.49\right)  $. When $K\sim0$, $\overset{\cdot}{K}$
cannot be neglected and $\left(  18.24\right)  -\left(  18.27\right)  $ is
dependent on $r.$ When $\underline{\rho}_{g}\sim0$ and $r\sim0,$ $K\sim-D_{0}$
and $\left(  18.24\right)  -\left(  18.25\right)  $ are reduced to $\left(
18.57\right)  $. When $\underline{\rho}_{g}\sim0$ and $r$ is large$,$
$K\sim-D_{0}/r^{2}$ and $\left(  18.24\right)  -\left(  18.25\right)  $ is
reduced to $\left(  18.52\right)  -\left(  18.53\right)  $. When
$\overset{\cdot}{R}\sim0,$ $0>K>-1-D_{0}$ and $\left(  18.25\right)  $ is
reduced to $\left(  18.59\right)  .$ The equations $\left(  18.52\right)
-\left(  18.53\right)  ,$ $\left(  18.57\right)  $ and $\left(  18.59\right)
$ are all independent of $r$.

After reheating, $\underline{\rho}_{g}$ and $K$ decrease as space expands. The
process is as follows. $K\sim1-D_{0},$ $\underline{\rho}_{g}\gg\underline
{\rho}_{g0}/A_{\rho}$ and $\overset{\cdot\cdot}{R}<0\longrightarrow
K=0,$\ $\underline{\rho}_{g}=\underline{\rho}_{g}^{\prime}>0$ and
$\overset{\cdot\cdot}{R}<0\longrightarrow K\sim-D_{0},$ $\underline{\rho}%
_{g}=0$ and $\overset{\cdot\cdot}{R}<0\longrightarrow K\gtrsim-1-D_{0},$
$\overset{\cdot}{R}=0$ and $\overset{\cdot\cdot}{R}>0\longrightarrow
K\sim-1-D_{0},$ $\overset{\cdot}{R}>0$ and $\overset{\cdot\cdot}{R}>0.$

In the early stage after reheating, $\rho_{g}$ is so large that $\eta\rho
_{g}R^{2}\gg K.$ Hence, in this case, $K$ cannot influence evolution of space
markedly so that $K$ may be taken as $-1$ for convenience. The two stages from
$K\sim0$ to $\rho_{g}\sim0$ and from $\rho_{g}\sim0$ to $\overset{\cdot}%
{R}\sim0$ are short. As mentioned above, $K\lesssim0$ in the two stages. As a
rough approximation for the whole evolving process, we take $K=-1$ in the two
stages. When $A_{\rho}\underline{\rho}_{g}+\underline{\rho}_{g0}\gtrsim0,$
$K\sim-1-D_{0},$ we may take $K=-1$ because $K/R^{2}$ describe the curvature
of space and is measurable. From the approximation we can see the main
evolving features of the universe. Based on the approximation, $\overset
{\cdot}{rK}=r\partial K/\partial r=0$ and $\left(  18.24\right)  -\left(
18.27\right)  $ is reduced to%

\begin{equation}
\dot{R}^{2}+K=\eta\rho_{g}R^{2},\text{ \ }K=-1, \tag{20.3}%
\end{equation}%
\begin{equation}
\ddot{R}=-\frac{\eta}{2}\left(  \rho_{g}+3p_{g}\right)  R,\text{ \ },
\tag{20.4}%
\end{equation}%
\begin{equation}
R^{2}\frac{d}{dt}\rho_{g}+3\left(  \rho_{g}+p_{g}\right)  R\overset{\cdot}%
{R}=0. \tag{20.5}%
\end{equation}
$\left(  20.3\right)  -\left(  20.5\right)  $ are the same as $\left(
4.5\right)  -\left(  4.6\right)  $ and $\left(  4.9\right)  $ in which
$V_{g}=\overset{\cdot}{V}_{g}=0$. Hence the inferences $\left(  7.10\right)
-\left(  7.23\right)  $ can be derived from $\left(  20.3\right)  -\left(
20.5\right)  .$

\section{Transformation of repulsive potential energy chiefly into the kinetic
energy of $SU(5)$ singlets}

In the section we consider the repulsion between s-matter and v-matter in the
flat space-time and the Newtonian mechanics.

The repulsion potential energy between $v-matter$ and $s-matter$ is determined
by the distributing mode of $s-matter$ and $v-matter.$ In $V-breaking,$
$v-particles$ with their masses can form $v-celestial$ bodies, but $s-SU(5)$
color single states cannot form any dumpling and must distribute loosely in
space$.$ Consequently, the huge repulsion potential energy must chiefly
transform into the kinetic energy of $s-SU(5)$ color single states when the
$v-celestial$ bodies form or space expands. In fact, when flat space expands
$N$ times, i.e., $R\longrightarrow NR,$ the repulsive-potential energy density
$V_{r}$ becomes $V_{r}/N$ and%
\begin{equation}
\bigtriangleup V_{r}=\left(  1-1/N\right)  V_{r}. \tag{21.1}%
\end{equation}
Consider a system in flat space which is composed of a $v-body$ with its mass
$M$ and a $s-colour$ single state with its mass $m.$ It is easy to get the
rate $\bigtriangleup E_{m}/\bigtriangleup E_{M}$ for static $M$ and $m$ at the
initial moment.
\begin{equation}
\frac{\bigtriangleup E_{m}}{\bigtriangleup E_{M}}=\frac{2M+\bigtriangleup
V_{r}}{2m+\bigtriangleup V_{r}}. \tag{21.2}%
\end{equation}
Because $M\gg m$, $\bigtriangleup E_{m}>\bigtriangleup E_{M}.$

Space expansion is not the necessary condition to transform repulsive
potential energy into kinetic energy. Supposing $\overset{\cdot}{R}=0$ and
some $v-matter$ gathers to a region and forms a galaxy, $s-color$ single
states, which are initially in the region, must be repulsed away from the
region by the celestial bodies in the galaxy. Consequently, the repulsive
potential energy chiefly transforms into the kinetic energy of the $s-matter$.

There are four of causes for $\overset{\cdot}{\underline{\rho}}_{g}\neq0:$

1. It is seen from $\left(  18.50\right)  $ that $\overset{\cdot}{K}\neq0$ can
cause $\overset{\cdot}{\underline{\rho}}_{g}\neq0.$ Of course, $\overset
{\cdot}{\underline{\rho}}_{g}\neq0$ is the cause of $\overset{\cdot}{K}\neq0.$

2. The change of $\underline{\rho}_{g}$ can be caused by the transformation of
$s-particles$ and $v-particles$ from one to another when temperature is high enough;

3. The expansion or contraction of space causes the change of $\underline
{\rho}_{g},$ because of $\rho_{sm}$ and that $\rho_{sm}\sim R^{-3}$ and
$\rho_{v\gamma}\sim R^{-4}.$

4. The repulsion potential chiefly transform to kinetic energy of the $SU(5)$
singlets when $\overset{\cdot}{R}\sim0$ and galaxies form.

One effect of the repulsion between $s-matter$ and $v-matter$ is that it can
prevent $v-galaxies$ are teared by expansion of space with an acceleration.
Any galaxy must undergo a pressure coming from s-matter surrounding the
galaxy. The pressure can prevent the $v-galaxy$ are teared by expansion of
space with an acceleration.

\section{New predictions}

\subsection{The second model can obtain the three predictions of the first
model and explain primordial, $CMBR$ and the large scale structure.}

The three predictions of the first model can be obtained by the second model
as well. For the third prediction, in contrast with the first model, after a
black hole with its mass and density big enough inflate, the black hole can
transit into either of the $S-breaking$ and the $V-breaking$.

Letting there be a $v-black$ hole with its mass and density to be so huge that
its temperature can arrive to $T_{v}\gtrsim T_{cr}=2\mu/\sqrt{\lambda}$ when
the black hole contracts by its self-gravitation, then the expectation values
of the Higgs fields inside the $v-black$ hole will change from $\varpi
_{v}=\varpi_{v0}$ and $\varpi_{s}=0$ into $\varpi_{v}=\varpi_{s}=0$.
Consequently, inflation must occur. After inflation, the most symmetric state
will transit into the $S-breaking$ or the $V-breaking$. No matter which
breaking appears, the energy of the black hole must transform into both
$v-energy$ and $s-energy$. Thus, a $v-observer$ will find that the black hole
disappears and a white hole appears.

As the same as the first model, the second model can explain primordial
nucleosynthesis, $CMBR$ and the large scale structure of space-time. As
mentioned in section $20.2$, after inflation and reheating, $V_{g}%
=\overset{\cdot}{V}_{g}=0,$ the evolution equation can approximately be
reduced to $\left(  20.3\right)  -\left(  20.5\right)  $ in which $K\sim-1.$
In the case, $\left(  20.3\right)  -\left(  20.5\right)  $ are the same as
$\left(  4.5\right)  -\left(  4.6\right)  $ and $\left(  4.9\right)  $ of the
first model. Hence the second model, as the same as the first model, can
explain primordial nucleosynthesis, $CMBR$ and the large scale structure as well.

\subsection{The universe is composed of infinite s-cosmic islands and v-cosmic
islands}

If there is one sort of spontaneous symmetry breaking, the hypersurfaces of
the universe must be complete. Hence $K$ cannot change and there is only one
sort of evolving mode in the whole universe. In contrast with the conventional
theory, there are two sorts of matter and the two sorts of breaking. The
$S-breaking$ and the $V-breaking$ are different in essence and symmetric.
Hence it is possible that some regions of the universe can exist in the
$S-breaking$, and the others can exist in the the -$V-breaking$. Consequently
the universe must be composed of $s-cosmic$ islands with the $S-breaking$ and
$v-cosmic$ islands with the $V-breaking$. Of course, a transitional region
($T-region$) must exist between a $s-island$ and a $v-island$.

Every cosmic island is only one part of the universe, must be finite and
evolves individually. Consequently its hypersurface of simultaneity must be
incomplete and its metric must be different with the RW-metric which describes
a complete hypersurface.

In the $s-islands,$ $\langle\omega_{s}\rangle=\langle\omega_{s}\rangle_{0}$
and $\langle\omega_{v}\rangle=0,$ $s-particles$ forms $s-galaxies$ and
$v-particles$ form $v-SU(5)$ singlets which corresponds to so-called dark
energy. The same results hold for $v-islands$ because of conjecture 1.

Through the transitional regions, $\langle\omega_{v}\rangle=\langle\omega
_{v}\rangle_{0}$ and $\langle\omega_{s}\rangle=0$ will transit to
$\langle\omega_{s}\rangle=\langle\omega_{s}\rangle_{0}$ and $\langle\omega
_{v}\rangle=0.$ Hence all $\langle\omega_{s}\rangle,$ $\langle\omega
_{v}\rangle,$ $V_{s}$ and $V_{v}$ are not equal to zero and the expectation
values $\langle\omega_{s}\rangle_{T}$ and $\langle\omega_{v}\rangle_{T}$
inside the $T-region$ must satisfy
\begin{equation}
0<\mid\langle\omega_{s}\rangle_{T}\mid<\mid\langle\omega_{s}\rangle_{0}%
\mid,\text{ \ \ }0<\mid\langle\omega_{v}\rangle_{T}\mid<\mid\langle\omega
_{v}\rangle_{0}\mid. \tag{22.1}%
\end{equation}
There must be only $v-islands$ neighboring a $s-island$. This is because that
if two $s-cosmic$ islands are neighboring, they must form one new larger
$s-cosmic$ island in order that $V=V_{\min}$.

Based on the following reasons, the probability is very little that a
$v-observer$ receives messages from a $s-island$:

$1$\textbf{.} The probability must be very small that a $s-particle$ (a quark,
a lepton or a photon) in a $s-island$ comes into a $v-island$, because a
$s-particle$ in the $s-island$ is non-color single state. If a $s-particle$
comes into the $v-island$, it would still be non-color single state. This is
impossible due to color confinement. Such a bound state as $\left(
u\overline{u}\mp d\overline{d}\right)  /\sqrt{2}$ is a $SU(5)$ color single
state in both $V-breaking$ and $S-breaking.$ It seems that it can comes into
the $v-cosmic$ island. In fact, it still hardly comes into the $v-island,$
because both masses are very different from each other. On the other hand,
such states must be unstable and will decay very fast in $s-island$.

$2$. The probability must be very small that a $v-particle$ in the $s-island$
come into the $v-island$ as well$.$ The $v-particle$ (a fermion or a gauge
boson) in the $s-cosmic$ island must be massless. If the $v-particle$ comes
into the $v-island$, its mass will change from $m_{0}=0$ to $m_{0}>0.$ Thus it
must suffer a strong-repulsive interaction, and hence it hardly comes into the
$v-cosmic$ island.

$3$. It is important that $\langle\omega_{s}\rangle_{T}\neq0$ and
$\langle\omega_{v}\rangle_{T}\neq0.$ Hence it is necessary that $m_{s}\neq
m_{T}$ and $m_{v}\neq m_{T}$, where $m_{s},$ $m_{v}$ and $m_{T}$ are the
masses of a particle when it is in a $s-island,$ a $v-island$ or a $T-region$,
respectively. Any particle coming into a $T-region$ must emit or absorb Higgs
particles in order to change its mass from $m_{s}$ or $m_{v}$ to $m_{T},$ i.e.%
\begin{equation}
f\longrightarrow f_{T}+h,\text{ \ }h\longrightarrow g_{T}+\overline{g}%
_{T},\text{ \ }f+h\longrightarrow f_{T}, \tag{22.2}%
\end{equation}
where $f$ is a $s-$ or $v-particle$ coming into a $T-region$, $f_{T}$ is the
particle $f$ with its mass $m_{T}$ in the $T-region,$ $h$ is a $s-$ or
$v-Higgs$ particle, and $g_{T}$ and $\overline{g}_{T}$ are any particle and
its anti-particle. The process hardly occur, because annihilation and creation
of $h$ will alter the Higgs potential.

$4$. Higgs particles in the $s-cosmic$ island must decay very fast into
fermions or gauge bosons, and hence they cannot come to the $v-cosmic$ island.

In summary, any particle hardly go to another cosmic island. Particles coming
to a $T-region$ will be reflected by the $T-region.$ Thus a galaxy inside a
cosmic island is similar to an electron static in a metal sphere shell.

A $v-cosmic$ island and a $s-cosmic$ island can influence each other by the
Higgs potential in the $T-region$ between both.

An observer in the cosmic island can regard the cosmic island as the whole
cosmos. It is possible that some cosmic islands are forming or expanding, and
the other cosmic islands are contracting.

Thus, according to the present model the cosmos as a whole is infinite and its
properties are always unchanging, and there is no starting point or end of
time. 

The sketch of the universe according to the second model is shown in figure 3.
Figure 3

\subsection{Mass redshifts}

Hydrogen spectrum is
\begin{align}
\omega_{nk}  &  =(E_{n}-E_{k})/\hbar=-\frac{\mu e^{4}}{2\hbar^{3}}(\frac
{1}{n^{2}}-\frac{1}{k^{2}}),\tag{22.3}\\
\mu &  =\frac{mM}{m+M}, \tag{22.4}%
\end{align}
where $m$ is the mass of an electron, and $M$ is the mass of a proton.
According the unified model, $m\propto\upsilon_{e}$, the mass of a quark
$m_{q}\propto\upsilon_{q},$ and $M\propto m_{q},$ where $\upsilon_{e}$ and
$\upsilon_{q}$ are the expectation values of the Higgs fields coupling with
the electron and the quark, respectively.

If there are some galaxies inside a $T-region,$ from $\left(  22.1\right)  $
we see that the mass $m_{T}$ of an electron and the mass $M_{T}$ of a proton
inside the $T-region$ must be
\begin{equation}
m_{T}<m,\text{ \ \ }M_{T}<M. \tag{22.5}%
\end{equation}
Thus we have
\begin{align}
\mu_{T}  &  <\mu,\tag{22.6}\\
\bigtriangleup\omega_{nk}  &  =\omega_{nk}-\omega_{nkT}=-\frac{(\mu-\mu
_{T})e^{4}}{2\hbar^{3}}(\frac{1}{n^{2}}-\frac{1}{k^{2}})<0. \tag{22.7}%
\end{align}

This sort of red-shifts is called $mass$ $redshift$. The mass redshift is
essentially different from the cosmological red-shift mentioned before. Thus,
the photons coming from the star in a $T-region$ must have larger red-shift
than that determined by the Hubble formula at the same distance. Thereby we
guess that some quasars are just the galaxies in the $T-region$ and they have
the mass redshifts. The fine-structure constant is considered to be changeable
based on the redshifts of some quasars. In contrast with the guess, we
consider that the phenomenon is possibly because the mass of electrons
changes, but is not because the fine-structure constant changes.

An ordinary $s-galaxy$ and a $s-quasar$ can be neighboring, because a
$T-region$ must be neighboring to an ordinary region.

\section{Conclusions of the model based on a new metric}

Based on the two conjectures in the first model and a new metric which
describes the incomplete hypersurface, the second sort of the cosmological
models without singularity is constructed. The inferences of the first model
can be derived from the present model as well. There is no singularity in the
model and the cosmological constant $\lambda=\lambda_{eff}=0$ is determined
although $\rho_{0}$ is still very large and there is no the fine tuning
problem, even if $\lambda_{eff}\neq0$.

Dark energy is explained as $SU(5)$ singlets. This is because the $SU(5)$
singlets can only distribute loosely in space or form huge superclusters,
cannot be observed and can cause space to expand with an acceleration$.$ In
contrast with the dark energy, the gravitational masses of the $SU(5)$
singlets is negative, although their masses are all positive.

Space in the $V-breaking$ has three evolving stages: Space first expands with
a deceleration because $\rho_{g}=\rho_{v}-\rho_{s}>0$ and $K>0$; then comes to
static; and finally expands with an acceleration up to now because $\rho
_{g}<0$ and $K<0$.

There are the critical temperature $T_{cr}$, the highest temperature $T_{\max
}$, the least scale $R_{\min}$ and the largest energy density $\rho_{\max}$ in
the universe. $V_{0}$ or $\left(  K_{cr}/R_{\min}^{2}\right)  $ and $T_{cr}$
are two new important constants, and $T_{\max}$ and $\rho_{\max}$ are
determined by $R\left(  T_{cr}\right)  $.

The three new predicts of the first model have been obtained as well. Besides,
there are two new predictions in the present model.

The universe is composed of infinite $s-cosmic$ islands and $v-cosmic$
islands. Every cosmic island is finite and evolves individually. The
hypersurface of every cosmic island must be incomplete. Hence the
$K-changeable$ metric is presented. An observer in the cosmic island can
regard the cosmic island as the whole cosmos. It is possible that some cosmic
islands are forming or expanding, and the other cosmic islands are
contracting. Thus, according to the present model the cosmos as a whole is
infinite and its properties are always unchanging, and there is no starting
point or end of time.

There is $mass$ $redshift,$ because $m_{eT}<m_{e0}$, here $m_{eT}$ is the mass
of an electron in a transitional region and $m_{e0}$ is the given mass of an
electron. We guess that some quasars are just the galaxies in a transitional
region and they have the mass redshifts. The fine-structure constant is
considered to change based on the redshifts of some quasars according to the
conventional theory. In contrast with the guess, we consider that the
phenomenon is possibly because the mass of electrons changes, but is not
because the fine-structure constant changes.

An ordinary $s-galaxy$ and a $s-quasar$ can be neighboring, because a
$T-region$ must be neighboring to an ordinary region.

There is no physical restriction for $T_{\mu\nu}$ originating from the field
equation. Hence we can suppose $T_{,\nu}^{\mu\nu}=0,$ where $T^{\mu\nu}$ may
contain contribution gravitational field.

\section{Summation}

Essential new conjecture 1 is presented. Although conjecture 1 corrects the
equivalence principle (the principle is violated by the $SU(5)$ singlets), but
it does not contradict experiments and observations up to now.

Based on conjecture 1, two cosmological models without singularity are
constructed. There is no singularity in the two models. Based on either of the
two models, the cosmological constant issue is solved, i.e. $\lambda
=\lambda_{eff}=0;$ Dark energy is explained as $SU(5)$ singlets; The evolution
of the universe is explained well and the distance-redshift relationship is
derived out which is consistent with observations up to now; The primordial
nucleosynthesis, $CMBR$ and the large scale structure of space-time are
explained; The five predictions are given (see parts 1 and 2). According to
either of the two models, there are the critical temperature $T_{cr}$, the
highest temperature $T_{\max}$, the least scale $R_{\min}$ and the largest
energy density $\rho_{\max}$ in the universe, and $V_{0}$ or $\left(
K_{cr}/R_{\min}^{2}\right)  $ and $T_{cr}$ are two new important constants in
the two models.

In contrast with the first model, according to the second model, the universe
is composed of infinite $s-cosmic$ islands and $v-cosmic$ islands; Every
cosmic island is finite and evolves individually; The cosmos as a whole is
infinite and its properties are always unchanging, and there is no starting
point or end of time. There is $mass$ $redshift$ which is different from the
cosmological redshift.

\acknowledgments I am very grateful to professor Zhao Zhan-yue, professor Wu
Zhao-yan, professor Zheng Zhi-peng and professor Zhao zheng-guo for their
helpful discussions and best support. I am very grateful to professor Liu
Yun-zuo, professor Lu Jingbin, doctor Yang Dong and doctor Ma Keyan for their
helpful discussions and help in the manuscript.

\clearpage


\begin{thebibliography}{99}                                                                                               %


\bibitem {[1]}Hawking S. W. and Ellis G. F. R., 1999, The Large Scale
Structure of Space-Time, Cambidge University Press, p7, 98, 101, 137, 256-298.

\bibitem {[2]}Caldwell R.R. 2004, Phys. World \textbf{17}, 37; Padmanabhan, T.
2003, Phys. Rep. \textbf{380} 325; Peebles P.J.E. and Ratra B. 2003, Rev. Mod.
Phys. \textbf{75} 559.

\bibitem {[3]}Weinberg S. 1987, Phys. Rev. Lett., \textbf{59} 2607; Martel H.,
Shapiro P.R. and Weinberg S. 1998, Astrophys. J \textbf{492} 29.

\bibitem {[4]}Peebles P.J.E. and Ratra B. 1988, Astrophys. J. \textbf{325}
L17; Ratra B. and Peebles, P.J.E. 1988, Phys. Rev. D \textbf{37} 3406; Peebles
P.J.E. and Ratra B. 2003, Rev. Mod. Phys. \textbf{75} 559.

\bibitem {[5]}Hall L.J., Nomura Y. and Oliver S.J., Phys. 2005, Rev. Lett.
\textbf{95, }141302.

\bibitem {[6]}Chen S-H , Quantum Field Theory Without Divergence A 2002,
hep-th/0203220; 2005, Quantum Field Theory: New Research, Kovras O. Editor,
Nova Science Publishers, Inc. 103-170.

\bibitem {[7]}Chen S-H. 2001, A Possible Candidate for Dark Matter,
hep-th/0103234; 2005, Progress in Dark Matter Research, Editor: J. Val Blain,
Nova Science Publishers, 65-72. Inc. arXiv: 1001.4221.

\bibitem {[8]}Peacock J. A. 1999, Cosmological Physics, Cambridge University
Press, 579, 458, 78, 90, 89, 460-464, 664, 296.

\bibitem {[9]}Chen S-H. 2009, Discussion of a Possible Universal Model without
Singularity, arXiv. 0908.1495v2; 2006, A Possible Universal Model without
Singularity and its Explanation for Evolution of the Universe, hep-ph/0611283.

\bibitem {[10]}Gibbons G. W. and Hawking, S. W. 1977, Phys. Rev D, 15, 2752.

\bibitem {[11]}Chaichian M. and Nelipa N.F. 1984, Introduction to Gauge Field
Theories, Springer-Verlag Berlin Heidelberg, 269; Ross G. G. 1984, Grand
Unified Theories, The Benjamin/Cummings Publishing Company, INC, 177-183.

\bibitem {[12]}Weinberg S. 1972, Gravitation and Cosmology, New York, Wiley
Chanter 12 section 3.

\bibitem {[13]}Coleman S. and Weiberg E.J. 1973, Phys. Rev. D7, 1888;
Brandenberg R.H.\ 1985, Rev. of Mod. Phys., 57. 1.

\bibitem {[14]}Liu L. Jiang Y. and Qian Z. 1989, Progress in Physics, V.9, No.
2, 159 [Use the language of Chinese].

\bibitem {[15]}Guth A. H., 1981, Phys. Rev. D 23, 347.

\bibitem {[16]}Ignatiev A. Y. and Volkas R. R., Phys. Rev. (2003) D 68,
023518; Berezhiani Z., Comelli D. and Villante F.L., Phys. Lett. (2001) B 503, 362.

\bibitem {[17]}Yu Yunqiang., Lectures in Cosmological Physics (Chinese),
Peking University Press, (2002), 151, 170-172.

\bibitem {[18]}Liddle A. R. and Lyth D. H., 2000, Cosmological Inflation and
Large-Scale Structure, Cambridge University Press, 20, 248.

\bibitem {[19]}Liang C.B and Zhou B, 2006, The Course of Differential Geometry
and General Relativity (Chinese), Second Edition, (China, Science Press) p 360-367.

\bibitem {[20]}Ohanian H.C. and Ruffini R., (1994), Gravitation and Spacetime
(2nd ed.), W.W. Norton and Company, Inc., Section 9.9.
\end{thebibliography}
\end{document}